\documentclass[10pt]{article}
\usepackage{graphicx} 
\usepackage{amsfonts}
\usepackage{amsmath}
\usepackage{amsthm}
\usepackage[letterpaper, margin=1in]{geometry}
\usepackage{setspace}
\usepackage{bbm}
\usepackage{xcolor}
\usepackage{comment}
\usepackage[
  colorlinks=true,
  linkcolor=blue,
  citecolor=blue,
  urlcolor=blue,
  pdfborder={0 0 0}
]{hyperref}
\usepackage{enumitem}
\usepackage[authoryear,round]{natbib}
\usepackage[ruled,vlined]{algorithm2e}
\usepackage{amssymb} 
\usepackage{dsfont}
\usepackage{subcaption}
\usepackage{graphicx}
\usepackage{subcaption}
\usepackage{arydshln}
\usepackage{authblk}

\usepackage{newtxtext} 
\usepackage{newtxmath} 

\usepackage[utf8]{inputenc} 
\usepackage[T1]{fontenc}    
\usepackage{hyperref}       
\usepackage{url}            
\usepackage{booktabs}       
\usepackage{amsfonts}       
\usepackage{nicefrac}       
\usepackage{microtype}      
\usepackage{xcolor}         

\usepackage{amsmath}
\usepackage{amsthm}
\usepackage{bbm}
\usepackage{enumitem}

\newtheorem{theorem}{Theorem}[section]
\newtheorem{corollary}[theorem]{Corollary}
\newtheorem{lemma}[theorem]{Lemma}

\newtheorem{proposition}[theorem]{Proposition}

\newtheorem{definition}[theorem]{Definition}

\usepackage{mathtools}

\usepackage{mathrsfs}
\usepackage{newtxmath}
\usepackage{multirow}
\usepackage{booktabs}
\DeclareMathAlphabet{\mathcalEU}{U}{eus}{m}{n}

\newcommand{\alg}{\mathrm{ALG}}
\newcommand{\opt}{\mathrm{OPT}}
\newcommand{\rr}{\textsf{RR}}

\newcommand{\mTSS}{\textsf{mTSS}}

\DeclareSymbolFont{myletters}{OML}{ntxmi}{m}{it}
\DeclareMathSymbol{\rob}{\mathord}{myletters}{13}
\DeclareMathSymbol{\con}{\mathord}{myletters}{12}

\newcommand{\eps}{\varepsilon}

\newcommand{\mA}{\mathcalEU{A}}
\newcommand{\mD}{\mathcal{D}}

\newcommand{\mI}{\mathcal{I}}
\newcommand{\mX}{\mathcal{X}}

\newcommand{\vg}{\Phi}

\newcommand{\mx}{\mathbf{x}}

\newcommand{\ml}{\mathscr{l}}

\title{%
  \normalfont\fontfamily{ptm}\fontsize{17}{21}\selectfont
  Near-Optimal Consistency--Robustness Bounds for\\
  Non-Clairvoyant Scheduling
}

\author{
\makebox[\textwidth][c]{%
\begin{minipage}[t]{0.32\textwidth}
\centering
Sizhe Li\thanks{Most of this work was done while the author was at CUHK(SZ).}\\
Johns Hopkins University\\
\texttt{sli357@jhu.edu}
\end{minipage}\hfill
\begin{minipage}[t]{0.32\textwidth}
\centering
Nicolas Christianson\\
Johns Hopkins University\\
\texttt{christianson@jhu.edu}
\end{minipage}\hfill
\begin{minipage}[t]{0.32\textwidth}
\centering
Tongxin Li\thanks{Corresponding author.}\\
CUHK(SZ)\\
\texttt{litongxin@cuhk.edu.cn}
\end{minipage}%
}
}

\date{}

\begin{document}

\maketitle

\begin{abstract}
We study the problem of characterizing the optimal consistency--robustness tradeoff for single-machine non-clairvoyant scheduling with predicted processing times.
For a fixed number of jobs $n$,~\cite{Wei2020} established a lower bound on the achievable consistency--robustness tradeoff which is known to be tight only in a few restrictive cases, leaving the optimal tradeoff for general $n$ unresolved.
A different \emph{worst-case-$n$ regime} was introduced
by~\citet{Benomar2025ProgressBars}, who applied the former work's construction to show that for $\lambda \in (0, 1)$, any $(1+\lambda)$-consistent algorithm must be $1 + \Omega(\sqrt{1/\lambda})$-robust. On the other hand, the best known upper bound gives $(1+\lambda)$-consistency and $1+O(1/\lambda)$-robustness, leaving a polynomial gap.
In this work, we develop a new lower bound using a multi-stage adversarial construction and a new algorithm that together give a near-optimal characterization of the
consistency--robustness tradeoff:
for every $n$, our upper bound is within a multiplicative factor $3/2$ of our lower bound, and thus our algorithm is within a multiplicative factor $3/2$ of the online optimal. In particular, our lower bound shows that any $(1+\lambda)$-consistent algorithm must have robustness at least $1/\sqrt{\lambda}$ as $n \to \infty$, while our algorithm achieves robustness
at most $\tfrac12+\sqrt{\smash{\tfrac54}+1/\lambda}$
for any $n$, matching the lower bound in the leading term as $\lambda \downarrow 0$;
our upper bound also closes the polynomial gap
in the worst-case-$n$ regime. 
We additionally show how our algorithm can be made error-tolerant, preserving the near-optimal consistency-robustness tradeoff while incurring only slightly degraded performance under bounded prediction errors.
\end{abstract}

\section{Introduction}

The framework of \emph{learning-augmented algorithms} (also known as \emph{algorithms with predictions})~\citep{mitzenmacherAlgorithmsPredictions2021} enables the integration of machine-learned predictions into the design and analysis of algorithms, extending the classic competitive analysis framework~\citep{Borodin2005}.
Central to this framework is the study of the Pareto frontier between \emph{consistency}, which measures algorithm performance when predictions are perfect, and \emph{robustness}, which measures performance under potentially adversarial prediction error.
While the optimal tradeoff between these metrics has been characterized (either exactly or asymptotically) for a number of problems including ski rental~\citep{Wei2020}, online search~\citep{Sun2021}, and metrical task systems \citep{Christianson2023}, it remains an open question for the classical problem of single-machine non-clairvoyant scheduling with total completion time objective \citep{Motwani1994}, despite the significant attention this problem has received in the learning-augmented algorithms literature (e.g., \citep{Kumar2018,Wei2020,Benomar2025ProgressBars}) and the problem's deep connections with applications in computer systems and beyond.

In this work, we seek to resolve this gap and develop near-optimal learning-augmented algorithms for this problem. In particular, we will consider both lower and upper bounds that depend on the number of jobs $n$. We parameterize the algorithm's consistency-robustness tradeoff (see Definition~\ref{def:consistency_robustness} for formal definitions) by a hyperparameter $\lambda > 0$; the open question is the characterization of the optimal robustness that can be achieved by any $(1+\lambda)$-consistent algorithm. 

Prior to this work, \citet{Wei2020} established the best known lower bound for this problem, showing that
any $(1+\lambda)$-consistent algorithm must have robustness at least
\begin{equation} \label{eq:wei-lower-bound}
\rob(n, \lambda) \geq \Phi_{\text{wz}} (n, \lambda) \coloneq \frac{n + n(n+1)\lambda}{1 + \frac{1}{2}\lambda (n+1)(n+2)} \qquad \text{for } \lambda \in \left[0, \tfrac{n-1}{n+1}\right].
\end{equation} 
They also proposed an algorithm, \textit{two-stage schedule}, which achieves $(1+\lambda)$-consistency and $(1+\tfrac{1}{1+6\lambda})$-robustness in the restricted setting of $n=2$ jobs, matching their lower bound \eqref{eq:wei-lower-bound} in this case; their lower bound is also tight at the two endpoints $\lambda = 0$ and $\lambda = \tfrac{n-1}{n+1}$ for any $n$. 
\textcolor{black}{However, the lower bound in~\eqref{eq:wei-lower-bound} fails to exactly characterize the optimal robustness for general $n$: we show in Section~\ref{sec:warmup} that this bound is loose in general. }

\begin{figure*}[t]
\centering
\includegraphics[width=\linewidth]
{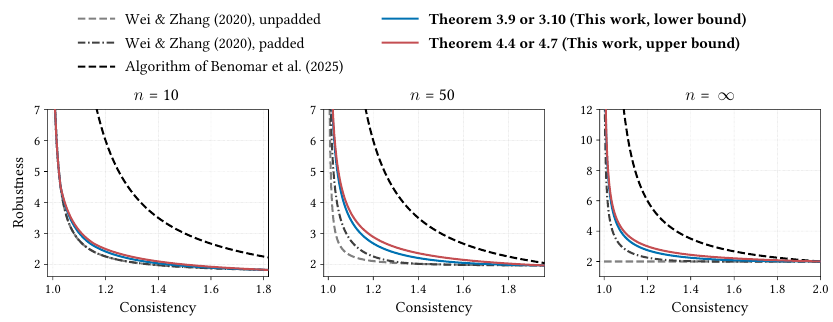}
    
\vspace{1em}

\begingroup
\small
\renewcommand{\arraystretch}{1.3}
\setlength{\tabcolsep}{8pt}
\begin{tabular*}{0.912\linewidth}{@{\extracolsep{\fill}}l*{7}{c}@{}}
\toprule
\textbf{Number of jobs}, $n$ &
$<5$ & 5 & 10 & 20 & 50 & 100 & $\infty$ \\
\midrule
Maximum gap ratio &
1.0000 & 1.0053 & 1.0357 & 1.0605 & 1.0789 & 1.0861 &
\textbf{1.0937} \\
\bottomrule
\end{tabular*}
\par
\endgroup
\vspace{-3pt}
    
    \caption{\textbf{Consistency-Robustness Bounds.} 
    \textbf{Top:} Comparison of our lower (Theorems~\ref{thm:piece-wise_lower_bound} and \ref{thm:lower_n->infty}) and upper (Theorems~\ref{thm:mTSS_conrob} and~\ref{thm:analytic_upper}) bounds against prior results for various $n$. 
    We show both the original lower bound of \citet{Wei2020}
and its optimally padded version; see
Section~\ref{sec:warmup} for details of padding.
    \textbf{Bottom:} The maximum multiplicative gap (ratio) between our upper and lower bounds across all $\lambda$, computed numerically. Notably, the gap for \boldmath{$n=\infty$} converges to approximately \textbf{1.0937}, reflecting the gap between our asymptotic lower and upper bounds in Theorems~\ref{thm:lower_n->infty} and~\ref{thm:analytic_upper}; we rigorously establish the looser bound $3/2$ on the gap for all $n$ in Theorem~\ref{thm:structural-three-halves}.} 
    \label{fig:combined_tradeoff}
\end{figure*}

On the other hand, for a general number of jobs $n$,~\citet{Kumar2018} proposed an algorithm called \textit{preferential round robin}, which achieves $(1+\lambda)$-consistency and $(2 + \frac{1}{\lambda})$-robustness;\footnote{The tradeoff is stated in~\cite{Kumar2018} as $(1+\lambda)/(2\lambda)$-consistency and $2/(1-\lambda)$-robustness, which is equivalent up to 
reparameterization.} this tradeoff was subsequently improved by \citet{Benomar2025ProgressBars}, who proposed an algorithm achieving $(1+\lambda)$-consistency and $(1+\tfrac{1}{\lambda})$-robustness.\footnote{The algorithm of \citet{Benomar2025ProgressBars} has an additional parameter $\rho\in(0,1]$. Taking $\rho=1$ in their Theorem~2.4 gives this bound and minimizes the stated robustness bound for fixed consistency.} 
\citet{Benomar2025ProgressBars} also initiated the study of a slightly different regime in which the number of jobs $n$ is allowed to depend on the consistency parameter $\lambda$ and is chosen to yield the worst possible tradeoff; we call this the \emph{worst-case-$n$ regime}. In this setting, they adapted the lower bound construction of \citet{Wei2020} to show that, for any $\lambda \in (0,1)$, there exists a choice of $n$ such that any $(1+\lambda)$-consistent algorithm incurs robustness at least $1 + \Omega\bigl(\sqrt{1/\lambda}\bigr)$. On the other hand, the best known upper bound in this regime remains $(1+\lambda)$-consistency and $1 + O(1/\lambda)$-robustness \citep{Kumar2018, Benomar2025ProgressBars}, leaving open a polynomial gap of $\Theta(\lambda^{-1/2})$.

Motivated by these gaps, our work seeks to answer the following questions:
\begin{center}
    \textit{Can we improve the lower and upper bounds on the consistency-robustness tradeoff for non-clairvoyant scheduling with predictions for any number of jobs $n$? In particular, can we develop an algorithm obtaining a near-optimal tradeoff for any $n$, and close the polynomial gaps on robustness in the worst-case-$n$ and $n \to \infty$ regimes?}
\end{center}

\subsection{Contributions}

This paper makes contributions along two fronts: improved lower and upper bounds on consistency and robustness for the non-clairvoyant scheduling problem, and new technical ideas that enable these improved bounds. 

\begin{table*}[t]
\centering
\caption{Summary of state-of-the-art bounds for non-clairvoyant scheduling with predictions. Entries denote lower/upper bounds on the robustness of a $(1+\lambda)$-consistent algorithm, where $\lambda \in [0, \frac{n-1}{n+1}]$ is a tradeoff parameter. All bounds in the table are contributions of this work except the worst-case-$n$ lower bound.
}
\label{tab:regime_bounds_tight}

\small
\renewcommand{\arraystretch}{1.3} 
\setlength{\tabcolsep}{8pt}       

\begin{tabular}{@{} l l l @{}}
\toprule
\textbf{Setting} & \textbf{Lower Bound} & \textbf{Upper Bound} \\ 
\midrule

\textbf{Fixed $n$} (Finite) & 
Implicit$^\star$ \hfill {(Theorem~\ref{thm:piece-wise_lower_bound})} & Explicit Finite Maximization$^\star$ \hfill {(Theorem~\ref{thm:mTSS_conrob})} \\ 

\textbf{Fixed $n$} ($n \to \infty$) & 
$\hat{\gamma}(\lambda) \geq \sqrt{{1/\lambda}}$ \textsuperscript{$\dagger$} \hfill {(Theorem~\ref{thm:lower_n->infty})} & 
$\frac{1}{2} + \sqrt{\frac{5}{4}+{1}/{\lambda}}
$ \hfill {(Theorem~\ref{thm:analytic_upper})} \\ 

\textbf{Worst-case-$n$} & 
$1+\Omega(\sqrt{1/\lambda})$ \hfill { \citep{Benomar2025ProgressBars}} & 
$1+O(\sqrt{1/\lambda})$ \hfill {(Theorem~\ref{thm:analytic_upper})} \\ 

\bottomrule
\addlinespace[3pt]
\multicolumn{3}{@{}l@{}}{\small
$^{\star}$ For every $n\geq2$ and admissible $\lambda$,
the two bounds differ by a factor of at most $3/2$
(Theorem~\ref{thm:structural-three-halves}).}\\
\multicolumn{3}{@{}l@{}}{\small \textsuperscript{$\dagger$} $\hat{\gamma}(\lambda)$ is the unique solution to the equation $1+\lambda = x - x^{\frac{2-x}{x+2}}((x-2)(x+1))^{\frac{x}{x+2}}$ for $x > 2$.}\\
\end{tabular}
\vspace{-6pt}
\end{table*}

\paragraph{\textbf{Consistency--robustness tradeoff.}}
This work provides a near-optimal characterization of the
consistency--robustness tradeoff for non-clairvoyant scheduling.
On the lower bound side, we prove that the bound of
\citet{Wei2020} is not tight for $n\geq5$, even when considering stronger \emph{padded} versions of their bound (Proposition~\ref{thm:wei_non_tight} and Corollary~\ref{cor:wei_padding_non_tight}). We then derive a stronger, implicitly-defined lower bound $\vg^{(n)}(\lambda)$ on
the robustness of any $(1+\lambda)$-consistent algorithm
(Theorem~\ref{thm:piece-wise_lower_bound}).
On the upper bound side, we propose a new algorithm,
\emph{Modified Two-Stage Schedule} ($\mTSS$), which achieves
$(1+\lambda)$-consistency and $\Psi_n(\lambda)$-robustness, where $\Psi_n(\lambda)$ is the optimal value of a certain finite optimization problem
(Theorem~\ref{thm:mTSS_conrob}).
We prove that, for every $n \geq 2$ and admissible $\lambda$, our upper bound is at most $3/2$ times our lower bound (Theorem~\ref{thm:structural-three-halves}), and thus our algorithm achieves within a factor $3/2$ of the optimal robustness-consistency tradeoff. Moreover, our bounds can be computed for any value of $n$, and appear to match empirically up to a much smaller multiplicative gap of $\lesssim 1.0937$. 

\textcolor{black}{
We also characterize the asymptotic behavior of the optimal tradeoff
as $n\to\infty$. Specifically, we show that any $(1+\lambda)$-consistent algorithm must have robustness at least $\hat\gamma(\lambda)\geq1/\sqrt{\lambda}$
in the large-$n$ limit, while $\mTSS$ achieves robustness
at most $\tfrac12+\sqrt{\smash{\tfrac54}+1/\lambda}$
for every $n$ and admissible $\lambda$
(Theorems~\ref{thm:lower_n->infty} and~\ref{thm:analytic_upper}).
These bounds match in their leading term $1/\sqrt{\lambda}$ as
$\lambda\downarrow0$.
In particular, our lower bound improves the leading constant
of the Wei--Zhang lower bound by a factor of $\sqrt2$ in this asymptotic regime
(cf. Proposition~\ref{prop:wei_padding_limit}).
Our universal upper bound also closes the polynomial gap
in the worst-case-$n$ regime.
We summarize our lower and upper bounds in Table~\ref{tab:regime_bounds_tight} and show plots comparing our bounds against the prior state-of-the-art bounds in Figure~\ref{fig:combined_tradeoff}.
}

As consistency only measures algorithm performance under perfect predictions, we complement these guarantees on consistency and robustness by additionally studying how to retain good performance when predictions have bounded error.
To this end, we develop a generic error-tolerant simulation method
that preserves consistency while providing local smoothness, i.e., smooth degradation in performance as prediction error increases, at the cost of only a small increase in the worst-case robustness bound (Theorem~\ref{thm:tolerance_simulation}).

\paragraph{\textbf{Technical contributions.}}
Our analysis departs from traditional combinatorial arguments by adopting an optimization-based perspective. Rather than analyzing specific hard instances or fixed algorithms, we formulate the search for consistency--robustness lower and upper bounds as variational problems over the space of adversarial attacks, using the following insights.

\textit{Lower bounds: from single-stage to multi-stage attacks.}
Existing lower bounds---in particular, the lower bound of \citet{Wei2020}---rely on adversaries that ``attack''
the algorithm at a single critical moment by revealing a prediction error.
We introduce the \emph{$\alpha$-Instance Attack Set}
(Definition~\ref{def:attack_set}), which allows the adversary
to choose among attacks at different stages---i.e., after varying numbers of jobs have been processed.
This forces the algorithm to balance its processing decisions
against multiple possible attack times and leads to a structured
nonlinear optimization problem characterizing the relationship between consistency and robustness on this instance class
(Problem~\ref{problem_1} or~\ref{refine_problem_1}).

Our key structural insight is that, within a certain range
of robustness values, optimality of a solution to this problem
forces equal processing across unfinished jobs and tightness
of all constraints
(Lemma~\ref{lem:optimal_profile_structure}). This yields a transformation of the optimization problem into a finite system of bilinear equations implicitly characterizing the finite-$n$ lower bound; in the asymptotic $n \to \infty$ limit, this system of equations yields the scaling \textcolor{black}{$\sqrt{1/\lambda}$}.
We anticipate that our general technique of developing and analyzing a multi-stage adversarial
construction as an optimization problem may be useful for studying
consistency--robustness tradeoffs in other online problems.



\textit{Upper bounds: reduction to the time of error detection.}
In developing our refined upper bounds, our algorithm $\mTSS$ refines the two-stage schedule of
\citet{Wei2020} by enforcing \emph{exact} calibration of consistency at the level $(1+\lambda)$. To analyze this strategy,
our main technical insight is that the high-dimensional,
continuous space of prediction errors admits a simple
worst-case structure for $\mTSS$ which reduces to algorithm performance at the time of error detection.
Leveraging the affine structure of the online cost and the concavity of the offline
optimum, we develop a unified reduction accounting for both over-predictions and under-predictions.
\textcolor{black}{Reparametrizing the remaining configurations with a collection of nonnegative ``gap'' variables, we show that the exact consistency calibration in $\mTSS$ imposes a single linear equality constraint on these variables, so the resulting feasible region is a cone. Elements of this cone can be decomposed in order to obtain a simplified optimization-based expression for the robustness.}
Together, these techniques make robustness analysis tractable
for general $n$, yielding a computable finite-$n$ guarantee and a
closed-form universal bound that closes the polynomial gap
in the worst-case-$n$ and $n \to \infty$ regimes.

\paragraph{\textbf{Note on independent concurrent work.}}
Our results were obtained independently of two concurrent works.
Shortly before we first posted our preprint, we became aware of an anonymous work currently under double-blind review for ICLR 2027 that also closes the polynomial gap in the consistency-robustness tradeoff for non-clairvoyant scheduling with worst-case-$n$,
obtaining $(1+\lambda)$-consistency and
$\min\{2/\sqrt{\lambda},\,2+1/\sqrt{\lambda}\}$-robustness; they
furthermore show that this tradeoff can be achieved using only ordinal predictions, rather than predicted processing
times~\citep{anonymousScheduling2026}.
Shortly after our preprint appeared, another paper by \citet{benomarPerchet2026} was posted, which
characterizes the exact deterministic consistency--robustness
frontier in the worst-case-$n$ regime for both ordinal and
processing-time predictions. In particular, their lower and upper bounds coincide with our asymptotic lower bound (Theorem~\ref{thm:lower_n->infty}).
They also showed that randomization can improve upon the deterministic frontier.
While both concurrent works focus on the worst case over the number
of jobs \(n\), our work additionally provides guarantees for each fixed \(n\), as well as a proof that there is a multiplicative gap of at most \(3/2\) between our upper and lower bounds.
For the worst case over $n$, our algorithm achieves the robustness
bound $\frac12+\sqrt{\smash{\frac{5}{4}}+1/\lambda}$,
which improves upon the upper bound of \citet{anonymousScheduling2026} and matches the
optimal leading-order behavior as $\lambda\to0$, though remains
slightly weaker than the exactly tight bound of \citet{benomarPerchet2026}.
Finally, we additionally show how to handle small prediction
errors while preserving a near-optimal consistency--robustness tradeoff.


\subsection{Related Work}
\label{sec:related_work}
Our work contributes to a long history of research on algorithms with predictions and the non-clairvoyant scheduling problem. We briefly review these areas below.

\paragraph{\textbf{Learning-Augmented Algorithms.}} \emph{Learning-augmented algorithms} (also known as \emph{algorithms with predictions}) is an emerging research area that incorporates predictions about future inputs into algorithm design, with the goal of combining the strong empirical performance of machine learning predictors with the worst-case guarantees of traditional online algorithms. Since the introduction of this framework by \citet{Kumar2018} and \citet{Lykouris2021}, it has been applied to a wide range of online problems such as caching~\citep{Lykouris2021, Rohatgi2019, wei2020better, im2022parsimonious}, ski rental~\citep{Kumar2018, Wei2020, antoniadis2021dpm}, online knapsack~\citep{Im2021, lechowicz2024time, daneshvaramoli2025okp}, online conversion~\citep{Sun2021,lechowiczOnlineConversionSwitching2024}, convex function chasing~\citep{christianson22chasing,lechowicz2024chasing},
and metrical task systems~\citep{Antoniadis2020, Christianson2023,coesterLearningAugmentedOnlineMinimization2026}. Beyond classic online algorithms problems, the learning-augmented algorithms framework has also been adopted in several neighboring areas, including mechanism design~\citep{agrawalLearningAugmentedMechanismDesign2022,balkanskiOnlineMechanismDesign2024}, control~\citep{TongxinLi2022Control,Li2024Disentangling}, and reinforcement learning~\citep{Golowich2022Can,Li2024Beyond}. While most works focus on developing algorithms with consistency and robustness guarantees, several recent works have proposed additional performance criteria that go beyond these standard notions, such as smooth degradation of performance in the face of prediction error \citep{elenter2024overcoming,Benomar2025smooth}, learnability of predictions \citep{dinitzFasterMatchingsLearned2021, khodakLearningPredictionsAlgorithms2022a}, and prediction-specific performance guarantees \citep{Li2026PredictionSpecific}. 

Since improving consistency typically comes at the expense of robustness, an overarching goal of this area is to characterize the Pareto-optimal tradeoff between consistency and robustness \citep{Wei2020}. 
A number of works have characterized exactly- or asymptotically-optimal consistency--robustness tradeoffs for online problems such as ski rental~\citep{Wei2020},
online conversion~\citep{Sun2021},
metrical task systems~\citep{Christianson2023}, and non-clairvoyant scheduling with $2$ jobs~\citep{Wei2020}. 
However, despite this progress, exact and asymptotically Pareto-optimal consistency--robustness tradeoffs remain open for many online problems, including non-clairvoyant scheduling with a general number of jobs $n$.

\paragraph{\textbf{Non-Clairvoyant Scheduling.}} Scheduling problems are a foundational class of optimization problems with a wide variety of applications spanning computer systems and operations research ~\citep{Panwalkar1977Survey,Lenstra1978Precedence,Pinedo2012}.
In its most basic form, scheduling considers a set of jobs to be processed on a single machine, with the objective of minimizing the total completion time of all jobs. In the non-clairvoyant setting, where job processing times are unknown a priori and are revealed only upon completion, the \emph{Round-Robin} ($\rr$) algorithm, which divides the
machine equally among all unfinished jobs, achieves the
optimal deterministic competitive ratio $2n/(n+1)$ for
$n$ jobs; its tight uniform bound over all $n$ is
$2$~\citep{Motwani1994}. In contrast, in the clairvoyant setting where job processing times are known a priori, the \emph{Shortest Job First} (\textsf{SJF}) algorithm, which schedules jobs in non-decreasing order of their sizes, is optimal. 

Beyond these classic settings, a growing body of work has studied scheduling models augmented with machine-learned predictions, with the goal of leveraging predicted job properties to improve performance while preserving worst-case guarantees.
The most standard prediction model assumes that predictions provide estimates of all job processing times.
Using this model, \citet{Kumar2018} proposed the \emph{Preferential Round-Robin} 
algorithm, which achieves $(1+\lambda)$-consistency and $(2+1/\lambda)$-robustness.
Subsequently, \citet{Wei2020} established a lower bound on the achievable consistency--robustness tradeoff in this setting, and showed that their bound is tight for $n=2$ jobs, as well as at the two endpoints of the consistency parameter $\lambda$. 

A number of works have also explored alternative prediction models and performance metrics, such as partial predictions~\citep{Benomar2024PartialPredictions}, permutation predictions~\citep{Lindermayr2022Permutation}, progress bar predictions~\citep{Benomar2025ProgressBars}, classification-based predictions~\citep{Dugois2026Classification}, different metrics of prediction error~\citep{Im2023NonClairvoyantPredictions}, and smoothness~\citep{Zhao2026Smoothness}. The results of \citet{Benomar2025ProgressBars} in particular specialize to the setting we consider, improving upon the consistency--robustness tradeoff of \citet{Kumar2018} by giving $(1+\lambda)$-consistency and $(1+1/\lambda)$-robustness. Their work also established a robustness lower bound of $1+\Omega(\sqrt{1/\lambda})$ for any $(1+\lambda)$-consistent algorithm in the aforementioned ``worst-case $n$'' regime.  

\section{Problem Formulation and Preliminaries}
\label{sec:problem}


In this section, we formally introduce the non-clairvoyant scheduling problem with and without predictions, as well as the performance metrics we consider in this work. Throughout the paper, we restrict our attention to deterministic algorithms.

\paragraph{Notation.}
$\mathbb{N}$, $\mathbb{N}_+$, and $\mathbb{R}$ denote the sets of natural numbers with and without zero, and real numbers, respectively. $\mathbb{R}_{\geq 0}^{n}$ is the set of vectors with nonnegative real coordinates. Define $[n] \coloneq \{1, \ldots, n\}$ for $n \in \mathbb{N}_+$.
In asymptotic notation involving the variable $\lambda$, the implied regime is $\lambda \to 0$.

\subsection{The Classic Non-Clairvoyant Scheduling Problem}
\label{sec:classic_nonclairvoyant_scheduling}
In the classic single-machine \emph{non-clairvoyant} scheduling problem,
a scheduler is given $n$ jobs. Job $i \in [n]$ requires $x_i$ units of
processing on the machine, but $x_i$ is unknown to the scheduler until the job completes. At each time, the machine has one unit of processing
capacity. A schedule may allocate this capacity among unfinished jobs,
and jobs may be preempted and resumed at any time at no cost;
equivalently, a fractional allocation can be implemented by
time-slicing on a single machine with arbitrarily small time quanta.
The objective is to  minimize the total completion time $\sum_{i \in [n]} C_i$, where $C_i$ is the \textit{clock time} at which job $i$ completes, i.e., the first time by which job $i$ has received its required $x_i$ units of processing. An instance of the problem is
specified by the vector of processing requirements
$I = (x_1,\ldots,x_n)$; we use $\mI$ to denote the set of all such instances.

As a benchmark, the \emph{clairvoyant} version of the problem is the case
in which the processing requirements $\{x_i\}_{i \in [n]}$ are known to
the scheduler in advance.
In this setting, it is well known that \emph{Shortest Remaining Processing
Time} (\textsf{SRPT}) minimizes the total completion time~\citep{Schrage1968}. Since all jobs are available at time $0$, this is
equivalent to \emph{Shortest Job First} (\textsf{SJF}), a special case of
Smith's rule \citep{Smith1956}. However, in the non-clairvoyant setting, the processing requirements are unknown
\emph{a priori}, so algorithms are evaluated by comparing their total completion
time to the clairvoyant optimum \citep{Motwani1994}.
This performance metric is formalized as the competitive ratio~\citep{Borodin2005}, defined as follows.






\begin{definition}[Competitive Ratio] Let $\mI$ denote the set of instances of the non-clairvoyant scheduling problem.
For an algorithm $\mA$ and instance $I \in \mI$, let $\alg(\mA,I)$ be the
total completion time achieved by $\mA$, and let $\opt(I)$ be the total
completion time of the optimal clairvoyant schedule. An algorithm $\mA$ is
$c$-competitive, or has competitive ratio $c$, if
\begin{align*}
    \alg (\mA, I) \leq c \cdot \opt(I) \quad \text{for all } I \in \mI.
\end{align*}
\end{definition}
It was established by \citet{Motwani1994} that the
Round-Robin ($\rr$) algorithm achieves the optimal
deterministic competitive ratio of $2n/(n+1)$ for $n$ jobs,
with a tight uniform bound of $2$ over all $n$.

\subsection{Non-Clairvoyant Scheduling with Predictions} 

The competitive ratio measures the worst-case performance of an algorithm on any problem instance; however, in practice, one may be able to do better when given predictions about the underlying instance. To move beyond the pessimism of worst-case algorithms for non-clairvoyant scheduling, we consider the ``prediction-augmented'' setting,  where the algorithm has access to a prediction for each job's size; let $y_i$ denote the \textit{predicted size} of job~$i$. 
Importantly, we place no assumptions on the accuracy of these predictions: they can be arbitrarily inaccurate or even adversarial. Thus, the algorithm seeks to take advantage of the predictions when they are good, while remaining robust in case of poor prediction quality. 
Consistent with the standard non-clairvoyant scheduling model, we continue to assume that the true processing time $x_i$ of each job $i$ is not known to the algorithm in advance and is revealed only once the job has received $x_i$ units of processing.


To evaluate the performance of algorithms for non-clairvoyant scheduling with predictions, we use the standard notions of \emph{consistency} and \emph{robustness} \citep{Kumar2018,Wei2020}.
To formalize these criteria, we define an \textit{instance} of the prediction-augmented problem as comprising both
the true processing requirements $\{x_i\}_{i \in [n]}$ and the predicted
processing requirements $\{y_i\}_{i \in [n]}$ of the jobs in an instance. Let $\mI$ denote the set
of all such instances, and let $\mI^* \subseteq \mI$ denote the set of
\textit{consistent instances}, i.e., those for which the predictions are correct, so $x_i = y_i$ for all
$i \in [n]$. Consistency then measures the worst-case performance of an algorithm over all consistent instances $\mI^*$, while \emph{robustness} considers the worst case over all prediction-augmented instances $\mI$, regardless of prediction quality. Formally, we have the following definitions:

\begin{definition}[Consistency and Robustness]
\label{def:consistency_robustness}
The \textbf{consistency} of an algorithm $\mA$ is its worst-case competitive
ratio over consistent instances. In particular, an algorithm $\mA$ is \textbf{$\con$-consistent}
if
\[
    \alg(\mA,I) \leq \con \cdot \opt(I)
    \quad \text{for all } I \in \mI^*.
\]
The \textbf{robustness} of $\mA$ is its worst-case competitive ratio over
all instances. In particular, an algorithm $\mA$ is \textbf{$\rob$-robust} if
\[
    \alg(\mA,I) \leq \rob \cdot \opt(I)
    \quad \text{for all } I \in \mI.
\]
\end{definition}

In learning-augmented algorithms, there is typically a tradeoff between consistency and robustness: a smaller consistency requires the algorithm to more closely follow the predictions, which can in turn lead to a larger robustness.
From a game-theoretic perspective, for a given parameter $\lambda > 0$,
the optimal consistency--robustness tradeoff can be viewed as the
equilibrium of a constrained minimax game in which the algorithm aims to
minimize its worst-case robustness subject to maintaining
$(1+\lambda)$-consistency, while an adversary chooses instances to
maximize the algorithm's robustness. This is the view that will guide our lower and upper bounds in this work as we seek to characterize this equilibrium.


\section{Lower Bounds}\label{sec:lb}

In this section, we prove a sharper lower bound on the
consistency--robustness tradeoff for non-clairvoyant scheduling with
predictions that significantly improves on the lower bound of \citet{Wei2020}. \textcolor{black}{As a warmup, we first revisit the lower bound construction of
\citet{Wei2020} and show that it is not tight in general.}
Motivated by this \textcolor{black}{limitation}, we then introduce a stronger multi-stage adversarial
strategy, express the robustness-consistency tradeoff that it induces as a nonlinear optimization problem, and show how this problem can be transformed into a form that gives an implicit lower bound for every finite $n$ and an asymptotic form as $n \to \infty$. 

In the analysis that follows, it suffices to restrict our
attention to work-conserving algorithms, which use the full
processing capacity whenever a job remains unfinished,
since eliminating unused capacity cannot increase completion
times. We let $d(i,j)$ denote the cumulative amount of
processing that has been received by job $i$ when job $j$ completes.

\vspace{-0.5em}
\subsection{Warmup: \textcolor{black}{Looseness} of the Wei and Zhang Lower Bound} \label{sec:warmup}
\vspace{-0.3em}
As mentioned in the introduction, \citet{Wei2020} show that any $(1+\lambda)$-consistent algorithm must have robustness lower bounded as in~\eqref{eq:wei-lower-bound}, and that this bound is tight (a) when $n = 2$ and (b) at the two endpoints $\lambda = 0$ and $\lambda = \tfrac{n-1}{n+1}$ for general $n$. In this section, we show that their lower bound is not tight in general, identifying a structural
limitation that makes it loose for $n \geq 5$. In
particular, we have the following result.

\begin{proposition}
\label{thm:wei_non_tight}
When $n \geq 5$, the lower bound in~\eqref{eq:wei-lower-bound} obtained by~\citet{Wei2020} is not tight. 
In particular, there exists a choice of $\lambda \in \bigl(0,\,\tfrac{n-1}{n+1}\bigr)$ such that any $(1+\lambda)$-consistent algorithm has robustness strictly greater than $\Phi_{\mathrm{wz}}(n,\lambda)$, where $\Phi_{\mathrm{wz}}(n,\lambda)$ is defined in~\eqref{eq:wei-lower-bound}.
\end{proposition}

\begin{figure}[t]
     \centering
\includegraphics[width= 0.8\linewidth]{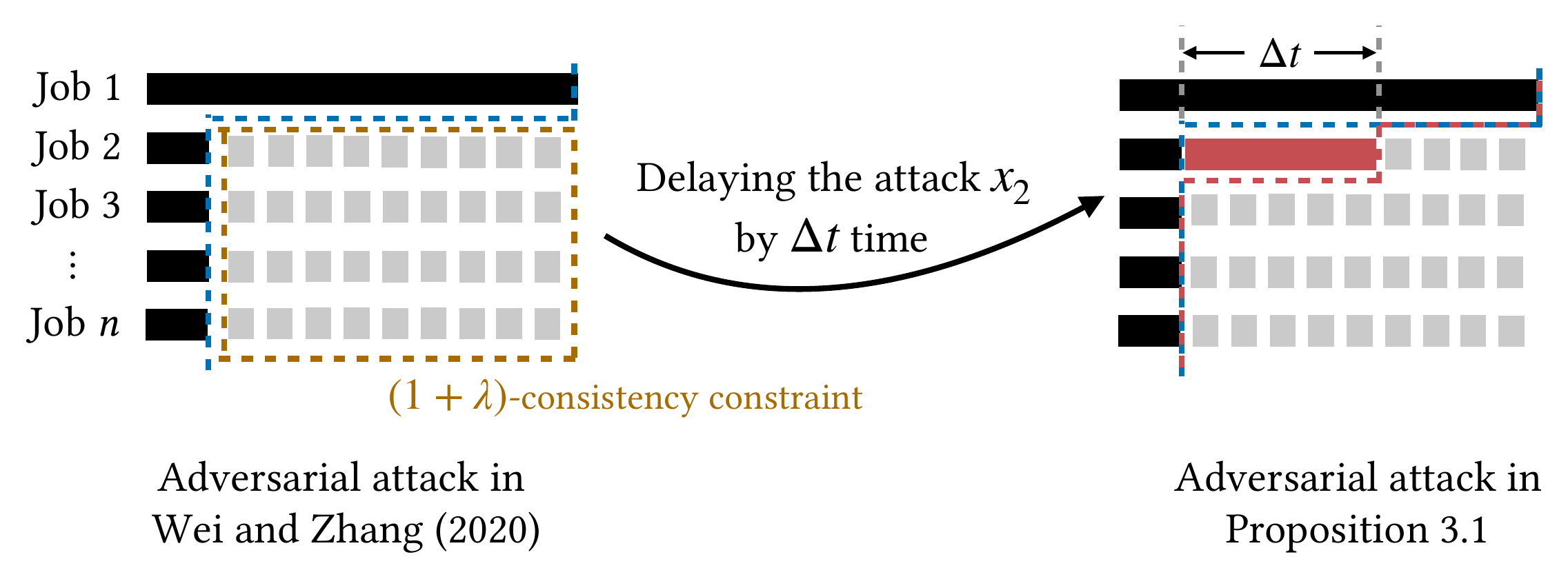}
     \caption{\textcolor{black}{\textbf{Comparison of the adversarial attacks used in Proposition~\ref{thm:wei_non_tight}}.
Left: the blue dashed line marks the single attack of \citet{Wei2020}.
Right: our construction retains this attack and adds the delayed attack (red dashed line) after an additional $\Delta t$ of processing.
Since the original attack exhausts the algorithm's flexibility under
the consistency constraint, the additional attack yields a strictly
stronger lower bound for $n\geq5$.}}\label{fig_non_tightness}
 \end{figure}

\paragraph{\textbf{Proof Sketch for Proposition~\ref{thm:wei_non_tight}.}} The lower bound construction in \cite{Wei2020} can be interpreted as fixing the prediction profile $y_1 = y_2 = \dots = y_n = 1$ and considering the adversarial input
\begin{equation}
\label{eq:wei_attack}
    x_1 = 1, \quad x_i = d(i,1) + \eps \quad \text{for all }i \geq 2, \quad \text{with} \quad \eps \downarrow 0,
\end{equation}
which ``attacks'' the algorithm with an incorrect prediction immediately upon the completion of the first job. Their analysis shows that for any algorithm to achieve the lower bound in~\eqref{eq:wei-lower-bound}, a necessary condition is
\begin{equation}
\label{eq:d(i,1)_main}
    d(2,1) = \cdots = d(n,1) = \tfrac{n+1}{\,n-1\,}\,\lambda.
\end{equation}
As illustrated in Figure~\ref{fig_non_tightness}, this forces the algorithm to exhaust its \emph{flexibility budget}---that is, the algorithm's flexibility to not directly follow predictions, offered by the $\lambda$ consistency buffer---when the first job finishes in order to minimize the competitive ratio over the adversarial input~\eqref{eq:wei_attack}.

We consider the following two cases. If \eqref{eq:d(i,1)_main} is not satisfied, then the lower bound is strictly greater than~\eqref{eq:wei-lower-bound}.
On the other hand, if \eqref{eq:d(i,1)_main} is satisfied, then since there is no remaining \emph{flexibility budget}, the algorithm must finish the remaining jobs in the predicted order. Thus, we can delay the ``attack'' $x_2$ by $\Delta t$---i.e., set $x_2 = d(2, 1) + \Delta t$ for some $\Delta t > \epsilon$---while keeping $x_i$ unchanged for all $i > 2$. As a result, the incurred cost of the online algorithm increases by $(n-1)\Delta t$, whereas the cost of the offline optimum increases by only $2\Delta t$.
In particular, when $n \geq 5$, there exists some $\lambda \in (0,\tfrac{n-1}{n+1})$ such that the bound in \eqref{eq:wei-lower-bound} is strictly smaller than $\tfrac{n-1}{2}$. Therefore, delaying the attack yields a strictly larger competitive ratio for these values of $\lambda$.
Combining the two cases, we obtain the non-tightness result; the full proof is presented in Appendix~\ref{app:wei_non_tight}. \qed \\

\color{black}
The above proposition shows that, for a fixed number of jobs $n$, the lower bound construction of \citet{Wei2020} is not tight. However, it is possible to \emph{pad} their construction to potentially strengthen the lower bound for a given $n$, as described in \cite[Section 3.2]{Wei2020}: a lower bound
for $k$ jobs also gives a lower bound for $n\geq k$ jobs by adding $n-k$ ``dummy'' jobs with zero predicted and actual processing times. Thus, the strongest lower bound obtained by padding their construction is
\begin{equation}
\label{eq:pad}
\Phi_{\text{wz}}^{\text{pad}}(n,\lambda)
\coloneq
\max_{\substack{2\leq k\leq n\\
\lambda\leq(k-1)/(k+1)}}
\Phi_{\text{wz}}(k,\lambda),
\end{equation}
where $\Phi_{\text{wz}}(k,\lambda)$
is simply the original lower bound in \eqref{eq:wei-lower-bound}; the restriction on $k$ ensures that $\lambda$ lies in the range of the corresponding $k$-job lower bound. Notably, as $n\to\infty$, the optimally padded lower bound of
\citet{Wei2020} converges to the supremum of their original
lower bound over the number of jobs, yielding the scaling $\Omega(\sqrt{1/\lambda})$ as $\lambda\downarrow 0$: 

\begin{proposition}
\label{prop:wei_padding_limit}
For every fixed $\lambda\in(0,1)$, the limit $\Phi_{\mathrm{wz}}^{\mathrm{pad}}(\infty,\lambda)
\coloneq
\lim_{n\to\infty}\Phi_{\mathrm{wz}}^{\mathrm{pad}}(n,\lambda)$
exists. Moreover, $\Phi_{\mathrm{wz}}^{\mathrm{pad}}(\infty,\lambda)=2$ for $\lambda\in[1/2,1)$,
and $\Phi_{\mathrm{wz}}^{\mathrm{pad}}(\infty,\lambda)
=
\frac{1+o(1)}{\sqrt{2\lambda}}$ as $\lambda\downarrow0$.
\end{proposition}

This result is proved in Appendix~\ref{app:padding}. This asymptotic limit yields two key observations: first, when $\lambda\in[1/2,1)$, the padded lower bound converges to the trivial bound of $2$ as $n \to \infty$, which is simply the robustness in the setting without predictions \citep{Motwani1994}. Second, as $\lambda\downarrow0$, the leading term of this lower bound is $1/\sqrt{2\lambda}$, which asymptotically matches the ``worst-case $n$'' lower bound of \citet{Benomar2025ProgressBars}. As we will show later, our strengthened lower bound can improve this leading term to $1/\sqrt{\lambda}$, matching our upper bound (Theorem~\ref{thm:lower_n->infty}).

Moreover, as we show in the following corollary, this padding remains insufficient to characterize the optimal worst-case robustness even for finite $n$.

\begin{corollary}
\label{cor:wei_padding_non_tight}
When $n\geq5$, the lower bound of \citet{Wei2020}
is not tight even after optimal padding.
In particular, for every
$\lambda\in[\tfrac12,\tfrac{n-1}{n+1})$,
any $(1+\lambda)$-consistent algorithm for $n$ jobs
has robustness strictly greater than
$\Phi_{\mathrm{wz}}^{\mathrm{pad}}(n,\lambda)$, as defined in \eqref{eq:pad}.
\end{corollary}
A proof of this result is given in Appendix~\ref{app:extend}. 
Altogether, Propositions~\ref{thm:wei_non_tight} and \ref{prop:wei_padding_limit} and Corollary~\ref{cor:wei_padding_non_tight} demonstrate that the construction of \citet{Wei2020}, which attacks the algorithm immediately upon the completion of the first job, is insufficient to characterize the worst-case behavior for general $n$. Thus, in the next section, we will introduce a multi-stage adversarial framework to obtain stronger lower bounds.

\color{black}

\vspace{-4pt}

\subsection{Designing a New Adversarial Strategy}

Motivated by the construction in the proof of Proposition~\ref{thm:wei_non_tight}, 
in this section, we introduce a multi-stage adversarial framework in which the adversary can choose when to reveal the prediction error among multiple times. We adopt the same prediction profile as in the previous construction and define our \emph{$\alpha$-Instance Attack Set} as follows.



\begin{definition}[$\alpha$-Instance Attack Set] \label{def:attack_set} Let $\alpha \in [n -1]$. Suppose all predictions are identical, i.e.,
$y_1 = y_2 = \cdots = y_n = 1$.
\textcolor{black}{
Let $\mA$ be an algorithm that completes jobs in increasing index order when the prediction is accurate (i.e., when $\mathbf x=\mathbf y =\mathbf{1}$), and
let $d(i,j)$ denote the cumulative amount of processing
received by job $i$ when job $j$ finishes under this execution, assuming accurate predictions.}
The \textbf{$\alpha$-Instance Attack Set} is defined as the set of processing time vectors
$
\mX_\alpha (\mA) \coloneq \{\mx_i\}_{i \in [\alpha]},
$
where the $i$-th attack instance's processing time vector $\mx_i$ is given by
\[\mx_i \coloneq (\underbrace{1, \ldots, 1}_{i \text{ copies}},\ d(i{+}1,i) + \eps,\ \ldots,\ d(n,i) + \eps),\]
where $\eps > 0$ denotes an infinitesimal quantity.
\end{definition}


\vspace{-5pt}
For a fixed $\alpha$, the $\alpha$-Instance Attack Set consists of a family of $\alpha$ different candidate instances, from which the adversary selects the one that induces the greatest robustness. For instances in this set,  it suffices to restrict our attention
to algorithms whose processing profile is monotone in the following sense:
for any fixed $k$, the quantities $d(i,k)$ are non-increasing as $i$ increases. That is, when job $k$ finishes, the algorithm has performed more processing on earlier jobs than later jobs. 
This assumption can be made without loss of generality, since for any algorithm $\mA$, there
exists another algorithm $\mA'$ with this monotonicity property that is
equivalent to $\mA$ with respect to its behavior on the attack set, in the sense that
attacking $\mA$ over $\mX_\alpha(\mA)$ yields the same outcomes as attacking
$\mA'$ over $\mX_\alpha(\mA')$. We formalize this equivalence in the following lemma, which is proved in Appendix~\ref{app:Monotonic_Equivalence_Lemma}. 

\vspace{-4pt}
\begin{lemma}
\label{Monotonic_Equivalence_Lemma}
Suppose all predictions are identical, i.e.,
$y_1 = y_2 = \cdots = y_n = 1$.
For any work-conserving algorithm $\mA$ that completes jobs in increasing index order,
there exists another algorithm $\mA'$ that also completes jobs in increasing index order
and satisfies
$
d(i,k) \ge d(j,k)$ for all $j \ge i > k,
$
and such that the two algorithms are equivalent in the sense that (a)
they have the same competitive ratio when predictions are perfect, and (b) attacking $\mA$ with an instance from $\mX_\alpha(\mA)$ yields the same competitive ratio as attacking $\mA'$ with an instance from $\mX_\alpha(\mA')$.
\end{lemma}

\vspace{-1.3em}
\subsection{A Strictly Improved Lower Bound}
\vspace{-0.4em}
In this section, we will establish that the $\alpha$-Instance Attack Set enables an improved lower bound. We begin by outlining the proof strategy before stating the refined lower bound formally.

\begin{figure}[t]
    \centering
    \includegraphics[width=0.8\linewidth]{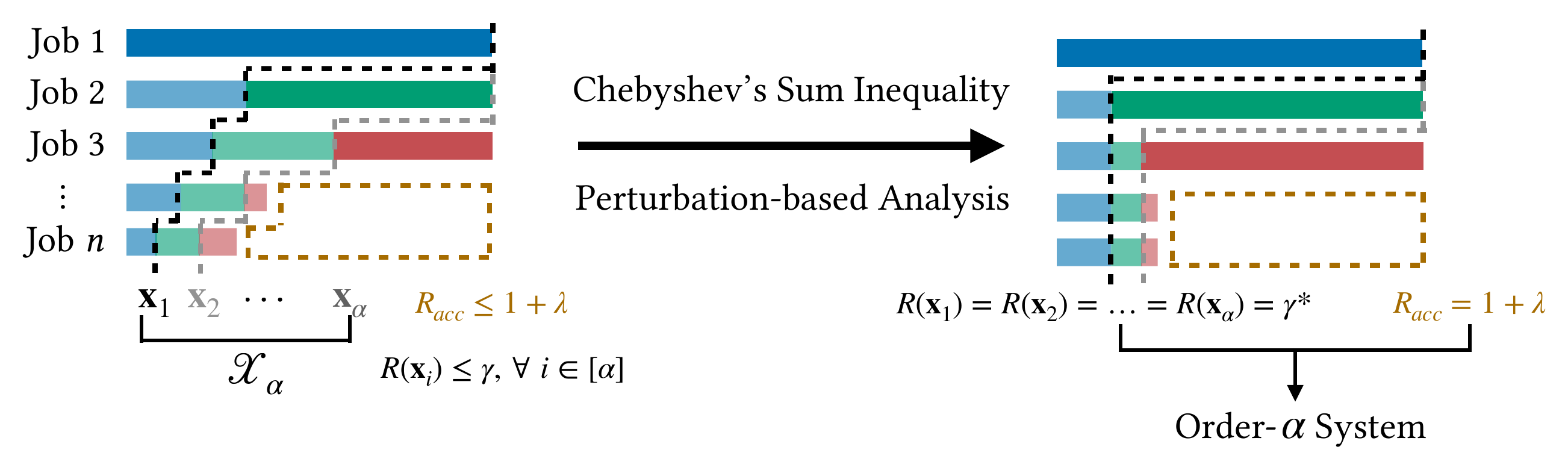}
    \caption{\textbf{Structural characterization of optimal processing profiles}
(Lemma~\ref{lem:optimal_profile_structure}).
Different colors correspond to processing snapshots when
successive jobs finish: darker shades highlight the completing
job, while lighter shades indicate the processing of the other
unfinished jobs.
Dashed lines mark the corresponding attack instances.
Here, $R(\mx_i)$ denotes the competitive ratio on the $i$-th
attack instance $\mx_i$ in the $\alpha$-Instance Attack Set $\mX_\alpha$, while
$R_{\rm acc}$ denotes the competitive ratio under perfect predictions.
At optimality, equal processing among unfinished jobs and tight
attack and consistency constraints yield the Order-$\alpha$ System.}
    \label{fig:overview}
\end{figure}
\vspace{-10pt}
\paragraph{\textbf{Overview of the Lower-Bound Construction.}}
Fix the prediction profile $y_1=\cdots=y_n=1$.
For an attack depth $\alpha$, the adversary chooses among the
$\alpha$ candidate instances in $\mX_\alpha(\mA)$
(Definition~\ref{def:attack_set}).
For any target robustness level $\rob$, the algorithm must therefore satisfy this robustness bound
on any attack from the set while maintaining
$(1+\lambda)$-consistency when the predictions are perfect.
We will eventually select $\alpha$ according to a prescribed rule, which we will specify later in this section; for now, assume that $\alpha \in [m]$, where $m$ is the maximum allowed attack depth.

By Lemma~\ref{Monotonic_Equivalence_Lemma}, we may restrict our
attention to monotone processing profiles.
For $i\in[\alpha]$ and $j>i$, let
$p_j^{(i)}\coloneq d(j,i)$ denote the processing received by an
unfinished job $j$ when job $i$ completes.

We can then formulate the following nonlinear optimization problem for minimizing the robustness $\rob$ while maintaining consistency:
\begin{subequations}
\label{problem_1}
\begin{align}
\min_{\rob,\mathbf p} \ & \quad  \rob\\
\text{s.t.}\
&\frac{i(2n+1-i)}2
+\sum_{j<i}\sum_{k=j+1}^{n}p_k^{(j)}
+(n+1-i)\sum_{j=i+1}^{n}p_j^{(i)}
\leq
\rob\left(
\frac{i(i+1)}2+\sum_{j=i+1}^{n}j p_j^{(i)}
\right),
 i\in[\alpha],
\label{problem_1_cons1}\\
&\frac{n(n+1)}2
+\sum_{i<\alpha}\sum_{j=i+1}^{n}p_j^{(i)}
+\sum_{j=\alpha+1}^{n}(j-\alpha)p_j^{(\alpha)}
\leq
(1+\lambda)\frac{n(n+1)}2,
\label{problem_1_cons2}\\
&0\leq p_n^{(i)}\leq\cdots\leq p_{i+1}^{(i)}\leq1,
\ \ i\in[\alpha];
\qquad
p_j^{(i)}\leq p_j^{(i+1)},
\ \ i\in[\alpha-1],\ \ i+1<j\leq n,
\label{problem_1_cons3}
\end{align}
\end{subequations}
where each constraint in \eqref{problem_1_cons1} enforces $\rob$-robustness on an attack in the $\alpha$-Instance Attack Set,
\eqref{problem_1_cons2} enforces the consistency budget,
and \eqref{problem_1_cons3} specifies the bounds and
monotonicity of the processing profiles.
We denote the optimal value of this optimization problem by
$\phi_{\alpha,n}(\lambda)$; existence of an optimal solution is established in
Lemma~\ref{lem:existence_fixed_alpha_optimizer}.

Our key structural insight is that, within a certain range
of optimal values, optimality forces both symmetry and tightness.
Specifically, an optimal processing profile assigns equal processing to
all unfinished jobs at each time of attack, makes every attack robustness
constraint tight, and exhausts the consistency budget when the $\alpha$-th job finishes. In particular, we have the following lemma.

\begin{lemma}
\label{lem:optimal_profile_structure}
Fix $\alpha\in[m]$ and
$\lambda\in(0,\frac{n-1}{n+1})$. Suppose that $\frac{2n}{n+1}
<
\phi_{\alpha,n}(\lambda)
<
\frac{n+1-\alpha}{\alpha}$.
Then every optimal solution to Problem~\ref{problem_1}
satisfies
\[
p_j^{(i)}=p^{(i)}
\qquad
\text{for every }i\in[\alpha]\text{ and }j>i,
\]
for some $p^{(1)},\ldots,p^{(\alpha)}$.
Moreover, all attack constraints
\eqref{problem_1_cons1} and the consistency constraint
\eqref{problem_1_cons2} hold with equality.
\end{lemma}

The proof of this lemma recursively constructs a symmetric comparison profile
with the same robustness and tight attack constraints.
A comparison using Chebyshev's inequality shows that it uses
no more consistency budget than the original optimum;
any unused budget would allow a feasible perturbation decreasing the robustness level
$\rob$, contradicting optimality.
The equality conditions then force the two profiles to coincide.
The full proof appears in
Appendix~\ref{app:piece-wise_lower_bound}.

The lemma reduces the processing variables for each job and each attack instance to the common values
$p^{(1)},\ldots,p^{(\alpha)}$, one for each instance.
Substituting them into the tight attack and consistency
constraints yields the following system of $\alpha+1$ bilinear equations characterizing the optimal solution.

\begin{definition}[Order-$\alpha$ System]
\label{def:order-m-system}
Fix integers $n\geq2$ and $\alpha\in[m]$, and a parameter
$\lambda\in[0,\frac{n-1}{n+1}]$.
The \textbf{Order-$\alpha$ System}
$\mathsf{Sys}_\alpha(n,\lambda)$ is a collection of $\alpha+1$
equations in the variables $(\rob,p^{(1)},\ldots,p^{(\alpha)})$
given by
\begin{equation}
\label{eq:order_alpha_1}
\frac{(2n+1-i)i}{2}
+\sum_{j=1}^{i-1}(n-j)p^{(j)}
+(n+1-i)(n-i)p^{(i)}
=
\rob\left(
\frac{i(i+1)}2+
\frac{(n+i+1)(n-i)}2p^{(i)}
\right)
\end{equation}
for every $i\in[\alpha]$, and
\begin{equation}
\label{eq:order_alpha_2}
\lambda\frac{n(n+1)}2
=
\sum_{j=1}^{\alpha-1}(n-j)p^{(j)}
+\frac{(n+1-\alpha)(n-\alpha)}2p^{(\alpha)}.
\end{equation}
\end{definition}

Equation~\eqref{eq:order_alpha_1} expresses equality in the
$i$-th attack constraint, while
\eqref{eq:order_alpha_2} expresses full use of the consistency
budget upon the completion of the $\alpha$-th job.
The common processing amounts $p^{(i)}$ reveal how a rational
online decision maker should balance its processing allocations
to hedge against the different possible attacks in the
$\alpha$-Instance Attack Set. Under the hypothesis of
Lemma~\ref{lem:optimal_profile_structure}, the optimal value
$\phi_{\alpha,n}(\lambda)$ is exactly the robustness coordinate
of the solution to this system.\\

\begin{figure}[t]
\centering
\includegraphics[width=\linewidth]{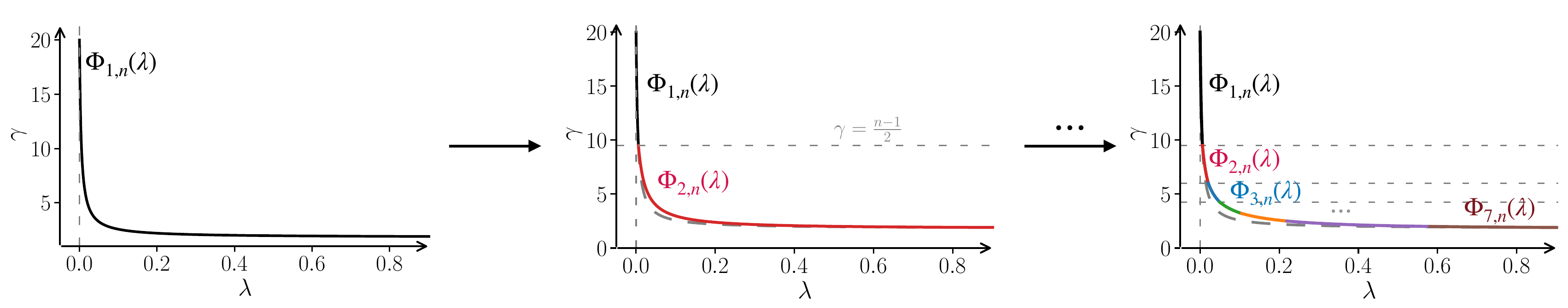}
\caption{Successive attack-depth transitions as $\lambda$
increases, illustrated for $n=20$.
The attack depth $\alpha$ increases from $i$ to $i+1$ when the
current bound falls below $\gamma=(n-i)/(i+1)$,
as indicated by the horizontal dashed lines.
Different colors indicate attack depths, while gray
horizontal dashed lines mark thresholds between consecutive
depths. The panels show the progression from depth one
to the maximal depth $m=7$.}
\label{fig:choose_alpha}
\end{figure}

\noindent
\textit{Selecting the attack depth $\alpha$.} \ 
Fix $n\geq2$. To understand when to increase the attack depth,
consider a symmetric optimal profile (as in Lemma~\ref{lem:optimal_profile_structure}) at depth $\alpha<n-1$.
Completing job $\alpha+1$ while leaving the processing of the
other unfinished jobs unchanged takes
$\Delta t=1-p^{(\alpha)}$.
Compared with the $\alpha$-th attack, the resulting attack
delays the completion of all $n-\alpha$ unfinished jobs by
$\Delta t$, increasing the online cost by $(n-\alpha)\Delta t$.
For the offline optimum, the enlarged job is the
$(\alpha+1)$-th largest, so its additional processing delays
itself and the $\alpha$ jobs scheduled after it, increasing
the offline cost by $(\alpha+1)\Delta t$.
When $\Delta t>0$, this increases the competitive ratio
precisely when
\[
\phi_{\alpha,n}(\lambda)<\frac{n-\alpha}{\alpha+1}.
\]
Thus, the additional attack exposes a weakness of the profile
that was optimal at depth $\alpha$, motivating an increase
in the attack depth.
Therefore, we start with $\alpha=1$ and iteratively increase
the attack depth from $i$ to $i+1$ whenever
$\lambda>
\phi_{i,n}^{-1}\!\left(\frac{n-i}{i+1}\right)$.
Figure~\ref{fig:choose_alpha} illustrates the resulting
attack-depth transitions as $\lambda$ increases for $n=20$.

For $0<\lambda<(n-1)/(n+1)$, we therefore choose the smallest
$\alpha\in[n-1]$ satisfying
\begin{equation}
\label{choose_alpha}
\phi_{\alpha,n}(\lambda)
\in
\left[
\frac{n-\alpha}{\alpha+1},
\frac{n+1-\alpha}{\alpha}
\right).
\end{equation}
Lemma~\ref{lem:justify_choice_of_alpha} verifies the validity of this selection rule. Since the universal lower bound
$\phi_{\alpha,n}(\lambda)\geq2n/(n+1)$ ensures that the lower
threshold is met once $\frac{n-\alpha}{\alpha+1}\leq\frac{2n}{n+1}$, it suffices to consider $\alpha\in[m]$, where $m$ is the maximum allowed attack depth, defined as
\begin{equation}
\label{eq:m_upper}
m\coloneq
\max\left\{
i\in[n-1]:
\frac{n+1-i}{i}>\frac{2n}{n+1}
\right\}.
\end{equation}

We now define the corresponding value functions and combine
them according to this selection rule.

\begin{definition}[Depth-$\alpha$ Value Functions]
\label{def:mth-solution}
For $\alpha\in[m]$ and
$\lambda\in[0,\frac{n-1}{n+1}]$, let
\[
\mathsf{Sol}_\alpha(n,\lambda)
\coloneq
\bigl(\vg_{\alpha,n}(\lambda),
p^{(1)},\ldots,p^{(\alpha)}\bigr)
\in
\left[\frac{2n}{n+1},\frac{n+1-\alpha}{\alpha}\right]
\times[0,1]^\alpha
\]
denote the solution to $\mathsf{Sys}_\alpha(n,\lambda)$
whenever it admits an admissible solution.
Its robustness coordinate $\vg_{\alpha,n}$ is called the
\textbf{depth-$\alpha$ value function}.

If $m=1$, define $\vg^{(n)}=\vg_{1,n}$ throughout
$[0,\frac{n-1}{n+1}]$.
For $m\geq2$, define
\begin{equation}
\label{Gamma}
\vg^{(n)}(\lambda)\coloneq
\begin{cases}
\vg_{1,n}(\lambda)
& \text{when }\lambda\in
\Bigl[0,\,
\vg_{1,n}^{-1}\!\left(\tfrac{n-1}{2}\right)\Bigr],
\\[4pt]
\vg_{2,n}(\lambda)
& \text{when }\lambda\in
\Bigl(
\vg_{1,n}^{-1}\!\left(\tfrac{n-1}{2}\right),\,
\vg_{2,n}^{-1}\!\left(\tfrac{n-2}{3}\right)
\Bigr],
\\
\quad\vdots & \quad\vdots
\\
\vg_{m,n}(\lambda)
& \text{when }\lambda\in
\Bigl(
\vg_{m-1,n}^{-1}\!\left(\tfrac{n+1-m}{m}\right),\,
\tfrac{n-1}{n+1}
\Bigr].
\end{cases}
\end{equation}
When $m=2$, only the first and last cases apply.
\end{definition}

Appendix~\ref{app:well_p} establishes the well-posedness of
these definitions, including existence and uniqueness,
continuity, strict monotonicity, and agreement of consecutive
value functions at their transition points.
Each $\vg_{\alpha,n}$ can be computed numerically by solving
the Order-$\alpha$ System on its admissible domain.

For $0<\lambda<(n-1)/(n+1)$, the selected depth satisfies $\phi_{\alpha,n}(\lambda)
=
\vg_{\alpha,n}(\lambda)
=
\vg^{(n)}(\lambda)$.
Since $\phi_{\alpha,n}(\lambda)$ is a robustness lower bound,
this yields the following theorem.
The full proof, including the endpoint cases and strict
improvement over the previous bound, appears in
Appendix~\ref{app:piece-wise_lower_bound}.

\begin{theorem}
\label{thm:piece-wise_lower_bound}
Let $\lambda\in[0,\frac{n-1}{n+1}]$.
Any $(1+\lambda)$-consistent algorithm for $n$-job
non-clairvoyant scheduling with predictions must have robustness
$\rob\geq\vg^{(n)}(\lambda)$.
For $n\geq5$, this strictly improves on the lower bound \eqref{eq:wei-lower-bound}
of \citet{Wei2020} whenever $\vg_{1,n}^{-1}(\frac{n-1}{2})
<
\lambda
<
\frac{n-1}{n+1}$,
where
$\vg_{1,n}^{-1}(\frac{n-1}{2})=\frac{2}{n^2-3n-2}$.
\end{theorem}
\subsection{An Asymptotic Lower Bound as $n \to \infty$}
In the previous section, we showed how to obtain an implicit lower bound on robustness in the fixed, finite-$n$ setting via the solution of the Order-$\alpha$ System of bilinear equations. Building on this characterization, we can derive the following explicit lower bound in the asymptotic regime $n \to \infty$.


\begin{theorem}
\label{thm:lower_n->infty}
    Fix $\lambda \in (0, 1)$. As $n \to \infty$, any $(1+\lambda)$-consistent algorithm for non-clairvoyant scheduling must have robustness at least $\hat{\gamma}(\lambda)$, where $\hat{\gamma}(\lambda)$ is the unique solution to the equation
    \begin{equation} \label{eq:lower_bound_closed_form_asymptotic}
        1+\lambda = x - x^{\frac{2-x}{x+2}}((x-2)(x+1))^{\frac{x}{x+2}}, \qquad x > 2.
    \end{equation}
    In particular, $\hat{\gamma}(\lambda) \geq \sqrt{1/\lambda}$.
\end{theorem}

The proof of Theorem~\ref{thm:lower_n->infty} is provided in
Appendix~\ref{app:lower_n->infty}.
Figure~\ref{fig:combined_tradeoff} illustrates this asymptotic
lower bound together with the finite-$n$ lower bounds from
Theorem~\ref{thm:piece-wise_lower_bound}. 






\section{Algorithms and Upper Bounds} \label{sec:ub}

With our refined lower bounds established, we now proceed to develop an improved upper bound for the optimal consistency--robustness tradeoff.
\textcolor{black}{We first introduce our \emph{Modified Two-Stage Schedule} ($\mTSS$) algorithm, which retains the two-stage structure of
\citet{Wei2020}'s algorithm, starting with a round-robin ($\rr$) stage that allocates equal processing time to all jobs before switching to processing the jobs in predicted order. The key difference in our approach is to calibrate the duration of the $\rr$ stage to exhaust the consistency budget.
We show that the robustness of this algorithm can be reduced to its performance at the time that any error is detected, which yields 
a tractable, optimization-based robustness bound for general $n$.
Relaxing this bound gives a closed-form analytic
bound that holds uniformly over all $n$.}

\vspace{-6pt}
\subsection{An Improved Algorithm: Modified Two-Stage Schedule}
\label{subsec:improved_alg}

In what follows, we assume that the jobs are indexed in nondecreasing order of their predicted processing times $y_i$, and let $\opt_{\mathbf y}$ denote the offline optimal scheduling cost under accurate predictions, namely, $\opt_{\mathbf y} = \sum_{i=1}^n (n+1-i)y_i$. The key idea that will guide our algorithm design is to exploit the consistency budget as fully as possible during the initial $\rr$ stage before any prediction error is revealed, thereby minimizing the worst-case robustness while preserving $(1+\lambda)$-consistency. Before presenting the full algorithm, we introduce the
\textit{round-commit truncated cost function}, which enables calibrating
the transition point at which the consistency budget has been exactly exhausted.


\paragraph{Round-Commit Truncated Cost Function.} The key challenge is to determine the exact point at which the algorithm should terminate the initial Round-Robin phase and commit to processing the jobs in the predicted order $(y_1, \ldots, y_n)$. Stopping too early leaves part of the consistency budget unused, thereby resulting in an unnecessarily large robustness, while stopping too late may violate the desired $(1+\lambda)$-consistency guarantee. To characterize the cost that has been accumulated at any potential transition point, we introduce the \emph{Round-Commit Truncated Cost Function}.

\begin{definition}[Round-Commit Truncated Cost Function]
\label{def:round_commit_truncated_cost}
For any prediction vector $\mathbf y=(y_1,\ldots,y_n)$, index $v\in[n]$,
service level $a\in[y_{v-1},y_v]$ (with $y_0=0$ by convention), and integer truncation
index $v-1\leq\ell\leq n$, consider the following schedule:
the algorithm first runs Round-Robin ($\rr$) until every
job $i$ has received $\min\{y_i,a\}$ units of processing;
it then switches to processing the remaining jobs in
the predicted order. Under this schedule, assuming that no misprediction is detected,
let $\varGamma_\ell^{(v)}(\mathbf y,a)$ denote the total
cost accumulated up to the completion of job $\ell$
when $\ell\geq v$.
If $\ell=v-1$, $\varGamma_\ell^{(v)}(\mathbf y,a)$ will instead denote the total cost
accumulated up until the transition from $\rr$ to \textcolor{black}{the predicted order processing phase.}
Formally,
\begin{equation}
\label{varGamma_ell}
\varGamma_\ell^{(v)}(\mathbf y,a)
\coloneq
\sum_{i=1}^{v-1}(2n+1-2i)y_i
+(n+1-v)^2a
+\sum_{i=v}^{\ell}(n+1-i)(y_i-a),
\end{equation}
where an empty sum is interpreted as zero.
\end{definition}

\begin{figure}[t]
     \centering
     \includegraphics[width=0.75 \linewidth]{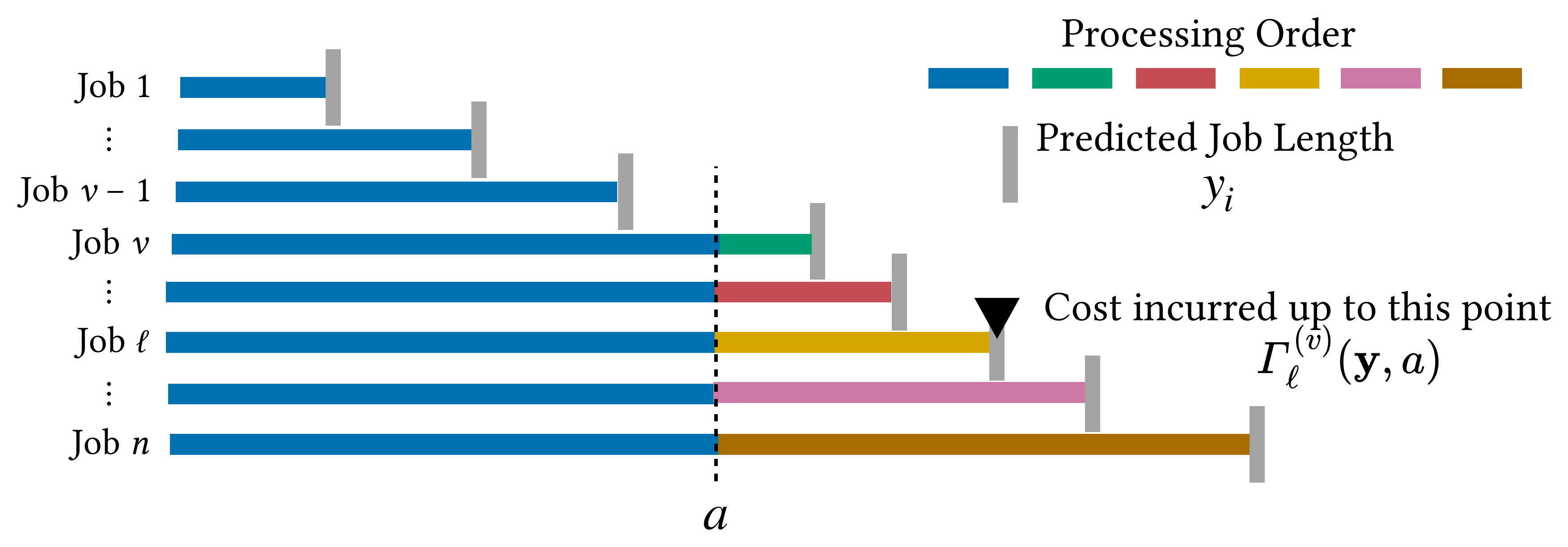}
     \caption{\textbf{An illustration of the Round-Commit Truncated Cost}
in Definition~\ref{def:round_commit_truncated_cost}.
Blue segments represent processing done under $\rr$;
the other colored segments represent processing done on jobs one at a time in the predicted order after $\rr$.}
     \label{fig:gamma-illustration}
 \end{figure}
    
\textcolor{black}{Figure~\ref{fig:gamma-illustration} illustrates
$\varGamma_\ell^{(v)}(\mathbf y,a)$. Then our algorithm \emph{Modified Two-Stage Schedule} ($\mTSS$) operates as follows:
we first evaluate $\varGamma_n^{(n)}(\mathbf y,y_n)$ to determine whether running only $\rr$ throughout will stay within the consistency budget $(1+\lambda)\opt_{\mathbf y}$ (assuming accurate predictions); if so, then we simply run $\rr$ the whole time.
Otherwise, we choose $v$ and $r\in[y_{v-1},y_v]$ such that $\smash{\varGamma_n^{(v)}(\mathbf y,r)}
=(1+\lambda)\opt_{\mathbf y}$.
We call this choice \emph{exact consistency calibration}:
the algorithm switches from $\rr$ to processing jobs in
the predicted order exactly when its consistency
budget has been exhausted. If at any time a prediction error is detected, we switch back to running $\rr$.
We present the formal description of $\mTSS$ in Algorithm~\ref{alg:mTSS}; in the remainder of this section, we will establish its consistency and robustness guarantees.}

\begin{algorithm}[t]
\DontPrintSemicolon
\SetKwInOut{Input}{Input}
\textbf{Input:}
$n$ jobs with predicted lengths $\mathbf{y}$; tradeoff parameter $\lambda \in \bigl[0,\tfrac{n-1}{n+1}\bigr]$

\caption{$\mTSS$: Modified Two-Stage Schedule}
\label{alg:mTSS}

\BlankLine
\textbf{Global interrupt rule.}
Whenever a prediction error is detected, immediately run $\rr$ until all jobs complete, and terminate\;

\BlankLine
\textbf{Stage 1. Round-Robin.}
\eIf{$\varGamma_n^{(n)}(\mathbf{y},y_n) < (1+\lambda)\opt_{\mathbf y}$}{
    Run $\rr$ until all jobs complete\;
}{
    $v \gets 1$\;
    \While{$\varGamma_n^{(v)}(\mathbf{y},y_v) < (1+\lambda)\opt_{\mathbf y}$}{
        $v \gets v+1$\;
    }
    Choose $r$ such that $\varGamma_n^{(v)}(\mathbf{y},r) = (1+\lambda)\opt_{\mathbf y}$\tcp*[r]{Saturate consistency}
    Run $\rr$ for $\sum_{i=1}^{v-1}y_i+(n+1-v)r$ time units\;
}

\BlankLine
\textbf{Stage 2. Commit.} Process the remaining jobs nonpreemptively, one at a time until completion, in nondecreasing order of predicted length\;
\end{algorithm}

\subsection{Reduction to the Time of Error Detection}
\label{subsec:detection_reduction}

The core challenge in analyzing the robustness of $\mTSS$
is handling general adversarial instances, since the space
of true processing times $\mathbf x$ and predictions
$\mathbf y$ can be high-dimensional.
We approach this challenge from the perspective of the
algorithm's execution.
Once a prediction error is detected, $\mTSS$ switches to $\rr$; we can show that the competitive ratio guarantee of $\rr$ controls the algorithm's performance on the remaining processing times.
This reduces the problem of robustness analysis to characterizing algorithm performance at the time of error detection.

To distinguish the processing already performed from the
remaining work, we extend the offline optimum to vectors
that may contain zero coordinates.
For any $\mathbf w\in\mathbb R_{\geq0}^n$, let
\[
\opt({\mathbf w})
\coloneq
\min_{\pi\in\mathfrak S_n}
\sum_{i=1}^n(n+1-i)w_{\pi(i)}
\]
denote the offline optimal total completion time for jobs
with processing times $\mathbf w$, where $\mathfrak S_n$
is the set of permutations of $[n]$. \textcolor{black}{Viewing $\pi$ as a permutation matrix, the objective is linear in $\pi$ (it is also linear in $\mathbf w$); thus $\opt({\mathbf w})$ can equivalently be expressed as the optimal value of a linear program over the Birkhoff polytope of doubly stochastic matrices.}
The following lemma reduces the worst-case competitive-ratio analysis
to the cost accumulated up to the transition to $\rr$,
which occurs at error detection in $\mTSS$.

\begin{lemma}
\label{lem:rr_continuation}
Suppose an algorithm $\mA$ for non-clairvoyant scheduling
switches to $\rr$ at time $T$, and then runs $\rr$ until all jobs complete.
Let $\mathbf a \in \mathbb{R}_{\geq0}^n$ denote the processing amounts received by
all jobs by time $T$, including the full processing times
of completed jobs, and let $\mathbf z=\mathbf x-\mathbf a$
denote their remaining processing times.
Let $C(t)$ denote the cost incurred by $\mA$ at time $t$.
If $\opt({\mathbf a})>0$, then the competitive ratio of $\mA$ on the instance $\mx$ satisfies
\[
\frac{\operatorname{ALG}(\mA, \mx)}{\opt({\mathbf x})}
\leq
\max\left\{
\frac{C(T)}{\opt({\mathbf a})},
\frac{2n}{n+1}
\right\}.
\]
\end{lemma}

\begin{proof}
The definition of the offline optimum gives
\[
\opt(\mathbf a+\mathbf z)
\geq
\opt(\mathbf a)+\opt(\mathbf z),
\]
since the minimum of a sum is at least the sum of the minima. The cost accumulated by $\mA$ after time $T$ is exactly the cost of
running $\rr$ on the positive coordinates of $\mathbf z$;
there are at most $n$ such jobs, so this cost is at most
$\frac{2n}{n+1}\opt(\mathbf z)$ \citep{Motwani1994}.
Therefore,
\[
\frac{\operatorname{ALG}(\mA,\mx)}{\opt(\mathbf x)}
\leq
\frac{
C(T)+\frac{2n}{n+1}\opt(\mathbf z)
}{
\opt(\mathbf a)+\opt(\mathbf z)
}
\leq
\max\left\{
\frac{C(T)}{\opt(\mathbf a)},
\frac{2n}{n+1}
\right\}.
\]
\end{proof}

For fixed predictions $\mathbf y$, we can use
Lemma~\ref{lem:rr_continuation} to reduce the robustness
analysis over adversarial true processing times $\mx$
to processing times drawn from a finite family of truncated prediction vectors.

\begin{lemma}
\label{lem:detection_reduction}
Suppose the calibration step in $\mTSS$ (Algorithm~\ref{alg:mTSS}) selects parameters $v$ and $r$.
For $v-1\leq s\leq n$, define $\bar{\mathbf y}^{(s)}
\coloneq
\bigl(
y_1,\ldots,y_s,
{r,\ldots,r}
\bigr) \in \mathbb{R}_{\geq 0}^n$.
Then, for every processing-time vector $\mx>0$,
\begin{equation}
\label{eq:goal}
\frac{\operatorname{ALG}(\mTSS \,,\mx)}{\opt(\mx)}
\leq
\max\left\{
\frac{2n}{n+1},
\max_{v\leq s\leq n}
\frac{\varGamma_s^{(v)}(\mathbf y,r)}
{\opt(\bar{\mathbf y}^{(s)})}
\right\}.
\end{equation}
\end{lemma}

\begin{proof}
If calibration selects $\rr$ throughout or the first error
is detected during the initial $\rr$ stage, $\mTSS$ only runs
$\rr$ and has competitive ratio at most $2n/(n+1)$.
We therefore consider the case where the first error is
detected during the commit stage.
Let $v$ and $r$ denote the parameters selected by calibration
in Algorithm~\ref{alg:mTSS}.

Let $M$ denote the right-hand side of \eqref{eq:goal}.
By definition,
\[
\varGamma_s^{(v)}(\mathbf y,r)
\leq M\cdot \opt(\bar{\mathbf y}^{(s)})
\qquad \text{for }v\leq s\leq n.
\]
This inequality also holds for $s=v-1$:
the cost of $\mTSS$ up to the transition
to the commit stage equals the total cost of running
$\rr$ throughout on the instance $\bar{\mathbf y}^{(v-1)}$.
Hence,
\[
\varGamma_{v-1}^{(v)}(\mathbf y,r)
\leq
\frac{2n}{n+1}\cdot\opt(\bar{\mathbf y}^{(v-1)})
\leq M\cdot \opt(\bar{\mathbf y}^{(v-1)}).
\]

Suppose the first error is detected while \textcolor{black}{running
job $s$ in the commit phase}.
Jobs $1,\ldots,s-1$ have completed with their predicted
processing times, while every job after $s$ has received
$r$ units of processing.

\emph{Under-prediction.}
If job $s$ has received $y_s$ units of processing and remains unfinished, the attained
processing vector is $\bar{\mathbf y}^{(s)}$, and the
accumulated cost is $\varGamma_s^{(v)}(\mathbf y,r)$.
Lemma~\ref{lem:rr_continuation} therefore bounds the
competitive ratio by $M$.

\emph{Over-prediction.}
Suppose job $s$ completes at service $a$, where $r<a<y_s$.
The attained processing vector and accumulated cost are
\[
\mathbf a
=
(y_1,\ldots,y_{s-1},a,r,\ldots,r),
\qquad
P
=
\varGamma_{s-1}^{(v)}(\mathbf y,r)
+(n+1-s)(a-r).
\]
Define $\theta=(a-r)/(y_s-r)\in(0,1)$.
Then
\[
\mathbf a
=(1-\theta)\bar{\mathbf y}^{(s-1)}
+\theta\bar{\mathbf y}^{(s)},\qquad 
P
=(1-\theta)\varGamma_{s-1}^{(v)}(\mathbf y,r)
+\theta\varGamma_s^{(v)}(\mathbf y,r).
\]
Since $\opt(\mathbf w)$ is the minimum of linear
functions of $\mathbf w$, it is concave. Consequently,
\[
P
\leq
M\cdot\bigl(
(1-\theta)\opt(\bar{\mathbf y}^{(s-1)})
+\theta\opt(\bar{\mathbf y}^{(s)})
\bigr)
\leq M\cdot \opt(\mathbf a).
\]
Applying Lemma~\ref{lem:rr_continuation} once again
bounds the competitive ratio by $M$.

Finally, if no error is detected, the predictions are
accurate, and the competitive ratio is
\[
\frac{\varGamma_n^{(v)}(\mathbf y,r)}
{\opt(\mathbf y)}
=1+\lambda\leq \frac{2n}{n+1} \leq M.
\]
This completes the proof.
\end{proof}

Thus both types of prediction error reduce to the same
family of truncated vectors $\bar{\mathbf y}^{(s)}$.
Since $y_{v-1}\leq r\leq y_v$, the offline costs of their induced instances are
\[
\opt(\bar{\mathbf y}^{(s)})
=
\sum_{i=1}^{v-1}(n+1-i)y_i
+
\frac{(n-s)(n+s-2v+3)}2\,r
+
\sum_{i=v}^{s}(s+1-i)y_i.
\]
For fixed $v$ and $s$, both this expression and
$\varGamma_s^{(v)}(\mathbf y,r)$ are linear in
$(\mathbf y,r)$.
Bounding the worst-case robustness therefore reduces to
a linear-fractional optimization problem, subject to the
ordering constraints and the calibration equality.

\subsection{A Nearly-Optimal Upper Bound for General $n$}

Building on the preceding reduction, we can exploit the
linear-fractional structure of the resulting cost ratios
and the calibration equality to derive a robustness bound
for $\mTSS$ for general $n$ that can be expressed as a maximization
over a single integer index.
We state the bound formally in the following theorem.




\begin{theorem}
\label{thm:mTSS_conrob}
For every $n\geq2$ and
$\lambda\in[0,\frac{n-1}{n+1}]$,
Algorithm $\mTSS$ (Algorithm~\ref{alg:mTSS}) is
$(1+\lambda)$-consistent and $\Psi_n(\lambda)$-robust, where
\[
\Psi_n(\lambda)
\coloneq
\max_{s\in[n]}
\frac{
2\lambda n^2
+\bigl(\tfrac{n-1}{n+1}-\lambda\bigr)s(2n+1-s)
}{
\lambda n(n+1)
+\bigl(\tfrac{n-1}{n+1}-\lambda\bigr)s(s+1)
}.
\]
\end{theorem}







Note that this bound improves empirically over previous
guarantees; see Figure~\ref{fig:combined_tradeoff}.
We sketch the proof of this theorem below; the full proof appears in Appendix~\ref{app:mTSS_conrob}.

\paragraph{\textbf{Proof Sketch for Theorem~\ref{thm:mTSS_conrob}}}
We first verify that, when $\lambda=0$, $\mTSS$ is
$1$-consistent and $n$-robust, matching $\Psi_n(0)=n$.

For $\lambda>0$, we establish $(1+\lambda)$-consistency
by checking that calibration respects the consistency budget.
For robustness, we first apply Lemma~\ref{lem:detection_reduction}
to reduce the analysis to computing cost ratios at error detection \eqref{eq:goal}.
For fixed calibration and detection indices $v$ and $s$,
the online cost in the reduced ratio is
$\varGamma_s^{(v)}(\mathbf y,r)$.
Since $v$ is fixed, the order of the entries in
$\bar{\mathbf y}^{(s)}$ is fixed.
Consequently, the corresponding offline cost
$\opt(\bar{\mathbf y}^{(s)})$ is linear in $(\mathbf y,r)$.
Thus both costs are linear in $(\mathbf y,r)$.
We seek to bound their ratio subject to the ordering
constraints and the calibration equality
\begin{equation}
\label{eq:calibration_equality}
\varGamma_n^{(v)}(\mathbf y,r)
=
(1+\lambda)\sum_{i=1}^n(n+1-i)y_i.
\end{equation}

The key step is to express these constraints and costs
through consecutive gaps.
Define
\[
\mathbf w
\coloneq
(y_1,\ldots,y_{v-1},r,y_v,\ldots,y_n),
\qquad
u_i\coloneq w_i-w_{i-1},
\quad i\in[n+1],
\]
where $w_0=0$.
Allowing zero predictions and replacing strict ordering
inequalities by weak inequalities gives an upper-bound relaxation
in which the ordering constraints are equivalent to
$\mathbf u\geq\mathbf0$.
Since every coordinate of $(\mathbf y,r)$ is linear in
$\mathbf u$, the two costs
\[
N(\mathbf u)\coloneq\varGamma_s^{(v)}(\mathbf y,r),
\qquad
D(\mathbf u)\coloneq\opt(\bar{\mathbf y}^{(s)})
\]
are linear, and calibration equality \eqref{eq:calibration_equality} becomes
$h^\top\mathbf u=0$ for a fixed coefficient vector
$h\in\mathbb R^{n+1}$.
The relaxed feasible set is therefore the cone
\[
K=\{\mathbf u\in\mathbb R_{\geq0}^{n+1}:
h^\top\mathbf u=0\}.
\]
The remaining task is thus to bound $N(\mathbf u)/D(\mathbf u)$
over $K\setminus\{\mathbf0\}$.
We prove a key structural property for this cone:
every nonzero feasible vector can be decomposed into
feasible vectors, each with at most two strictly positive coordinates.
By linearity of both cost expressions, the original cost ratio is thus
a weighted average of the component ratios, and therefore
cannot exceed their maximum. Specifically, we have the following result:

\begin{lemma}
\label{lem:gap-decomposition}
Let $h\in\mathbb R^q$ and define the cone
$K=\{\mathbf u\in\mathbb R_{\geq0}^q:h^\top\mathbf u=0\}$.
For every $\mathbf u\in K\setminus\{\mathbf0\}$,
there exist an integer $L\geq1$ and vectors
$\mathbf u^{(1)},\ldots,\mathbf u^{(L)}
\in K\setminus\{\mathbf0\}$ such that
\[
\mathbf u=\sum_{\ell=1}^L\mathbf u^{(\ell)},
\qquad
\bigl|\operatorname{supp}(\mathbf u^{(\ell)})\bigr|\leq2
\quad\text{for every }\ell\in\{1,\ldots,L\},
\]
where $\operatorname{supp}(\mathbf w)=\{i:w_i\neq0\}$.
Moreover, if $A,B:\mathbb R^q\to\mathbb R$ are linear and
$B(\mathbf w)>0$ for every
$\mathbf w\in K\setminus\{\mathbf0\}$, then
\[
\frac{A(\mathbf u)}{B(\mathbf u)}
\leq
\max_{1\leq\ell\leq L}
\frac{A(\mathbf u^{(\ell)})}{B(\mathbf u^{(\ell)})}.
\]
\end{lemma}

Applying Lemma~\ref{lem:gap-decomposition} with $A=N$ and
$B=D$, we reduce the analysis to feasible vectors with
at most two positive gaps. Each such vector can be written as
\[
\mathbf u=a\mathbf e_i+b\mathbf e_j,
\qquad
a,b\geq0,\quad a+b>0,
\]
where $\mathbf e_i$ and $\mathbf e_j$ are standard basis vectors.
Combining this representation with the calibration equality
and the scale invariance of the cost ratio yields an
optimization-based bound involving only a few integer indices.
A sequence of monotonicity and weighted-average comparisons eliminates the auxiliary indices, leaving a maximization
over the single integer variable $s$:
\[
\max_{s\in[n]}
\frac{
2\lambda n^2
+\bigl(\tfrac{n-1}{n+1}-\lambda\bigr)s(2n+1-s)
}{
\lambda n(n+1)
+\bigl(\tfrac{n-1}{n+1}-\lambda\bigr)s(s+1)
}
=
\Psi_n(\lambda).
\]
Finally, we verify that the case $s=1$ in the maximization is at least
$2n/(n+1)$, so this bound also covers the $\rr$ alternative
and completes the robustness analysis.\qed\\

We next show that $\mTSS$ approximates the optimal
consistency--robustness tradeoff within a factor of $3/2$,
uniformly over all $n$ and all admissible $\lambda$.
Specifically, its robustness is at most $3/2$ times the
best robustness that can be achieved by any $(1+\lambda)$-consistent online
algorithm.
The following theorem establishes this guarantee by comparing
$\Psi_n(\lambda)$ with our lower bound $\Phi^{(n)}(\lambda)$. 

\begin{theorem}\label{thm:structural-three-halves}
For every $n\ge2$ and $0\le\lambda\le(n-1)/(n+1)$,
\[
\Phi^{(n)}(\lambda)\leq \Psi_n(\lambda)\le\frac32 \cdot \Phi^{(n)}(\lambda),
\]
where $\Psi_n(\lambda)$ is defined in Theorem~\ref{thm:mTSS_conrob} and $\Phi^{(n)} (\lambda)$ is defined in Theorem~\ref{thm:piece-wise_lower_bound}.
\end{theorem}

A proof of Theorem~\ref{thm:structural-three-halves} is provided in Appendix~\ref{appendix:approximation}. The key step is to show that the smallest maximizing index
$s$ in the upper bound $\Psi_n(\lambda)$ does not exceed
the attack depth $\alpha$ selected in our lower-bound
construction.
We then use the corresponding attack constraint,
consistency budget, and monotonicity of the processing
amounts to derive an explicit lower bound on
$\Phi^{(n)}(\lambda)$.
Combining this lower bound on $\Phi^{(n)}(\lambda)$
with the weighted-average relation governing consecutive
terms in the maximization defining $\Psi_n(\lambda)$ yields the factor of $3/2$.

\vspace{-1.2em}
\subsection{A Universal Closed-Form Upper Bound}


Theorem~\ref{thm:mTSS_conrob} provides a finite-\(n\) robustness guarantee through the one-dimensional maximization defining \(\Psi_n(\lambda)\), which can be evaluated by enumerating its \(n\) candidate values in \(O(n)\) arithmetic operations.
Although this expression is readily computable, its dependence
on $\lambda$ is not immediately clear.
To obtain a more interpretable guarantee, we bound the maximization
using a quadratic inequality.
The resulting closed-form upper bound holds uniformly across all $n$,
makes the dependence on $\lambda$ explicit, and enables a direct
comparison with our lower bounds. We provide a proof of the bound in Appendix~\ref{app:analytic_upper}.

\begin{theorem} \label{thm:analytic_upper}
For any number of jobs $n \geq 2$ and any $\lambda \in (0, \tfrac{n-1}{n+1}]$, Algorithm~$\mTSS$ (Algorithm~\ref{alg:mTSS}) is $(1+\lambda)$-consistent and ${\left(\tfrac12+\sqrt{{\tfrac54}+1/\lambda}\right)}$-robust.

\end{theorem}

\subsection{Ensuring Smoothness via Error-Tolerant Simulation}
\label{sec:tolerance_smoothness}

The standard consistency--robustness framework controls
performance under perfect predictions and in the worst case,
but does not guarantee smooth degradation in performance under small
prediction errors. While prior work has developed smooth algorithms for
non-clairvoyant scheduling~\citep{
Im2023NonClairvoyantPredictions,Benomar2025ProgressBars,Zhao2026Smoothness},
we propose a generic tolerance-based
approach that preserves consistency and provides
\emph{local smoothness}: within a prescribed error range,
algorithm performance degrades smoothly as the
prediction error increases; this smoothness comes at the cost of only a small, controlled increase in robustness.

We define the coordinatewise relative prediction error as
\[
\varepsilon(\mathbf x,\mathbf y)
\coloneq
\max_{i\in[n]}
\left\{
\frac{x_i}{y_i},\frac{y_i}{x_i}
\right\}-1.
\]
The key idea is to hide prediction errors within a specified tolerance
range from $\mA$.
Whenever $\varepsilon(\mathbf x,\mathbf y)\leq\tau$,
the simulated execution of $\mA$ is identical to its execution
under perfectly accurate predictions.
Algorithm~\ref{alg:tolerance_simulation} implements this idea without knowing the true processing times in advance. As we show in the following theorem, this simulation preserves
consistency and provides local smoothness while only modestly increasing robustness.

\begin{algorithm}[t]
\DontPrintSemicolon

\caption{Prediction-Tolerant Simulation
($\mA^\tau_{\mathrm{tol}}$)}
\label{alg:tolerance_simulation}
\KwIn{Deterministic algorithm $\mA$,
predictions $\mathbf y>0$, and tolerance $\tau>0$.}

Initialize a simulated copy of $\mA$ with predictions $\mathbf y$. In the simulated execution, a job remains unfinished until we report its completion to $\mA$\;
Initialize the processing time allocated by the simulated
$\mA$ to each job: $s_i\gets0$ for all $i\in[n]$\;

\While{some job remains unfinished}{
    Follow $\mA$'s allocations,
increasing each $s_i$ accordingly and
leaving allocations to already completed jobs idle,
until a job completes or first reaches $s_i=y_i$\;

    \uIf{job $i$ first reaches $s_i=y_i$}{
        Pause $\mA$'s simulated execution, keeping its clock
and all $s_j$ unchanged\;

        \If{job $i$ is unfinished}{
            Process $i$ alone until it completes
            or $\tau y_i$ additional time elapses\;
        }

        \If{job $i$ has completed}{
            Report job $i$'s completion to $\mA$
at simulated service $y_i$\;
        }

        Resume $\mA$\;
    }
    \Else{
        \eIf{$s_i<y_i/(1+\tau)$ or $s_i>y_i$}{
            Report job $i$'s completion to $\mA$
at simulated service $s_i$\;
        }{
            Withhold job $i$'s completion report until $s_i=y_i$\;
        }
    }
}
\end{algorithm}

\begin{theorem}[Prediction-tolerant simulation]
\label{thm:tolerance_simulation}
Suppose $\mA$ is a deterministic, $\beta$-consistent
and $\gamma$-robust algorithm for non-clairvoyant scheduling.
For every $\tau>0$, Algorithm~\ref{alg:tolerance_simulation}
is $\beta$-consistent and $(1+\tau)^2\gamma$-robust.
Moreover, whenever
$\varepsilon\coloneq\varepsilon(\mathbf x,\mathbf y)\leq\tau$,
\[
\frac{
\operatorname{ALG}(\mA^\tau_{\mathrm{tol}},I)
}{
\operatorname{OPT}(\mathbf x)
}
\leq\beta(1+\varepsilon)^2.
\]
\end{theorem}

A proof of this theorem is presented in Appendix~\ref{app:tolerance_simulation_proof}. Letting $\mTSS_{\lambda,\mathrm{tol}}^\tau$ denote the
prediction-tolerant simulation of $\mTSS$ via Algorithm~\ref{alg:tolerance_simulation}, we immediately obtain the following corollary showing that we can maintain the near-optimal robustness-consistency guarantees of $\mTSS$ while ensuring smoothness to small prediction errors.

\begin{corollary}
\label{cor:smooth_mtss}
For every $n\geq2$,
$\lambda\in[0,\frac{n-1}{n+1}]$, and $\tau>0$,
algorithm $\mTSS_{\lambda,\mathrm{tol}}^\tau$ is
$(1+\lambda)$-consistent and
$(1+\tau)^2\Psi_n(\lambda)$-robust.
Moreover, whenever
$\varepsilon\coloneq\varepsilon(\mathbf x,\mathbf y)\leq\tau$,
\[
\frac{
\operatorname{ALG}(\mTSS_{\lambda,\mathrm{tol}}^\tau,I)
}{
\operatorname{OPT}(\mathbf x)
}
\leq(1+\lambda)(1+\varepsilon)^2.
\]
\end{corollary}

\section{Conclusion and Future Directions}
In this work, we provide a nearly-optimal characterization of
the consistency--robustness tradeoff for non-clairvoyant
scheduling with predictions.
We show that the lower bound of \citet{Wei2020} is not tight
for general $n$ and introduce a multi-stage adversarial
strategy that yields a strictly stronger piecewise lower bound
for $n\geq5$ over a nontrivial range of $\lambda$.
Complementing this lower bound, we propose the \emph{Modified Two-Stage
Schedule} algorithm ($\mTSS$), which achieves a nearly-optimal consistency--robustness tradeoff, with an upper bound within a factor of $3/2$ of our
lower bound for every $n\geq2$ and any admissible $\lambda$.

These two developments allow us to paint a complete asymptotic picture of the optimal consistency--robustness tradeoff for this problem. As the number of jobs $n \to \infty$, our results establish that the best achievable robustness for a $(1+\lambda)$-consistent algorithm scales as $\sqrt{1/\lambda}$. Furthermore, our results resolve (up to constant factors) the polynomial gap between the upper and lower bounds on robustness in the worst-case-$n$ regime~\citep{Benomar2025ProgressBars}. Together, our results suggest that the key to optimal scheduling with predictions lies in the effective management of the algorithm's \textit{flexibility budget}, i.e., balancing following the predictions directly with preserving robustness with a more conservative strategy. We also show that these guarantees can be made error-tolerant, developing a generic approach to enable learning-augmented algorithms to provide locally smooth degradation to small prediction errors. 

Our work suggests several potentially promising directions for future work.
First, a natural extension is to characterize near-optimal
consistency--robustness tradeoffs for non-clairvoyant scheduling
with known job weights and predicted processing times,
complementing existing results for weighted scheduling with
other forms of predictions~\citep{Lindermayr2022Permutation}.
Second, our results could potentially serve as a foundation for refining
consistency--robustness tradeoffs under alternative prediction
models, including partial predictions~\citep{Benomar2024PartialPredictions},
permutation predictions~\citep{Lindermayr2022Permutation},
and progress bar predictions~\citep{Benomar2025ProgressBars}. Third, it would be interesting to explore whether the proposed optimization-based formulation of multi-stage lower bounds can be extended to other online problems such as metrical task systems in order to obtain tighter, instance-dependent lower bounds. 
Finally, closing the small gap for finite $n \ge 5$ may require new technical ideas, particularly on the upper bound front. To this end, we conjecture that our lower bound is tight for every \(n\), i.e., the optimal robustness among \((1+\lambda)\)-consistent algorithms for \(n\)-job non-clairvoyant scheduling is exactly \(\Phi^{(n)}(\lambda)\).\footnote{The asymptotic case of this conjecture, as \(n \to \infty\), was resolved in concurrent work by \cite{benomarPerchet2026}.} Achieving this bound may require an algorithm that repeatedly alternates between \(\rr\) and committing to the predicted order, with the duration of each \(\rr\) phase determined by the Order-\(\alpha\) system.

\section*{AI Use Acknowledgment}
The authors used AI tools including OpenAI's
ChatGPT and Codex in this work (with models ranging from GPT-5.1 to
GPT-6 Astra) to assist with initial proof development and drafting based on the authors' proof ideas, to perform limited editing for language and grammar, and to help simplify and check the correctness of mathematical proofs. 
All AI-assisted content included in this paper has been reviewed,
verified, and revised by the authors, who take
full responsibility for the content of this paper.

\bibliographystyle{plainnat}
\bibliography{bibliography}

\newpage

\clearpage
\appendix




\section{Properties of the Value Function $\vg$}
In this section, we establish the well-definedness of $\vg^{(n)}$  (Appendix~\ref{app:well_p}) and prove a monotonicity property for the function ${\vg}_{\cdot, n}$ (Appendix~\ref{app:improve}).

\subsection{Well-Posedness of the Function $\vg^{(n)}$} \label{app:well_p}

To establish the well-posedness of $\vg^{(n)}$, we first introduce the \emph{Reversed Order-$\alpha$ System}, which replicates the equations in the Order-$\alpha$ System (see Definition~\ref{def:order-m-system}) while treating $\lambda$ as the decision variable instead of $\rob$.

\begin{definition}[Reversed Order-$\alpha$ System]\label{def:rev_order-m-system}
Fix integers $n\ge 2$ and $\alpha \in [m]$,
and $\rob \in [\frac{2n}{n+1} ,\frac{n+1-\alpha}{\alpha}]$, where $m = \max\left\{ \alpha \in [n-1] : \frac{n+1-\alpha}{\alpha} > \frac{2n}{n+1} \right\}$. 
The \textbf{Reversed Order-$\alpha$ System} $\mathsf{Sys}^{\mathsf{rev}}_\alpha(n,\rob)$ is a collection of $\alpha + 1$ equations in the variables $(\lambda,p^{(1)},\dots,p^{(\alpha)})$ 
given by \eqref{eq:order_alpha_1} and \eqref{eq:order_alpha_2}.
\end{definition}

We next establish that the reversed order-$\alpha$ system is well-posed, in the sense that it admits a unique solution for every admissible parameter value.

\begin{lemma}[Existence and Uniqueness of the Reversed Order-$\alpha$ System]\label{clm:unique}
Let $n \ge 2$ and $\alpha \in [m]$, where
$m = \max\left\{ \alpha \in [n-1] : \frac{n+1-\alpha}{\alpha} > \frac{2n}{n+1} \right\}$.
For any
$\rob \in \left[\frac{2n}{n+1}, \frac{n+1-\alpha}{\alpha} \right]$,
the reversed system $\mathsf{Sys}^{\mathsf{rev}}_\alpha(n,\rob)$ admits a unique solution
$(\lambda,p^{(1)},\dots,p^{(\alpha)})$
in $[0,\tfrac{n-1}{n+1}] \times [0,1]^\alpha$.
\end{lemma}

\begin{proof}
By rearranging the terms in~\eqref{eq:order_alpha_1}, we have for all $i \in [\alpha]$,
\begin{equation}
\label{eq:order_alpha_1_re}
(n-i) \left( \frac{(n+1+i) \rob}{2} -(n+1-i) \right) p^{(i)} = \sum_{j=1}^{i-1} (n-j) p^{(j)} + \frac{i}{2}\cdot \bigr( (2n+1-i) - (i+1) \rob\bigr).
\end{equation}
Since $\rob \in \left[\frac{2n}{n+1},
\;\frac{n+1-\alpha}{\alpha}
\right] \subseteq
\bigcup_{\alpha=1}^{m}
\left[\frac{2n}{n+1},
\frac{n+1-\alpha}{\alpha}
\right]
=
\left[\frac{2n}{n+1},\, n\right]$, we have
\begin{equation}
\label{positive}
    \frac{(n+1+i) \rob}{2} -(n+1-i)  \geq \frac{(n+2)\rob}{2} - n \geq \frac{n+2}{2}\cdot \frac{2n}{n+1} - n =\frac{n}{n+1} > 0.
\end{equation}
Setting $i=1$ in~\eqref{eq:order_alpha_1_re}, we obtain
\begin{equation}
    (n-1)\left( \frac{(n+2)\rob}{2} - n \right) p^{(1)} = n - \rob.
\end{equation}
It follows immediately that $p^{(1)}$ is a strictly decreasing function of $\rob$. Now suppose that, for all $i \in [k]$, $p^{(i)}$ is a strictly decreasing function of $\rob$. 
Consider~\eqref{eq:order_alpha_1_re} with $i = k+1$.
By~\eqref{positive}, the coefficient
\[
(n-(k+1))\left(\frac{(n+k+2)\rob}{2}-(n-k)\right)
\]
on the left-hand side is strictly positive and strictly increasing in $\rob$.
By the assumption, the summation term $\sum_{j=1}^{k}(n-j)p^{(j)}$ on the right-hand side is strictly decreasing in $\rob$, while the remaining term
$
\tfrac{k+1}{2}\bigl((2n-k)- (k+2)\rob\bigr)
$
is also strictly decreasing in $\rob$. Therefore, using the non-negativity established below, $p^{(k+1)}$ is strictly decreasing in $\gamma$.
By induction on $i$, it follows that $p^{(i)}$ is a strictly decreasing function of $\rob$ for all $i \in [\alpha]$.

We define
\begin{equation}
\label{eq:S_def}
S_\alpha(\rob)
\;\coloneq\;
\sum_{j=1}^{\alpha-1}(n-j)\,p^{(j)}(\rob)
+\frac{(n+1-\alpha)(n-\alpha)}{2}\,p^{(\alpha)}(\rob),
\end{equation}
where we emphasize the dependence of each $p^{(i)}$ on $\rob$ induced by~\eqref{eq:order_alpha_1_re}.
Since each $p^{(i)}(\rob)$ is strictly decreasing in $\rob$ and all coefficients in~\eqref{eq:S_def} are positive, it follows that $S_\alpha(\rob)$ is strictly decreasing in $\rob$.
Since $p^{(i)}$ is uniquely determined by $\rob$ through \eqref{eq:order_alpha_1_re}, for any fixed $\rob\in[\tfrac{2n}{n+1},\frac{n+1-\alpha}{\alpha}]$, there exists at most one $\lambda\in[0,\frac{n-1}{n+1}]$ such that 
$S_\alpha (\rob) = \tfrac{n(n+1)}{2}\lambda$, thereby satisfying \eqref{eq:order_alpha_2}.
This proves the uniqueness.

We next prove the existence result. We claim that $p^{(i)}\in[0,1]$ for all $i\in[\alpha]$. We prove this by induction on $i$. Suppose that $p^{(j)}\in[0,1]$ for all $j<i$, where the case $i=1$ is understood with the empty sum equal to zero. Then
$0\le \sum_{j=1}^{i-1}(n-j)p^{(j)}\le \sum_{j=1}^{i-1}(n-j)$.
Moreover, since $i\le \alpha$, we have
$\frac{n+1-\alpha}{\alpha}\le \frac{2n+1-i}{i+1}$.
Indeed, this is equivalent to $(i+1)(n+1-\alpha)\le \alpha(2n+1-i)$, and the difference between the right-hand side and the left-hand side is $(n+1)(2\alpha-i-1)\ge 0$. Therefore,
$\rob\le \frac{2n+1-i}{i+1}$, and hence
$\frac{i}{2}\bigl((2n+1-i)-(i+1)\rob\bigr)\ge 0$.
It follows from~\eqref{eq:order_alpha_1_re} and~\eqref{positive} that $p^{(i)}\ge 0$.

It remains to show that $p^{(i)}\le 1$. By the induction hypothesis,
\[
\sum_{j=1}^{i-1}(n-j)p^{(j)}
+\frac{i}{2}\bigl((2n+1-i)-(i+1)\rob\bigr)
\le
\sum_{j=1}^{i-1}(n-j)
+\frac{i}{2}\bigl((2n+1-i)-(i+1)\rob\bigr).
\]
On the other hand, a direct calculation gives
\begin{align*}
& (n-i)\left(\frac{(n+1+i)\rob}{2}-(n+1-i)\right)
-\sum_{j=1}^{i-1}(n-j)
-\frac{i}{2}\bigl((2n+1-i)-(i+1)\rob\bigr)  \\
&\qquad =
\frac{n}{2}\bigl((n+1)\rob-2n\bigr)
\ge 0,
\end{align*}
where the last inequality follows from $\rob\ge \frac{2n}{n+1}$. Thus the right-hand side of~\eqref{eq:order_alpha_1_re} is at most the coefficient of $p^{(i)}$ on the left-hand side. Since this coefficient is strictly positive, we obtain $p^{(i)}\le 1$. Therefore $p^{(i)}\in[0,1]$. By induction, $p^{(i)}\in[0,1]$ for all $i\in[\alpha]$.

Now set
\[
\lambda
=
\frac{2}{n(n+1)}
S_\alpha (\rob).
\]
Then~\eqref{eq:order_alpha_2} holds by construction. Moreover, since $p^{(i)}\in[0,1]$ for all $i\in[\alpha]$,
\[
0
\le
S_\alpha (\rob)
\le
\sum_{j=1}^{\alpha-1}(n-j)
+\frac{(n+1-\alpha)(n-\alpha)}{2}
=
\frac{n(n-1)}{2}.
\]
Hence $\lambda\in\left[0,\frac{n-1}{n+1}\right]$. Therefore
$(\lambda,p^{(1)},\dots,p^{(\alpha)})\in[0,\frac{n-1}{n+1}]\times[0,1]^\alpha$
satisfies~\eqref{eq:order_alpha_1} and~\eqref{eq:order_alpha_2}. This proves the existence.

\end{proof}

Lemma~\ref{clm:unique} implies that, for every admissible $\rob$, the reversed order-$\alpha$ system admits a unique solution, thereby inducing a well-defined mapping from $\rob$ to $\lambda$.

\begin{definition}[$\Lambda$-Function]\label{def:Lambda_func}
Fix integers $n\geq 2$ and $\alpha \in [m]$ and a parameter $\rob \in \left[ \frac{2n}{n+1}, \frac{n+1-\alpha}{\alpha} \right]$, where $m = \max\left\{ \alpha \in [n-1] : \frac{n+1-\alpha}{\alpha} > \frac{2n}{n+1} \right\}$. Let $(\lambda,\,p^{(1)},\dots,p^{(\alpha)}) \in [0, \tfrac{n-1}{n+1}]\times[0,1]^\alpha$ be the solution to $\mathsf{Sys}^{\mathsf{rev}}_\alpha(n,\rob)$.
The \textbf{value function}
$
\Lambda_{\alpha,n}
$ is defined as the map from $\rob$ to the unique admissible value of $\lambda$.
\end{definition}

The following lemma establishes regularity properties of the $\Lambda$-function, including continuity and monotonicity with respect to $\rob$.

\begin{lemma} \label{lem:con_de}
    Fix integers $n\geq 2$ and $\alpha \in [m]$, where $m = \max\left\{ \alpha \in [n-1] : \frac{n+1-\alpha}{\alpha} > \frac{2n}{n+1} \right\}$. Then $\Lambda_{\alpha, n} (\rob)$ is continuous and strictly decreasing in $ \rob \in \left[\frac{2n}{n+1}, \frac{n+1-\alpha}{\alpha} \right]$.
\end{lemma}

\begin{proof}

By Lemma~\ref{clm:unique}, for each admissible $\rob$, the reversed order-$\alpha$ system admits a unique solution, so the map $\Lambda_{\alpha,n}$ is well-defined.

As shown in the proof of Lemma~\ref{clm:unique},
$\Lambda_{\alpha,n}(\rob)=\tfrac{2}{n(n+1)}S_\alpha(\rob)$
is strictly decreasing in $\rob$. 
Moreover, the triangular recurrence and the strictly positive
denominators established in that proof imply, by induction,
that each $p^{(i)}(\rob)$ is continuous in $\rob$.
Since $S_\alpha(\rob)$ is a weighted sum of these processing
amounts, it follows that $\Lambda_{\alpha,n}(\rob)$ is
continuous in $\rob$.
\end{proof}

We next prove continuity of $\Lambda_{\alpha, n}$ across adjacent selections of $\alpha$.

\begin{lemma}
\label{lem:Lambda_eq}
Let $m = \max\left\{ \alpha \in [n-1] : \frac{n+1-\alpha}{\alpha} > \frac{2n}{n+1} \right\}$. Then
\[
\Lambda_{\alpha, n} (\tfrac{n-\alpha}{\alpha+1}) = \Lambda_{\alpha+1, n} (\tfrac{n-\alpha}{\alpha+1}), \qquad \text{for all } \alpha \in [m-1],
\]
where $\Lambda_{\alpha, n} (\cdot)$ is defined in Definition~\ref{def:Lambda_func}.
\end{lemma}

\begin{proof}
Consider $\alpha\in [m-1]$. Let $\lambda_\alpha \coloneq \Lambda_{\alpha, n} (\tfrac{n-\alpha}{\alpha+1})$. Let 
\[
(\lambda_\alpha, p^{(1)}, p^{(2)}, \ldots, p^{(\alpha)})\in [0,\tfrac{n-1}{n+1}]\times[0,1]^\alpha
\]
be a solution to $\mathsf{Sys}^{\mathsf{rev}}_\alpha(n, \tfrac{n-\alpha}{\alpha+1})$. By Definition~\ref{def:rev_order-m-system}, we have
\begin{equation}
\label{eq:order_alpha_1_B}
\frac{(2n+1-i)\,i}{2}
+\sum_{j=1}^{i-1}(n-j)\,p^{(j)}
+(n+1-i)(n-i)\,p^{(i)}
=\frac{n-\alpha}{\alpha+1}\! \cdot \!\left(\frac{i(i+1)}{2}
+\frac{(n+i+1)(n-i)}{2}\,p^{(i)}\right)
\end{equation}
for every $i\in [\alpha]$, and\vspace{-1em}
\begin{equation}
\label{eq:order_alpha_2_B}
\lambda_\alpha\cdot \frac{n(n+1)}{2}
=\sum_{j=1}^{\alpha-1}(n-j)\,p^{(j)}
+\frac{(n+1-\alpha)(n-\alpha)}{2}\,p^{(\alpha)}.
\end{equation}

We construct $\hat{p}^{(i)} = p^{(i)}$ for all $i \in [\alpha]$ and $\hat{p}^{(\alpha+1)} = p^{(\alpha)}$. By construction, together with \eqref{eq:order_alpha_1_B}, we have
\begin{equation}
\label{eq:order_alpha_1_(1-alpha)}
\frac{(2n+1-i)\,i}{2}
+\sum_{j=1}^{i-1}(n-j)\,\hat{p}^{(j)}
+(n+1-i)(n-i)\,\hat{p}^{(i)}
=\frac{n-\alpha}{\alpha+1}\! \cdot \!\left(\frac{i(i+1)}{2}
+\frac{(n+i+1)(n-i)}{2}\,\hat{p}^{(i)}\right)
\end{equation}
for every $i\in [\alpha]$. Invoking \eqref{eq:order_alpha_1_(1-alpha)} with $i = \alpha$, we have
\begin{equation*}
\frac{(2n+1-\alpha)\,\alpha}{2}
+\sum_{j=1}^{\alpha-1}(n-j)\,\hat{p}^{(j)}
+(n+1-\alpha)(n-\alpha)\,\hat{p}^{(\alpha)}
=\frac{n-\alpha}{\alpha+1}\! \cdot \!\left(\frac{\alpha(\alpha+1)}{2}
+\frac{(n+\alpha+1)(n-\alpha)}{2}\,\hat{p}^{(\alpha)}\right).
\end{equation*}
Adding $(n-\alpha)- (n-\alpha) \hat{p}^{(\alpha)}$ to both sides of the above equation and using $\hat{p}^{(\alpha+1)} = p^{(\alpha)} = \hat{p}^{(\alpha)}$ gives
\begin{align}
&\frac{(2n-\alpha)\,(\alpha+1)}{2}
+\sum_{j=1}^{\alpha}(n-j)\,\hat{p}^{(j)}
+(n-\alpha)(n-\alpha-1)\,\hat{p}^{(\alpha+1)} \notag\\
&\qquad \qquad \qquad 
=\frac{n-\alpha}{\alpha+1}\! \cdot \!\left(\frac{(\alpha+1)(\alpha+2)}{2}
+\frac{(n+\alpha+2)(n-\alpha-1)}{2}\,\hat{p}^{(\alpha+1)}\right)  \label{eq:order_alpha_1_(alpha+1)}.
\end{align}
Also note that 
\begin{align}
    \sum_{j=1}^{\alpha} (n-j) \hat{p}^{(j)} &+ \frac{(n-\alpha)(n-\alpha-1)}{2} \hat{p}^{(\alpha+1)}
    =\sum_{j=1}^{\alpha-1} (n-j) \hat{p}^{(j)} + (n-\alpha)\hat{p}^{(\alpha)} + \frac{(n-\alpha)(n-\alpha-1)}{2} \hat{p}^{(\alpha+1)} \notag\\
    &=\sum_{j=1}^{\alpha-1} (n-j)p^{(j)} + \frac{(n+1-\alpha)(n-\alpha)}{2} p^{(\alpha)} = \lambda_\alpha \cdot \frac{n(n+1)}{2} \label{eq:order_alpha_2_}
\end{align}

Since $\alpha + 1 \leq m$, we have $\tfrac{n-\alpha}{\alpha+1} > \tfrac{2n}{n+1}$. Combining \eqref{eq:order_alpha_1_(1-alpha)}, \eqref{eq:order_alpha_1_(alpha+1)} and \eqref{eq:order_alpha_2_}, it follows that 
\[
(\lambda_\alpha, \hat{p}^{(1)}, \ldots, \hat{p}^{(\alpha+1)}) \in [0, \tfrac{n-1}{n+1}]\times[0,1]^{\alpha+1}
\] 
is a solution to $\mathsf{Sys}^{\mathsf{rev}}_{\alpha+1}(n,\tfrac{n-\alpha}{\alpha+1})$. By Definition~\ref{def:Lambda_func}, $\lambda_\alpha = \Lambda_{\alpha+1, n}(\tfrac{n-\alpha}{\alpha+1})$. Combining this with the definition of $\lambda_\alpha$ concludes the proof.
\end{proof}

Based on these foundations, we establish the well-definedness of $\Phi^{(n)}$.

\begin{lemma}
    For fixed integers $n\geq 2$, the function $\vg^{(n)} (\lambda)$ is well-defined, where $\vg^{(n)} (\cdot)$ is defined in \eqref{Gamma}.
\end{lemma}

\begin{proof}
By Definition~\ref{def:Lambda_func} and Lemmas~\ref{clm:unique} and~\ref{lem:con_de}, the map $\Lambda_{\alpha,n} (\rob)$ is well-defined, continuous, and strictly decreasing for $\rob \in \left[\frac{2n}{n+1}, \frac{n+1-\alpha}{\alpha} \right]$. This ensures that its inverse, $\Phi_{\alpha,n}(\lambda)$, as defined in Definition~\ref{def:mth-solution}, is well-defined for $\lambda \in [\Lambda_{\alpha,n}(\frac{n+1-\alpha}{\alpha}), \Lambda_{\alpha,n}( \frac{2n}{n+1})]$, where $\Lambda_{\alpha, n}(\tfrac{2n}{n+1}) = \tfrac{n-1}{n+1}$ and $\Lambda_{1, n}(n) = 0$. 
Since $m =  \max\left\{\, i \in [n-1] : \frac{n+1-i}{i} > \frac{2n}{n+1} \,\right\}$, we have
\begin{equation*}
    [\Lambda_{\alpha,n}(\tfrac{n+1-\alpha}{\alpha}), \Lambda_{\alpha,n}( \tfrac{n-\alpha}{\alpha+1})] \subseteq [\Lambda_{\alpha,n}(\tfrac{n+1-\alpha}{\alpha}), \Lambda_{\alpha,n}( \tfrac{2n}{n+1})]
\end{equation*}
for all $\alpha \in [m-1]$, and 
\begin{equation*}
    [\Lambda_{m,n}(\tfrac{n+1-m}{m}), \tfrac{n-1}{n+1}] = [\Lambda_{m,n}(\tfrac{n+1-m}{m}), \Lambda_{m,n}( \tfrac{2n}{n+1})].
\end{equation*}
Therefore, $\Phi_{\alpha,n}(\lambda)$ is also well defined for $\lambda$ in those ranges.

Since $\Lambda_{\alpha,n} (\rob)$ is strictly decreasing for $\rob \in \left[ \frac{2n}{n+1}, \frac{n+1-\alpha}{\alpha} \right]$, Lemma~\ref{lem:Lambda_eq} gives
\[
\Lambda_{\alpha,n} (\tfrac{n+1-\alpha}{\alpha}) < 
\Lambda_{\alpha,n} (\tfrac{n-\alpha}{\alpha+1}) = \Lambda_{\alpha+1,n} (\tfrac{n-\alpha}{\alpha+1}), \qquad \forall \ \alpha \in [m-1].
\]

Combining the above results with Lemma~\ref{lem:Lambda_eq} and the construction of $\vg^{(n)}$ establishes its well-definedness.
\end{proof}

\subsection{Monotonicity Property of Function $\Phi_{\alpha, n}$} 
\label{app:improve}

\begin{lemma}
\label{lem:Phi_order_monotone}
For a fixed integer $n\geq 2$, let $\alpha'>\alpha$, where
$\alpha',\alpha\in[m]$, and 
$$m=\max\left\{
\alpha\in[n-1]:
\frac{n+1-\alpha}{\alpha}>\frac{2n}{n+1}
\right\}.$$
Then $\Phi_{\alpha',n}(\lambda)>\Phi_{\alpha,n}(\lambda)$ for all $\lambda\in
\left(
\Lambda_{\alpha',n}\!\left(\frac{n+1-\alpha'}{\alpha'}\right),
\frac{n-1}{n+1}
\right)$, where $\Phi_{\cdot,n}$ is defined as in Definition~\ref{def:mth-solution}.
\end{lemma}

\begin{proof}
Define
\[
    \gamma_0:=\frac{2n}{n+1},
    \qquad
    r_k:=\frac{n+1-k}{k},
    \qquad
    \bar\lambda_n:=\frac{n-1}{n+1}.
\]
We have shown in the proof of Lemma~\ref{clm:unique} that, for every $i\in[m]$ and 
$\gamma\in[\gamma_0,r_i]$, the solution variable $p^{(i)}$ satisfies $0\le p^{(i)}(\gamma)\le1$,
and that $p^{(i)}(\gamma)$ is strictly decreasing in $\gamma$.

By Definition~\ref{def:Lambda_func} and \eqref{eq:order_alpha_2}, $\Lambda_{\alpha, n}$ satisfies
\[
\Lambda_{\alpha,n}(\gamma)
:=
\frac{2}{n(n+1)}
\left[
\sum_{j=1}^{\alpha-1}(n-j)p^{(j)}(\gamma)
+
\frac{(n+1-\alpha)(n-\alpha)}{2}p^{(\alpha)}(\gamma)
\right],
\qquad
\gamma\in[\gamma_0,r_\alpha].
\]
Recall that $\Lambda_{\alpha,n}$ is continuous and strictly decreasing, and $\Phi_{\alpha,n}=\Lambda_{\alpha,n}^{-1}$ is well-defined on $[\Lambda_{\alpha,n}(r_\alpha),\bar\lambda_n]$. 

Note that \eqref{eq:order_alpha_1_re} is obtained through rearranging \eqref{eq:order_alpha_1}.
For each $i\in[m]$, let $p^{(i)}(\gamma)$ be recursively determined by
\eqref{eq:order_alpha_1_re}, i.e.,
\begin{equation}
\label{eq:recur}
a_i(\gamma)p^{(i)}(\gamma)
=
\sum_{j=1}^{i-1}(n-j)p^{(j)}(\gamma)
+
\frac{i}{2}\bigl((2n+1-i)-(i+1)\gamma\bigr),
\end{equation}
where
\[
a_i(\gamma):=
(n-i)\left(
\frac{(n+1+i)\gamma}{2}-(n+1-i)
\right).
\]
For $\gamma\ge \gamma_0$, we have $a_i(\gamma)>0$.
For $j\ge2$, subtracting \eqref{eq:recur} with $i=j-1$ from that with $i=j$ gives
\begin{equation}
\label{ap_1}
a_j p^{(j)} (\rob)-a_{j-1}p^{(j-1)}(\rob)
=
(n+1-j)(p^{(j-1)} (\rob)+1)-j\gamma
\end{equation}
By definition of $a_j (\rob)$
\begin{equation}
\label{ap_2}
a_{j-1}-a_j=j\gamma-2(n+1-j).
\end{equation}
Then, adding \eqref{ap_1} to $p^{({j-1})} (\rob)$ times \eqref{ap_2} gives
\[
a_j(\gamma)
\bigl(p^{(j)}(\gamma)-p^{(j-1)}(\gamma)\bigr)
=
\bigl(j\gamma-(n+1-j)\bigr)
\bigl(p^{(j-1)}(\gamma)-1\bigr).
\]

We now compare adjacent branches. Fix $k\in[m-1]$. Applying the difference
identity above with $j=k+1$ gives
\begin{equation}
\label{eq:diff}
a_{k+1}(\gamma)
\bigl(p^{(k+1)}(\gamma)-p^{(k)}(\gamma)\bigr)
=
\bigl((k+1)\gamma-(n-k)\bigr)
\bigl(p^{(k)}(\gamma)-1\bigr).
\end{equation}
For $\gamma\in(\gamma_0,r_{k+1})$, both factors on the right-hand side of \eqref{eq:diff} are
strictly negative. Hence $p^{(k+1)}(\gamma)>p^{(k)}(\gamma)$.
Therefore,
\[
\begin{aligned}
\Lambda_{k+1,n}(\gamma)-\Lambda_{k,n}(\gamma)=
\frac{2}{n(n+1)}
\cdot
\frac{(n-k)(n-k-1)}{2}
\bigl(p^{(k+1)}(\gamma)-p^{(k)}(\gamma)\bigr)
>0
\end{aligned}
\]
for all $\gamma\in(\gamma_0,r_{k+1})$. At $\gamma=r_{k+1}$, the first factor
on the right-hand side of \eqref{eq:diff} vanishes, so $p^{(k+1)}(r_{k+1})=p^{(k)}(r_{k+1})$,
and hence $\Lambda_{k+1,n}(r_{k+1})
    =
    \Lambda_{k,n}(r_{k+1})$.
Since $r_{k+1}<r_k$ and $\Lambda_{k,n}$ is strictly decreasing,
\[
    \Lambda_{k+1,n}(r_{k+1})
    =
    \Lambda_{k,n}(r_{k+1})
    >
    \Lambda_{k,n}(r_k).
\]
Thus the lower endpoints $\Lambda_{k,n}(r_k)$ are strictly increasing
in $k$.

Now take any $\alpha'>\alpha$ and $\lambda\in
    \left(
    \Lambda_{\alpha',n}(r_{\alpha'}),
    \bar\lambda_n
    \right)
    $.
By the monotonicity of the lower endpoints, both
$\Phi_{\alpha,n}(\lambda)$ and $\Phi_{\alpha',n}(\lambda)$ are well-defined.
Set
\[
    \gamma_\alpha:={\Phi}_{\alpha,n}(\lambda),
    \qquad
    \gamma_{\alpha'}:={\Phi}_{\alpha',n}(\lambda).
\]
Since $\lambda<\bar\lambda_n=\Lambda_{\alpha,n}(\gamma_0)$,
and $\Lambda_{\alpha,n}$ is strictly decreasing, we have $\gamma_\alpha>\gamma_0$.

Applying the adjacent comparison successively at $\gamma=r_{\alpha'}$ gives $\Lambda_{\alpha',n}(r_{\alpha'})
    \ge
    \Lambda_{\alpha,n}(r_{\alpha'})$.
Hence $\lambda>
    \Lambda_{\alpha',n}(r_{\alpha'})
    \ge
    \Lambda_{\alpha,n}(r_{\alpha'})$.
Since $\Lambda_{\alpha,n}$ is strictly decreasing, this implies $\gamma_\alpha<r_{\alpha'}$.
Thus $\gamma_\alpha\in(\gamma_0,r_{\alpha'})$.

For every $k=\alpha,\ldots,\alpha'-1$, we have $\gamma_\alpha<r_{\alpha'}\le r_{k+1}$.
Therefore, $\Lambda_{k+1,n}(\gamma_\alpha)
    >
    \Lambda_{k,n}(\gamma_\alpha)$.
Iterating over $k=\alpha,\ldots,\alpha'-1$ yields $\Lambda_{\alpha',n}(\gamma_\alpha)
    >
    \Lambda_{\alpha,n}(\gamma_\alpha)
    =
    \lambda$.
On the other hand, $\Lambda_{\alpha',n}(\gamma_{\alpha'})
    =
    \lambda$.
Since $\Lambda_{\alpha',n}$ is strictly decreasing, we conclude that $\gamma_{\alpha'}>\gamma_\alpha$.
That is, $\Phi_{\alpha',n}(\lambda)>\Phi_{\alpha,n}(\lambda)$.
This completes the proof.
\end{proof}

\section{Proofs for Section~\ref{sec:lb}}

In this section, we present proofs of the results in Section~\ref{sec:lb}.

\subsection{Proof of Proposition~\ref{thm:wei_non_tight}} \label{app:wei_non_tight}

\begin{proof}
    Consider a $(1+\lambda)$-consistent algorithm $\mA$. Let the predictions be $y_1 = y_2 = \dots = y_n = 1$. Let $d (i, j)$ denote the processing time of job $i$ when job $j$ finishes. We assume without loss of generality that job $1$ is the first job finished by $\mA$ and that $d (i, 1) \geq d(j ,1)$ for all $i < j$. We consider the scenario where the prediction $y_1$ is accurate (i.e. $x_1 = y_1$). A necessary condition for maintaining
$(1+\lambda)$-consistency is
    \begin{equation*}
        (1+\lambda) \cdot \frac{n(n+1)}{2} \geq n \cdot \left( 1 + \sum_{i=2}^n d(i,1)\right) + \sum_{i=2}^n (n+1-i)\cdot (1 - d(i,1)),
    \end{equation*}
    where the first term represents the costs incurred up to the moment when job~1 completes, while the second term represents the minimum cost required to finish from this state. Simplifying, we obtain
    \begin{equation}
    \label{eq:c_1}
        \frac{n(n+1)}{2} \lambda \geq \sum_{i=2}^n (i-1) \cdot d(i,1).
    \end{equation}

    We first consider an adversarial setup where $x_1 = y_1 = 1$ and $x_i = d(i, 1) + \eps$, where $\eps$ is a small positive number. For this setup, we have
    \begin{equation}
    \label{eq:o_1}
        \opt = 1 + \sum_{i=2}^n i \cdot d(i,1) + O(\eps).
    \end{equation}
    We also have
    \begin{equation}
    \label{eq:a_1}
        \alg = n \cdot \left(1 + \sum_{i=2}^n d(i,1)\right) + O(\eps)
    \end{equation}
    Combining \eqref{eq:o_1} and \eqref{eq:a_1} and taking $\eps \to 0$, we have
    \begin{equation}
    \label{eq:opt/alg}
        \frac{\opt}{\alg} - \frac{1}{n} = \frac{1}{n} \cdot \left(\frac{\sum_{i=2}^n (i-1)\cdot d(i,1)}{1+\sum_{i=2}^nd(i,1)}\right).
    \end{equation}
    We set the numerator $\sum_{i=2}^n (i-1)\cdot d(i,1)=C$. To maximize the quantity on the right-hand side,\footnote{Since we want to minimize the ratio $\alg/\opt$.} we would want to have $\sum_{i=2}^n d(i,1)$ as small as possible subject to the constraints $d(i,1) \geq d(j,1) \geq 0$ for all $j > i \geq2$ and $\sum_{i=2}^n (i-1)\cdot d(i,1)=C$. It's clear that the linear program is bounded and feasible, so an optimum exists. Furthermore, if we have a feasible solution with $d(i,1) > d(i+1,1)$, then this solution cannot be optimal. This is because we can set $d(i,1) \gets d(i,1) - \frac{\vartheta}{i-1}$ and $d(i+1,1) \gets d(i+1,1) + \frac{\vartheta}{i}$ for sufficiently small $\vartheta >0$,  reducing the objective while remaining feasible. Thus, if an optimal solution exists, it must have $d(2,1) = d(3,1) = \ldots = d(n,1) = \tfrac{2C}{n(n-1)}$. 
Then, \eqref{eq:opt/alg} yields
    \begin{equation*}
        \frac{\opt}{\alg} - \frac{1}{n} \leq \frac{1}{n} \cdot \left(\frac{C}{1+2C/n}\right),
    \end{equation*}
    whose right-hand side is strictly increasing in $C$. Since \eqref{eq:c_1} gives
$C\leq \tfrac{n(n+1)}2\lambda$, we obtain
\[
\frac{\opt}{\alg}
\leq
\frac1n+\frac1n \cdot\frac{C}{1+2C/n}
\leq
\frac1n+
\frac1n \cdot \frac{n(n+1)\lambda/2}{1+(n+1)\lambda}
=
\frac{1+\tfrac{(n+1)(n+2)}2\lambda}
{n+n(n+1)\lambda}.
\]

    Inverting the numerator and denominator, we recover exactly the
lower bound of \cite{Wei2020} in~\eqref{eq:wei-lower-bound}. The equality is attained only when 
\begin{equation}
\label{eq:d(i,1)}
    d(2,1) = \cdots = d(n,1) = \tfrac{n+1}{\,n-1\,}\,\lambda.
\end{equation}
We distinguish the following two cases.\\

\noindent \textbf{Case 1}: If \eqref{eq:d(i,1)} is not satisfied, then algorithm $\mA$ will necessarily incur a robustness strictly larger than the bound given in \eqref{eq:wei-lower-bound}. 

\noindent \textbf{Case 2}: If \eqref{eq:d(i,1)} is satisfied, then we consider a positive number $\Delta t$ with $\Delta t < 1 -  \tfrac{n+1}{\,n-1\,}\,\lambda$. Suppose that, within the time interval of length $\Delta t$ immediately following the completion of job~1, 
the processing time allocated to job~$i$ ($i \geq 2$) is denoted by $\Delta t_i$. We assume without loss of generality that $\Delta t_i \geq \Delta t_j$ for all $i < j$.\footnote{This assumption is justified, as under \eqref{eq:d(i,1)} all remaining unfinished jobs become equivalent once job~1 is completed.} Note that 
\begin{equation*}
    d(i, 1) + \Delta t_i \, \leq \, d(i, 1) + \Delta t \, <\, 1 = y_i, \quad \forall \, i \geq 2.
\end{equation*}
No job is completed during this period when all predictions are accurate. Consistency requires
\begin{equation}
\label{consistency}
    (1 + \lambda) \cdot \frac{n(n+1)}{2} \geq n \cdot \left( 1 + \sum_{i=2}^n d(i, 1) \right) + (n-1) \cdot \Delta t + \sum_{i = 2}^n (n+1-i)\cdot (1 - d(i , 1) -\Delta t_i ).
\end{equation}
\eqref{eq:d(i,1)} and \eqref{consistency} give
\begin{equation*}
    (n-1)\cdot \Delta t - \sum_{i=2}^n (n+1-i)\cdot\Delta t_i \leq 0.
\end{equation*}
Since $\sum_{i=2}^n \Delta t_i=\Delta t$ and $\Delta t_i\ge 0 $, we have $\Delta t_2 = \Delta t$ and $\Delta t_i = 0$ for all $i \geq 3$. Therefore, $\mA$ must allocate all processing time to job~2 during the $\Delta t$ interval immediately following the completion of job~1.

We then consider another adversarial setup where $x_1 = y_1 = 1$, $x_2 = \tfrac{n+1}{\,n-1\,}\,\lambda + \Delta t$ and $x_i = \tfrac{n+1}{\,n-1\,}\,\lambda + \eps$ for $i \geq 3$, where $\eps$ is a small positive number. In this setup, 
\begin{equation}
    \label{eq:disprove_alg}
    \alg = n \cdot \left( 1 + \sum_{i=2}^n d(i, 1) \right) + (n-1) \cdot \Delta t + O(\eps).
\end{equation}
\begin{equation}
    \label{eq:disprove_opt}
    \opt = 1 + 2\cdot (d(2,1) + \Delta t) + \sum_{i=3}^n i \cdot d(i, 1) + O(\eps).
\end{equation}
Combining \eqref{eq:d(i,1)}, \eqref{eq:disprove_alg} and \eqref{eq:disprove_opt}, we have
\begin{equation*}
    \frac{\alg}{\opt} = \frac{n+n(n+1) \lambda + (n-1)\Delta t + O(\eps)}{1 + (n+1)(n+2)\lambda / 2 + 2 \Delta t + O(\eps)}.
\end{equation*}
Taking $\eps \to 0$ yields
\begin{equation*}
    \frac{\alg}{\opt} = \frac{n+n(n+1) \lambda + (n-1)\Delta t}{1 + (n+1)(n+2)\lambda / 2 + 2 \Delta t}.
\end{equation*}

Since $n \geq 5$, we have $2/(n^2 - 3n-2) < (n-1)/(n+1)$. For $\lambda \in \left(\tfrac{2}{n^2-3n-2},\,  \tfrac{n-1}{\,n+1\,}\right)$, 
it follows that 
\[
   \frac{\alg}{\opt} \;>\; \frac{n + n(n+1)\lambda}{1 + \tfrac{(n+1)(n+2)}{2}\,\lambda}.
\]
Therefore, within this interval of~$\lambda$, the ratio $\alg/\opt$ is strictly larger than the value in~\eqref{eq:wei-lower-bound}, 
thereby showing that the bound in~\citet{Wei2020} is not tight.
\end{proof}

\subsection{Proof of Proposition~\ref{prop:wei_padding_limit}}
\label{app:padding}
\begin{proof}
Fix $\lambda\in(0,1)$.
For every admissible $k$,
\[
\Phi_{\mathrm{wz}}(k,\lambda)
=
\frac{k}{1+\frac12\lambda(k+1)(k+2)}
+
\frac{\lambda k(k+1)}
{1+\frac12\lambda(k+1)(k+2)}
\leq
\frac{k}{1+\lambda k^2/2}+2
\leq
\frac1{\sqrt{2\lambda}}+2,
\]
where the last inequality follows from
$1+\lambda k^2/2\geq\sqrt{2\lambda}\,k$.
For all sufficiently large $n$, the maximum defining
$\Phi_{\mathrm{wz}}^{\mathrm{pad}}(n,\lambda)$ is nonempty.
It is nondecreasing in $n$ and bounded above, so its limit exists.

For $\lambda\in[1/2,1)$,
\[
\Phi_{\mathrm{wz}}(k,\lambda)-2
=
\frac{(1-2\lambda)k-2(1+\lambda)}
{1+\frac12\lambda(k+1)(k+2)}
<0.
\]
Since all sufficiently large $k$ are admissible and
$\Phi_{\mathrm{wz}}(k,\lambda)\to2$ as $k\to\infty$,
we obtain
$\Phi_{\mathrm{wz}}^{\mathrm{pad}}(\infty,\lambda)=2$.

Finally, let
$k_\lambda=\lfloor\sqrt{2/\lambda}\rfloor$,
which is admissible for all sufficiently small $\lambda$.
Since $\sqrt{\lambda}\,k_\lambda\to\sqrt2$,
\[
\sqrt{\lambda}\,
\Phi_{\mathrm{wz}}(k_\lambda,\lambda)
=
\frac{
\sqrt{\lambda}\,k_\lambda
\bigl(1+\lambda(k_\lambda+1)\bigr)
}{
1+\frac12\lambda(k_\lambda+1)(k_\lambda+2)
}
\longrightarrow\frac1{\sqrt2}.
\]
Combining this with
\[
\Phi_{\mathrm{wz}}(k_\lambda,\lambda)
\leq
\Phi_{\mathrm{wz}}^{\mathrm{pad}}(\infty,\lambda)
\leq
\frac1{\sqrt{2\lambda}}+2
\]
proves the claimed asymptotic expression.
\end{proof}

\subsection{Proof of Corollary~\ref{cor:wei_padding_non_tight}}
\label{app:extend}
\begin{proof}
Fix $n\geq5$ and
$\lambda\in[\frac12,\frac{n-1}{n+1})$.
For $\lambda\geq1/2$, the expression in
\eqref{eq:wei-lower-bound} is strictly increasing in the
number of jobs. Hence, $\Phi_{\mathrm{wz}}^{\mathrm{pad}}(n,\lambda)
=\Phi_{\mathrm{wz}}(n,\lambda)$.
Since
\[
\frac{2}{n^2-3n-2}\leq\frac14<\frac12\leq\lambda
<\frac{n-1}{n+1},
\]
the proof of Proposition~\ref{thm:wei_non_tight}
in Appendix~\ref{app:wei_non_tight} shows that every
$(1+\lambda)$-consistent algorithm for $n$ jobs
has robustness strictly greater than
$\Phi_{\mathrm{wz}}(n,\lambda)$,
and therefore its robustness is also strictly greater than the padded lower bound.
\end{proof}

\subsection{Proof of Lemma~\ref{Monotonic_Equivalence_Lemma}} \label{app:Monotonic_Equivalence_Lemma}
\begin{proof}
Run $\mA$ on the accurate all-unit instance. Let $C_j$ denote the
clock completion time of job $j$ under $\mA$, and let $D_i(t)$ denote
the cumulative amount of processing received by job $i$ by time $t$.
Thus,
\[
d(i,j)=D_i(C_j).
\]

For every $t\geq0$, let $D^{(k)}(t)$ denote the $k$-th largest
element of the multiset $\{D_i(t)\}_{i\in[n]}$, so that
\[
D^{(1)}(t)\geq D^{(2)}(t)\geq\cdots\geq D^{(n)}(t).
\]
In particular, if $d^{(k)}(j)$ denotes the $k$-th largest element of
$\{d(i,j)\}_{i\in[n]}$, then
\[
d^{(k)}(j)=D^{(k)}(C_j).
\]

We construct $\mA'$ so that, as long as the observed completion
history agrees with that of the accurate all-unit instance, the
cumulative amount of processing received by job $k$ at time $t$ is
\[
D'_k(t)\coloneq D^{(k)}(t).
\]
This defines a feasible processing profile. Indeed, since every
$D_i(t)$ is nondecreasing in $t$, each order statistic $D^{(k)}(t)$
is also nondecreasing. Moreover, for every $0\leq s<t$,
\[
\begin{aligned}
\sum_{k=1}^n\bigl(D'_k(t)-D'_k(s)\bigr)
&=
\sum_{k=1}^n\bigl(D^{(k)}(t)-D^{(k)}(s)\bigr)=
\sum_{i=1}^n\bigl(D_i(t)-D_i(s)\bigr)
\leq t-s,
\end{aligned}
\]
where the last inequality follows from the unit-capacity constraint.
Hence the sorted cumulative profile can be implemented by a feasible
single-machine schedule. If the observed completion history departs
from the accurate all-unit execution, we complete the definition of
$\mA'$ by processing the remaining jobs according to any fixed
feasible work-conserving rule.

Let $C'_j$ denote the clock completion time of job $j$ under $\mA'$
on the accurate all-unit instance. Since $\mA$ completes jobs in
increasing index order, the $j$-th largest cumulative processing
value first reaches one at time $C_j$. Therefore, $\mA'$ also
completes job $j$ at time $C_j$, with simultaneous completions, if
any, ordered by increasing index. Consequently,
\[
C'_j=C_j
\qquad
\text{for every }j\in[n],
\]
and
\[
d'(k,j)
=
D'_k(C'_j)
=
D^{(k)}(C_j)
=
d^{(k)}(j).
\]
Thus,
\[
\{d(i,j)\}_{i\in[n]}
=
\{d^{(i)}(j)\}_{i\in[n]}
=
\{d'(i,j)\}_{i\in[n]}
\qquad
\text{for every }j\in[n].
\]
Since $d'(1,j),\ldots,d'(n,j)$ are arranged in non-increasing order,
\[
d'(i,k)\geq d'(j,k)
\qquad
\text{for all }j\geq i>k.
\]

Moreover, $C'_j=C_j$ for every $j\in[n]$, so the two algorithms incur
the same total completion time on the accurate all-unit instance:
\[
\operatorname{ALG}(\mA,\mathbf 1)
=
\sum_{j=1}^n C_j
=
\sum_{j=1}^n C'_j
=
\operatorname{ALG}(\mA',\mathbf 1).
\]
Since the offline optimum depends only on the instance, $\mA$ and
$\mA'$ have the same competitive ratio under perfect predictions.

It remains to compare their corresponding attack instances. Fix
$i\in[\alpha]$, and let $\mx_i\in\mX_\alpha(\mA)$ and
$\mx'_i\in\mX_\alpha(\mA')$ be the corresponding $i$-th attacks.
At the $i$-th completion time, sorting preserves the multiset of
processing received by the unfinished jobs:
\[
\{d(j,i):j>i\}
=
\{d^{(j)}(i):j>i\}
=
\{d'(j,i):j>i\}.
\]
Hence $\mx_i$ and $\mx'_i$ are permutations of each other, and
therefore
\[
\operatorname{OPT}(\mx_i)
=
\operatorname{OPT}(\mx'_i).
\]

By non-clairvoyance, the two algorithms reproduce their respective
accurate executions through the $i$-th completion time. Indeed, before
that time, none of the unfinished jobs can reveal that its true
processing time is only $\eps$ larger than the amount of processing
it has received. At time $C_i=C'_i$, every remaining job has exactly
$\eps$ units of processing left. Their total remaining workload is
therefore $(n-i)\eps$. Since $\mA$ is work-conserving, the remaining processing time is $O(\eps)$. Therefore,
\[
\operatorname{ALG}(\mA,\mx_i)
=
\sum_{j=1}^i C_j+(n-i)C_i+O(\eps),
\]
while
\[
\operatorname{ALG}(\mA',\mx'_i)
=
\sum_{j=1}^i C'_j+(n-i)C'_i+O(\eps)
=
\sum_{j=1}^i C_j+(n-i)C_i+O(\eps).
\]
Because $\eps$ is infinitesimal in Definition~\ref{def:attack_set},
the two corresponding attacks yield the same outcome. Since this
holds for every $i\in[\alpha]$, attacking $\mA$ over
$\mX_\alpha(\mA)$ yields the same outcomes as attacking $\mA'$ over
$\mX_\alpha(\mA')$.
\end{proof}


\subsection{Proof of Theorem~\ref{thm:piece-wise_lower_bound}}
\label{app:piece-wise_lower_bound}

\begin{proof}[Proof of Theorem~\ref{thm:piece-wise_lower_bound}]

The proof has four steps. We first formulate the constraints
imposed by consistency and the adversarial attacks. We then
establish the structure of optimal solutions and identify the
resulting equations with the Order-$\alpha$ System. Finally,
we verify the attack-depth selection rule and establish the
claimed lower bound.

\paragraph{Step 1: The attack-based optimization problem.}
Fix $n\geq2$, $\lambda\in[0,\frac{n-1}{n+1}]$, and a
$(1+\lambda)$-consistent algorithm $\mA$ with robustness $\rob$.
Run $\mA$ on the accurate all-unit instance, and index the jobs
by their completion order. By
Lemma~\ref{Monotonic_Equivalence_Lemma}, we may assume that
\[
d(j,i)\leq d(k,i)
\qquad\text{whenever }i<k<j,
\]
for the analysis of the accurate instance and the corresponding
adversarial attacks.

For an attack depth $\alpha\in[n-1]$, write $p_j^{(i)}\coloneq d(j,i)$, 
$\mathbf p\coloneq
\bigl(p_j^{(i)}\bigr)_{1\leq i\leq\alpha,\ i<j\leq n}$, and we define the aggregates
\[
P^{(i)}\coloneq\sum_{j=i+1}^n p_j^{(i)},
\qquad
I_p^{(i)}\coloneq\sum_{j=i+1}^n j p_j^{(i)},
\qquad i\in[\alpha].
\]

Consider the $i$-th attack, whose processing-time vector is
\[
\mx_i=
\bigl(\underbrace{1,\ldots,1}_{i\text{ copies}},
      p_{i+1}^{(i)}+\varepsilon,\ldots,p_n^{(i)}+\varepsilon\bigr),
\qquad \varepsilon\downarrow0.
\]
By non-clairvoyance, the execution agrees with the accurate
execution through the $i$-th completion. At the $j$-th completion, the total amount of processing performed
is $j+P^{(j)}$, which is also the completion time.
In the $i$-th attack instance, each remaining job has only
$\varepsilon$ units of processing left and therefore completes
within $(n-i)\varepsilon$ additional time. Thus, up to an additive
$O(\varepsilon)$ term, the online cost equals
\[
\begin{aligned}
&\sum_{j=1}^{i-1}\bigl(j+P^{(j)}\bigr)
+(n+1-i)\bigl(i+P^{(i)}\bigr)=
\frac{i(2n+1-i)}2
+\sum_{j=1}^{i-1}P^{(j)}
+(n+1-i)P^{(i)}.
\end{aligned}
\]
In the limit, the processing times are ordered as
$1,\ldots,1,p_{i+1}^{(i)},\ldots,p_n^{(i)}$ in nonincreasing
order. By the shortest-processing-time rule, the offline cost
converges to $i(i+1)/2+I_p^{(i)}$ as $\varepsilon\downarrow0$.
Consequently,
\[
\frac{i(2n+1-i)}2+\sum_{j=1}^{i-1}P^{(j)}+(n+1-i)P^{(i)}
\leq\rob\left(\frac{i(i+1)}2+I_p^{(i)}\right).
\]

On the accurate instance, the remaining processing times after
the $\alpha$-th completion are
$1-p_{\alpha+1}^{(\alpha)}\leq\cdots\leq1-p_n^{(\alpha)}$.
Completing them in this order minimizes their additional cost.
Hence the total completion time is at least
\[
\begin{aligned}
&\frac{\alpha(2n+1-\alpha)}2+\sum_{j=1}^{\alpha-1}P^{(j)}
+(n+1-\alpha)P^{(\alpha)}
+\sum_{j=\alpha+1}^n(n+1-j)\bigl(1-p_j^{(\alpha)}\bigr)\\
&\qquad=
\frac{n(n+1)}2+
\sum_{j=1}^{\alpha-1}P^{(j)}+I_p^{(\alpha)}
-\alpha P^{(\alpha)}.
\end{aligned}
\]
Consistency bounds the excess over $n(n+1)/2$ by
$\lambda n(n+1)/2$.
Together with the bounds and monotonicity of cumulative
processing, these observations yield the following problem:
\begin{subequations}
\label{refine_problem_1}
\begin{align}
\min_{\rob,\mathbf p}\quad &\rob\\
\text{s.t.}\quad
&\frac{i(2n+1-i)}2+\sum_{j=1}^{i-1}P^{(j)}
 +(n+1-i)P^{(i)}\leq\rob\left(\frac{i(i+1)}2+I_p^{(i)}\right),
 \qquad i\in[\alpha],
 \label{refine_problem_1_cons1}\\
&\sum_{j=1}^{\alpha-1}P^{(j)}+I_p^{(\alpha)}
 -\alpha P^{(\alpha)}\leq\lambda\frac{n(n+1)}2,
 \label{refine_problem_1_cons2}\\
&0\leq p_j^{(i)}\leq1,
 \qquad 1\leq i\leq\alpha,\quad i<j\leq n,
 \label{refine_problem_1_box}\\
&p_j^{(i)}\leq p_k^{(i)},
 \qquad 1\leq i\leq\alpha,\quad i<k<j\leq n,
 \label{refine_problem_1_cons3}\\
&p_j^{(i)}\leq p_j^{(k)},
 \qquad 1\leq i<k\leq\alpha,\quad k<j\leq n.
\end{align}
\end{subequations}
Here $P^{(i)}$ and $I_p^{(i)}$ are the aggregates defined above,
not additional decision variables. Every algorithm under
consideration induces a feasible point, so the optimal value
$\phi_{\alpha,n}(\lambda)$ is a lower bound on its robustness.

\begin{lemma}
\label{lem:existence_fixed_alpha_optimizer}
For every $\lambda\in[0,\frac{n-1}{n+1}]$ and
$\alpha\in[n-1]$, Problem~\ref{refine_problem_1} admits a globally
optimal solution.
\end{lemma}

\begin{proof}
Let $\mathcal F_{\lambda,\alpha}$ denote the feasible region of
Problem~\ref{refine_problem_1}. A direct substitution shows that
$(\rob,\mathbf p)=(n,\mathbf0)$ is feasible.

Consider the objective sublevel set
\[
\mathcal K_{\lambda,\alpha}
\coloneq
\left\{
(\rob,\mathbf p)\in\mathcal F_{\lambda,\alpha}:
\rob\leq n
\right\}.
\]
This set is nonempty because it contains $(n,\mathbf0)$.
The box constraints give
\[
0\leq p_j^{(i)}\leq1
\qquad
\text{for every }i\in[\alpha]\text{ and }i<j\leq n,
\]
so all coordinates of $\mathbf p$ are bounded. Moreover, in
Constraint~\eqref{refine_problem_1_cons1}, both the left-hand side
and the coefficient multiplying $\rob$ are strictly positive.
Hence $\rob>0$. Together with $\rob\leq n$, this shows that
$\mathcal K_{\lambda,\alpha}$ is bounded.
It is also closed because all its defining constraints are
non-strict inequalities of continuous functions.
Therefore, $\mathcal K_{\lambda,\alpha}$ is compact.

By the Weierstrass extreme value theorem, the continuous objective
attains a minimum on $\mathcal K_{\lambda,\alpha}$.
This minimum is at most $n$, since $(n,\mathbf0)$ belongs to
$\mathcal K_{\lambda,\alpha}$. Every feasible point outside
$\mathcal K_{\lambda,\alpha}$ has objective value greater than $n$
and therefore cannot improve this minimum.
Hence it is also a global minimum over
$\mathcal F_{\lambda,\alpha}$.
\end{proof}

\paragraph{Step 2: Structure of optimal solutions.}
Let $m$ be the maximal depth defined in \eqref{eq:m_upper}.
We next show that, when the optimal value lies in a certain
range, every optimal solution has equal processing across the
unfinished jobs and saturates both the attack constraints and
the consistency budget.
\begin{lemma}[Restatement of Lemma~\ref{lem:optimal_profile_structure}]
Fix $\alpha\in[m]$ and
$\lambda\in(0,\frac{n-1}{n+1})$. Suppose that
\[
\frac{2n}{n+1}
<
\phi_{\alpha,n}(\lambda)
<
\frac{n+1-\alpha}{\alpha}.
\]
Then every optimal solution to Problem~\ref{refine_problem_1}
satisfies
\[
p_j^{(i)}=p^{(i)}
\qquad
\text{for every }i\in[\alpha]\text{ and }j>i,
\]
for some $p^{(1)},\ldots,p^{(\alpha)}$.
Moreover, all constraints in
\eqref{refine_problem_1_cons1} and
\eqref{refine_problem_1_cons2} hold with equality.
\end{lemma}

\begin{proof}
Fix an arbitrary optimal solution $(\rob,\mathbf p)$, where
$\rob=\phi_{\alpha,n}(\lambda)$.
Throughout the proof, the unbarred aggregates
$P^{(i)}$ and $I_p^{(i)}$ refer to this fixed optimal solution.

\smallskip
\noindent\emph{A symmetric comparison profile.}
We seek a symmetric comparison profile at the same robustness
$\rob$, with every attack constraint tight.
Let $\bar p^{(i)}$ denote the common processing time of the
unfinished jobs when job~$i$ finishes, and define the corresponding aggregates
\[
\bar P^{(i)}
\coloneq
\sum_{j=i+1}^n\bar p^{(i)}
=
(n-i)\bar p^{(i)},
\qquad
\bar I_p^{(i)}
\coloneq
\sum_{j=i+1}^n j\bar p^{(i)}
=
\frac{n+i+1}{2}\bar P^{(i)}.
\]
Equality in the $i$-th attack constraint requires
\[
\frac{i(2n+1-i)}2
+\sum_{j=1}^{i-1}\bar P^{(j)}
+(n+1-i)\bar P^{(i)}
=
\rob\left(
\frac{i(i+1)}2+\bar I_p^{(i)}
\right).
\]
Substituting
$\bar I_p^{(i)}=\frac{n+i+1}{2}\bar P^{(i)}$
and rearranging gives the recurrence
\begin{equation}
\label{eq:lower_canonical_recursion}
\left(\frac{n+i+1}{2}\rob-(n+1-i)\right)\bar P^{(i)}
=
\frac{i(2n+1-i)}2-\rob\frac{i(i+1)}2
+\sum_{j=1}^{i-1}\bar P^{(j)},
\qquad i\in[\alpha].
\end{equation}
The coefficient on the left is strictly positive, since
\[
\frac{n+i+1}{2}\rob-(n+1-i)
\geq\frac{n+2}{2}\rob-n
>\frac{n}{n+1}>0.
\]
Starting with $i=1$, this recurrence uniquely determines
$\bar P^{(1)},\ldots,\bar P^{(\alpha)}$, and hence
$\bar p^{(i)}=\bar P^{(i)}/(n-i)$.
These quantities depend continuously on $\rob \in (2n/(n+1), (n+1-\alpha)/\alpha)$.
By construction, the resulting symmetric profile satisfies
every constraint in \eqref{refine_problem_1_cons1} with equality.

We verify that this profile also satisfies the box and
monotonicity constraints. The first equation gives
\[
\bar p^{(1)}
=
\frac{n-\rob}{(n-1)((n+2)\rob/2-n)}
\in(0,1),
\]
where positivity follows from $2n/(n+2)<\rob<(n+1-\alpha)/\alpha  \leq  n$, and the upper bound follows
from
\[
(n-1)\left(\frac{n+2}{2}\rob-n\right)-(n-\rob)
=
\frac{n(n+1)}2
\left(\rob-\frac{2n}{n+1}\right)>0.
\]
Subtracting consecutive equations in
\eqref{eq:lower_canonical_recursion} gives, for $i<\alpha$,
\[
\bar p^{(i+1)}
=
\bar p^{(i)}
+
\frac{n-i-(i+1)\rob}
{(n-i-1)((n+i+2)\rob/2-(n-i))}
\bigl(1-\bar p^{(i)}\bigr).
\]
The fraction lies strictly between zero and one.
Indeed, its numerator is positive because
$\rob<(n+1-\alpha)/\alpha\leq(n-i)/(i+1)$,
and its denominator exceeds its numerator by
\[
(n-i)\left(\frac{n+i+1}{2}\rob-(n-i)\right)>0.
\]
Induction therefore yields
\begin{equation}
\label{eq:lower_symmetric_profile_monotonicity}
0<\bar p^{(1)}<\cdots<\bar p^{(\alpha)}<1.
\end{equation}

\smallskip
\noindent\emph{Comparison of the consistency budgets.}
Monotonicity across the processing times of unfinished jobs and Chebyshev's sum inequality give
\begin{equation}
\label{chebyshev_sum}
I_p^{(i)}
\leq\frac{n+i+1}{2}P^{(i)},
\qquad i\in[\alpha].
\end{equation}
Since $\rob>0$, the $i$-th attack constraint and
\eqref{chebyshev_sum} give
\[
\begin{aligned}
&\frac{i(2n+1-i)}2
+\sum_{j=1}^{i-1}P^{(j)}
+(n+1-i)P^{(i)}\leq
\rob\left(\frac{i(i+1)}2+I_p^{(i)}\right)
\leq
\rob\left(
\frac{i(i+1)}2+\frac{n+i+1}{2}P^{(i)}
\right).
\end{aligned}
\]
Moving the terms involving $P^{(i)}$ to the left yields
\[
\left(\frac{n+i+1}{2}\rob-(n+1-i)\right)P^{(i)}
\geq
\frac{i(2n+1-i)}2-\rob\frac{i(i+1)}2
+\sum_{j=1}^{i-1}P^{(j)}.
\]
The comparison profile satisfies the corresponding equality
\eqref{eq:lower_canonical_recursion}.
Subtracting that equality from this inequality cancels the
constant terms and gives
\[
\left(\frac{n+i+1}{2}\rob-(n+1-i)\right)
\bigl(P^{(i)}-\bar P^{(i)}\bigr)
\geq
\sum_{j=1}^{i-1}\bigl(P^{(j)}-\bar P^{(j)}\bigr).
\]
Since the coefficient on the left is positive, induction gives
\begin{equation}
\label{eq:lower_aggregate_comparison}
P^{(i)}\geq\bar P^{(i)}
\qquad\text{for every }i\in[\alpha].
\end{equation}

The final attack constraint implies
\[
\begin{aligned}
&\rob\left(
\sum_{j=1}^{\alpha-1}P^{(j)}
+I_p^{(\alpha)}-\alpha P^{(\alpha)}
\right)\geq
(\rob+1)\sum_{j=1}^{\alpha-1}P^{(j)}
+\frac{\alpha(2n+1-\alpha)}2
-\rob\frac{\alpha(\alpha+1)}2
+\bigl(n+1-\alpha-\alpha\rob\bigr)P^{(\alpha)}.
\end{aligned}
\]
For the symmetric comparison profile, the final attack constraint
holds with equality. Since
$\bar I_p^{(\alpha)}=\frac{n+\alpha+1}{2}\bar P^{(\alpha)}$,
we have
\[
\begin{aligned}
&
\rob\left(
\sum_{j=1}^{\alpha-1}\bar P^{(j)}
+\frac{n+1-\alpha}{2}\bar P^{(\alpha)}
\right) = \rob\left(
\sum_{j=1}^{\alpha-1}\bar P^{(j)}
+\bar I_p^{(\alpha)}-\alpha\bar P^{(\alpha)}
\right)\\
&\quad=
(\rob+1)\sum_{j=1}^{\alpha-1}\bar P^{(j)}
+\frac{\alpha(2n+1-\alpha)}2
-\rob\frac{\alpha(\alpha+1)}2
+\bigl(n+1-\alpha-\alpha\rob\bigr)\bar P^{(\alpha)}.
\end{aligned}
\]
Subtracting the two relations gives
\begin{align}
&\rob\left[
\sum_{j=1}^{\alpha-1}P^{(j)}
+I_p^{(\alpha)}-\alpha P^{(\alpha)}
-\sum_{j=1}^{\alpha-1}\bar P^{(j)}
-\frac{n+1-\alpha}{2}\bar P^{(\alpha)}
\right] \notag \\
&\quad\geq
(\rob+1)\sum_{j=1}^{\alpha-1}
\bigl(P^{(j)}-\bar P^{(j)}\bigr)
+\bigl(n+1-\alpha-\alpha\rob\bigr)
\bigl(P^{(\alpha)}-\bar P^{(\alpha)}\bigr)
\geq0.\label{eq:lower_budget_comparison}
\end{align}
The last inequality follows from
\eqref{eq:lower_aggregate_comparison} and
$n+1-\alpha-\alpha\rob>0$.
Consequently, the comparison profile uses no more consistency
budget than the original optimum:
\begin{equation}
\label{consistency_chain}
\sum_{j=1}^{\alpha-1}\bar P^{(j)}
+\frac{n+1-\alpha}{2}\bar P^{(\alpha)}
\leq
\sum_{j=1}^{\alpha-1}P^{(j)}
+I_p^{(\alpha)}-\alpha P^{(\alpha)}
\leq
\lambda\frac{n(n+1)}2.
\end{equation}

\smallskip
\noindent\emph{Tightness and symmetry.}
Suppose that the first expression in this chain is strictly less
than $\lambda n(n+1)/2$.
Since the comparison profile depends continuously on $\rob$,
we may replace $\rob$ by a sufficiently close, smaller value
$\widetilde{\rob}$ within the same open interval, while keeping
its consistency budget below $\lambda n(n+1)/2$.
The resulting profile still satisfies all attack constraints
with equality, and the preceding recurrence argument guarantees
its box and monotonicity constraints.
It would therefore be feasible with objective value
$\widetilde{\rob}<\rob$, contradicting optimality.
Hence
\begin{equation}
\label{eq:lower_canonical_budget}
\sum_{j=1}^{\alpha-1}\bar P^{(j)}
+\frac{n+1-\alpha}{2}\bar P^{(\alpha)}
=
\lambda\frac{n(n+1)}2.
\end{equation}
Combining \eqref{consistency_chain} with
\eqref{eq:lower_canonical_budget} shows that both inequalities
in \eqref{consistency_chain} are equalities.
Hence the left-hand side of \eqref{eq:lower_budget_comparison}
is zero. Since $P^{(i)}\geq\bar P^{(i)}$ for every $i$ and
all coefficients on the right are strictly positive, we obtain
$P^{(i)}=\bar P^{(i)}$ for every $i$. The original optimum
and the comparison profile then have the same left-hand side in each
attack constraint. The comparison profile satisfies that constraint
with equality, so feasibility of the original optimum and
$\rob>0$ imply $\bar I_p^{(i)}\leq I_p^{(i)}$.
Together with \eqref{chebyshev_sum}, this gives
\[
\bar I_p^{(i)}
\leq I_p^{(i)}
\leq\frac{n+i+1}{2}P^{(i)}
=\frac{n+i+1}{2}\bar P^{(i)}
=\bar I_p^{(i)}.
\]
Hence every attack constraint is tight, and equality holds in
Chebyshev's inequality. Since the weights $i+1,\ldots,n$ are
strictly increasing, equality forces the processing amounts
of all unfinished jobs at the $i$-th completion to be equal
(trivially so when only one job remains).
Their common value is
$P^{(i)}/(n-i)=\bar P^{(i)}/(n-i)=\bar p^{(i)}$.
Since the chosen optimum was arbitrary, these conclusions hold
at every optimum.
\end{proof}

\paragraph{Step 3: Identification with the Order-$\alpha$ System.}
Suppose that the optimal value lies in the interval of
Lemma~\ref{lem:optimal_profile_structure}.
At any optimum, let $p^{(i)}$ be the common processing amount of unfinished jobs when job~$i$ finishes. Then
\[
P^{(i)}=(n-i)p^{(i)},
\qquad
I_p^{(i)}=\frac{(n-i)(n+i+1)}2p^{(i)}.
\]
Substituting these expressions into the tight attack and
consistency constraints gives
\[
\begin{aligned}
&\frac{i(2n+1-i)}2
+\sum_{j=1}^{i-1}(n-j)p^{(j)}
+(n+1-i)(n-i)p^{(i)}\\
&\qquad=
\rob\left(
\frac{i(i+1)}2+\frac{(n-i)(n+i+1)}2p^{(i)}
\right),
\qquad i\in[\alpha],
\end{aligned}
\]
and
\[
\sum_{j=1}^{\alpha-1}(n-j)p^{(j)}
+\frac{(n-\alpha)(n+1-\alpha)}2p^{(\alpha)}
=
\lambda\frac{n(n+1)}2.
\]
These are precisely the equations of
$\mathsf{Sys}_\alpha(n,\lambda)$.
The box constraints make this an admissible solution.
By the uniqueness established in Appendix~\ref{app:well_p},
\begin{equation}
\label{eq:lower_optimum_order_system}
\phi_{\alpha,n}(\lambda)=\vg_{\alpha,n}(\lambda).
\end{equation}

\paragraph{Step 4: Selecting the attack depth and concluding the proof.}
We now justify the selection rule in \eqref{choose_alpha}
and apply the structural characterization established above.
The following lemma provides the key step in establishing
the validity of this rule.
Recall that $\phi_{\alpha,n}(\lambda)$ denotes the optimal
objective value of Problem~\ref{refine_problem_1} for fixed
$n$, $\lambda$, and attack depth $\alpha$.

\begin{lemma}
\label{lem:justify_choice_of_alpha}
For every $\alpha\in[m-1]$, if $\phi_{\alpha,n}(\lambda)<\frac{n-\alpha}{\alpha+1}$, then $\phi_{\alpha+1,n}(\lambda)<\frac{n-\alpha}{\alpha+1}$.
\end{lemma}

\begin{proof}
Fix an optimal solution at depth $\alpha$, with
$\rob=\phi_{\alpha,n}(\lambda)$, and let
$P^{(i)}$ and $I_p^{(i)}$ denote its aggregates.
Extend this processing profile to depth $\alpha+1$ by defining
\[
p_j^{(\alpha+1)}\coloneq p_j^{(\alpha)}
\qquad
\text{for }j>\alpha+1,
\]
while keeping all previously defined coordinates unchanged.
The box and monotonicity constraints remain satisfied.
The new aggregates are
\[
P^{(\alpha+1)}
=
P^{(\alpha)}-p_{\alpha+1}^{(\alpha)},
\qquad
I_p^{(\alpha+1)}
=
I_p^{(\alpha)}
-(\alpha+1)p_{\alpha+1}^{(\alpha)}.
\]
Consequently, the consistency budget is unchanged:
\[
\begin{aligned}
&\sum_{j=1}^{\alpha}P^{(j)}
+I_p^{(\alpha+1)}-(\alpha+1)P^{(\alpha+1)}=
\sum_{j=1}^{\alpha-1}P^{(j)}
+I_p^{(\alpha)}-\alpha P^{(\alpha)}
\leq\lambda\frac{n(n+1)}2.
\end{aligned}
\]

We check the attack constraints at robustness
$(n-\alpha)/(\alpha+1)$.
The first $\alpha$ constraints hold strictly because they
already hold at
$\rob=\phi_{\alpha,n}(\lambda)<(n-\alpha)/(\alpha+1)$.
For the new attack constraint, we have
\[
\begin{aligned}
&\frac{(2n-\alpha)(\alpha+1)}2
+\sum_{j=1}^{\alpha}P^{(j)}
+(n-\alpha)P^{(\alpha+1)}\\
= \ &
\left(\frac{\alpha(2n+1-\alpha)}2
+\sum_{j=1}^{\alpha-1}P^{(j)}
+(n+1-\alpha)P^{(\alpha)}\right)
+(n-\alpha)\bigl(1-p_{\alpha+1}^{(\alpha)}\bigr)\\
< \ &
\frac{n-\alpha}{\alpha+1}
\left(\frac{\alpha(\alpha+1)}2+I_p^{(\alpha)}\right)
+(n-\alpha)\bigl(1-p_{\alpha+1}^{(\alpha)}\bigr)=
\frac{n-\alpha}{\alpha+1}
\left(
\frac{(\alpha+1)(\alpha+2)}2+I_p^{(\alpha+1)}
\right).
\end{aligned}
\]
The strict inequality follows from the original
$\alpha$-th attack constraint and
$\phi_{\alpha,n}(\lambda)<(n-\alpha)/(\alpha+1)$.

Thus all attack constraints hold strictly at robustness
$(n-\alpha)/(\alpha+1)$.
Since there are finitely many such constraints, they remain
satisfied at a sufficiently close, smaller robustness value.
The extended profile is therefore feasible at depth $\alpha+1$
with robustness strictly below $(n-\alpha)/(\alpha+1)$.
This proves the claim.
\end{proof}

\smallskip
\noindent\emph{Validity of the selection rule.}
First suppose $0<\lambda<(n-1)/(n+1)$.
For every attack depth, the first attack constraint and
\eqref{chebyshev_sum} give
\begin{equation}
\label{eq:lower_first_attack_baseline}
\rob
\geq
\frac{n(1+P^{(1)})}{1+\frac{n+2}{2}P^{(1)}}
\geq
\frac{2n}{n+1}.
\end{equation}
The last inequality follows because the fraction is strictly
decreasing in $P^{(1)}\in[0,n-1]$.
If an optimum attained $2n/(n+1)$, then $P^{(1)}=n-1$,
so every unfinished job would have received one unit of
processing at the first completion.
The box and monotonicity constraints would then force
$p_j^{(i)}=1$ for every $i\in[\alpha]$ and $j>i$.
Its consistency budget would be
\[
\sum_{j=1}^{\alpha-1}(n-j)
+\frac{(n-\alpha)(n+1-\alpha)}2
=
\frac{n(n-1)}2
>
\lambda\frac{n(n+1)}2,
\]
a contradiction. Hence
\[
\phi_{\alpha,n}(\lambda)>\frac{2n}{n+1}
\qquad
\text{for every }\alpha\in[m].
\]

At depth one, setting $p_j^{(1)}=({n+1})\lambda / (n-1)$ for every $j>1$
satisfies the consistency constraint with equality.
The first attack constraint is tight at the robustness value
appearing below, so
\[
\phi_{1,n}(\lambda)
\leq
\frac{n(1+(n+1)\lambda)}
{1+\frac{(n+1)(n+2)}2\lambda}
<n.
\]

We now verify that the prescribed rule for choosing $\alpha$
is well-defined. Fix $\lambda$ and consider the sequence
$\{\phi_{k,n}(\lambda)\}_{k=1}^m$. If $\phi_{1,n}(\lambda)\geq(n-1)/2$, then the previously
established bound $\phi_{1,n}(\lambda)<n$ gives $\phi_{1,n}(\lambda)\in
\left[\frac{n-1}{2},n\right)$,
so the rule selects $\alpha=1$.
Otherwise, if $m\geq2$, Lemma~\ref{lem:justify_choice_of_alpha}
gives $\phi_{2,n}(\lambda)<\frac{n-1}{2}$.
If $\phi_{2,n}(\lambda)\geq(n-2)/3$, then $\phi_{2,n}(\lambda)\in
\left[\frac{n-2}{3},\frac{n-1}{2}\right)$,
and the rule selects $\alpha=2$. Continuing in the same manner, whenever a depth $\alpha<m$
fails the lower bound in \eqref{choose_alpha},
Lemma~\ref{lem:justify_choice_of_alpha} guarantees the required
strict upper bound at depth $\alpha+1$.
The rule therefore advances until both bounds are satisfied. This must happen no later than depth $m$.
Indeed, the definition of $m$ and the lower bound established
above give $\frac{n-m}{m+1}
\leq\frac{2n}{n+1}
<\phi_{m,n}(\lambda)$.
Thus, if the rule reaches depth $m$, its lower bound is
necessarily satisfied, while its strict upper bound has
already been ensured by the preceding argument.
Consequently, there exists a smallest $\alpha\in[m]$ such that
\begin{equation}
\label{choice of alpha}
\phi_{\alpha,n}(\lambda)
\in
\left[
\frac{n-\alpha}{\alpha+1},
\frac{n+1-\alpha}{\alpha}
\right),
\end{equation}
proving that \eqref{choose_alpha} is well-defined.

For this selected depth, we have $\frac{2n}{n+1}
<
\phi_{\alpha,n}(\lambda)
<
\frac{n+1-\alpha}{\alpha}$.
Lemma~\ref{lem:optimal_profile_structure} therefore applies,
and \eqref{eq:lower_optimum_order_system} gives $\phi_{\alpha,n}(\lambda)=\vg_{\alpha,n}(\lambda)$.
By strict monotonicity and agreement of adjacent branches
in Appendix~\ref{app:well_p}, the interval in
\eqref{choice of alpha} corresponds precisely to the
depth-$\alpha$ branch in \eqref{Gamma}.
Consequently,
\[
\rob
\geq
\phi_{\alpha,n}(\lambda)
=
\vg_{\alpha,n}(\lambda)
=
\vg^{(n)}(\lambda).
\]

\smallskip
\noindent\emph{Endpoint cases.}
At $\lambda=0$, the consistency constraint for $\alpha=1$
forces $p_j^{(1)}=0$ for every $j>1$.
The first attack then gives
\[
\rob\geq n=\vg^{(n)}(0).
\]
At $\lambda=(n-1)/(n+1)$,
\eqref{eq:lower_first_attack_baseline} gives
\[
\rob\geq\frac{2n}{n+1}
=
\vg^{(n)}\!\left(\frac{n-1}{n+1}\right),
\]
where the final identity follows by substituting
$p^{(1)}=\cdots=p^{(m)}=1$ and $\rob=2n/(n+1)$
into the order system.

\smallskip
\noindent\emph{Strict improvement.}
Suppose $n\geq5$ and $\frac{2}{n^2-3n-2}<\lambda<\frac{n-1}{n+1}$.
The transition from depth one occurs at
\[
\vg_{1,n}^{-1}\!\left(\frac{n-1}{2}\right)
=
\frac{2}{n^2-3n-2},
\]
so the selected depth is at least two.
Since the lower bound of \cite{Wei2020} equals
$\vg_{1,n}(\lambda)$, Lemma~\ref{lem:Phi_order_monotone} yields
\[
\vg^{(n)}(\lambda)
=
\vg_{\alpha,n}(\lambda)
>
\vg_{1,n}(\lambda).
\]
This proves the strict-improvement claim and completes the proof.
\end{proof}

\subsection{Proof of Theorem~\ref{thm:lower_n->infty}} \label{app:lower_n->infty}

\begin{proof}
Define
\[
F(x)
\coloneq
x-
x^{\frac{2-x}{x+2}}
\bigl((x-2)(x+1)\bigr)^{\frac{x}{x+2}},
\qquad x\geq2.
\]
We first derive this function from the finite-$n$ order systems,
then establish its monotonicity and the claimed asymptotic
lower bound.

\paragraph{Step 1: A product formula for the finite-$n$ systems.}
Fix $x>2$. For all sufficiently large $n$, define
\[
\alpha_n\coloneq
\left\lfloor\frac{n+1}{x+1}\right\rfloor,
\qquad
\gamma_n\coloneq\frac{n+1-\alpha_n}{\alpha_n},
\qquad
\lambda_n\coloneq
\bigl(\Phi^{(n)}\bigr)^{-1}(\gamma_n).
\]
Then $\gamma_n\to x$. For sufficiently large $n$, or more precisely for $n \geq 2(x+1)-1$, we have $\frac{2n}{n+1}<2<x\leq\gamma_n
=\frac{n+1}{\alpha_n}-1
\leq\frac{n-1}{2}<n$. 
Combining this with continuity and strict monotonicity of $\Phi^{(n)}$, it follows that $\lambda_n$ is well-defined.

Since $\gamma_n=(n+1-\alpha_n)/\alpha_n$,
Lemma~\ref{lem:Lambda_eq} and the well-posedness results in
Appendix~\ref{app:well_p} guarantee that the Order-$\alpha_n$
System at $\lambda_n$ admits an admissible solution with
robustness $\gamma_n$.
Write its processing coordinates as
$p^{(1)},\ldots,p^{(\alpha_n)}$, suppressing their dependence on $n$.

Define
\[
D_{n,i}\coloneq
(n+1)(\gamma_n-2)+i(\gamma_n+2),
\qquad 0\leq i\leq\alpha_n,
\]
and set $p^{(0)}\coloneq0$.
Subtracting consecutive order equations gives
\[
(n-i)D_{n,i}(1-p^{(i)})
=
(n-i+1)(D_{n,i}-\gamma_n)(1-p^{(i-1)}),
\qquad 1\leq i\leq\alpha_n,
\]
where the case $i=1$ follows directly from the first order
equation. Since
$D_{n,i}-\gamma_n=D_{n,i-1}+2>0$,
iteration yields
\begin{equation}
\label{eq:asymptotic_deficit_product}
\frac{n-\alpha_n}{n}(1-p^{(\alpha_n)})
=
\prod_{i=1}^{\alpha_n}
\left(1-\frac{\gamma_n}{D_{n,i}}\right).
\end{equation}

The budget equation gives
\[
\sum_{j=1}^{\alpha_n-1}(n-j)p^{(j)}
=
\lambda_n\frac{n(n+1)}2
-
\frac{(n-\alpha_n)(n+1-\alpha_n)}2p^{(\alpha_n)}.
\]
Substituting this expression into the last order equation and
rearranging yields
\[
\begin{aligned}
\lambda_n n(n+1)
={}&
\alpha_n\bigl[
\gamma_n(\alpha_n+1)-(2n+1-\alpha_n)
\bigr]\\
&+
(n-\alpha_n)\bigl[
\gamma_n(n+\alpha_n+1)-(n+1-\alpha_n)
\bigr]p^{(\alpha_n)}\\
={}&
(n+1)\bigl[
1-\alpha_n+\gamma_n(n-\alpha_n)p^{(\alpha_n)}
\bigr],
\end{aligned}
\]
where the last equality uses
$\alpha_n\gamma_n=n+1-\alpha_n$.
Dividing by $n(n+1)$ and using the same identity gives
\[
\begin{aligned}
\lambda_n
&=
\frac{1-\alpha_n+
\gamma_n(n-\alpha_n)p^{(\alpha_n)}}{n}=
\gamma_n-1-
\gamma_n\frac{n-\alpha_n}{n}(1-p^{(\alpha_n)}).
\end{aligned}
\]
Combining this with \eqref{eq:asymptotic_deficit_product}, we obtain
\begin{equation}
\label{eq:asymptotic_boundary_product}
1+\lambda_n
=
\gamma_n\left[
1-\prod_{i=1}^{\alpha_n}
\left(1-\frac{\gamma_n}{D_{n,i}}\right)
\right].
\end{equation}

\paragraph{Step 2: Taking the limit as $n\to\infty$.}
Because $\gamma_n\to x>2$, we have
$\gamma_n/D_{n,i}=O(n^{-1})$ uniformly for
$1\leq i\leq\alpha_n$.
Using $\log(1-u)=-u+O(u^2)$, we obtain
\[
\begin{aligned}
\log\prod_{i=1}^{\alpha_n}
\left(1-\frac{\gamma_n}{D_{n,i}}\right)
&=
-\frac1{n+1}\sum_{i=1}^{\alpha_n}
\frac{\gamma_n}
{\gamma_n-2+(\gamma_n+2)i/(n+1)}
+O(n^{-1})\\
&\longrightarrow
-\int_0^{1/(x+1)}
\frac{x}{x-2+(x+2)t}\,dt=
\frac{x}{x+2}
\log\frac{(x-2)(x+1)}{x^2},
\end{aligned}
\]
where the limit follows from a Riemann sum and
$\alpha_n/(n+1)\to1/(x+1)$.
Substituting into
\eqref{eq:asymptotic_boundary_product} yields
\begin{equation}
\label{eq:asymptotic_boundary_limit}
1+\lambda_n
\longrightarrow
x\left[
1-\left(\frac{(x-2)(x+1)}{x^2}\right)^{\frac{x}{x+2}}
\right]
=
F(x).
\end{equation}

\paragraph{Step 3: Monotonicity and the asymptotic lower bound.}
\begin{lemma}
\label{lem:G(x)}
With the convention $F(\infty)\coloneq1$, the function
$F:[2,\infty]\longrightarrow[1,2]$ is continuous and strictly
decreasing, with $F(2)=2$.
\end{lemma}

\begin{proof}
For $x\geq2$, define $\delta(x)\coloneq(x+2)/x^2$.
Direct integration gives
\begin{equation}
\label{eq:asymptotic_threshold_integral}
F(x)
=
\int_0^1
\bigl(1-\delta(x)t\bigr)^{-\frac2{x+2}}\,dt.
\end{equation}
Both $\delta(x)$ and $2/(x+2)$ strictly decrease with $x$.
Consequently, the integrand strictly decreases with $x$ for
every $t\in(0,1)$, so $F$ is strictly decreasing on $[2,\infty)$.

Moreover, for every finite $x\geq2$ and $t\in(0,1)$,
\[
1<
\bigl(1-\delta(x)t\bigr)^{-\frac2{x+2}}
\leq
(1-t)^{-1/2}.
\]
Since $(1-t)^{-1/2}$ is integrable, dominated convergence
shows that $F$ is continuous on $[2,\infty)$ and that
$\lim_{x\to\infty}F(x)=1$.
Also,
\[
F(2)=\int_0^1(1-t)^{-1/2}\,dt=2.
\]
Since $F(x)>1$ for every finite $x\geq2$, the extension
$F(\infty)=1$ is continuous and preserves strict monotonicity.
\end{proof}

By Lemma~\ref{lem:G(x)}, for every $\lambda\in(0,1)$,
there exists a unique $\hat\gamma(\lambda)>2$ satisfying
\[
F(\hat\gamma(\lambda))=1+\lambda.
\]

Fix $\lambda\in(0,1)$ and any $c<\hat\gamma(\lambda)$.
Choose $\max\{2,c\}<x<\hat\gamma(\lambda)$.
Since $F$ is strictly decreasing and $x < \hat \rob(\lambda)$, we have
$F(x)>1+\lambda$.
By \eqref{eq:asymptotic_boundary_limit} and
$\gamma_n\to x$, for all sufficiently large $n$,
\[
\lambda<\lambda_n,
\qquad
c<\gamma_n.
\]
The finite-$n$ lower-bound theorem and strict monotonicity of
$\Phi^{(n)}$ therefore imply that every
$(1+\lambda)$-consistent algorithm for $n$ jobs has robustness
at least
\[
\Phi^{(n)}(\lambda)
>
\Phi^{(n)}(\lambda_n)
=
\gamma_n
>
c.
\]
Since $c<\hat\gamma(\lambda)$ was arbitrary, this proves the
asserted asymptotic lower bound.

Finally, convexity gives
\[
(1-u)^{-\frac2{x+2}}
\geq
1+\frac{2u}{x+2},
\qquad 0\leq u<1.
\]
Applying this inequality in
\eqref{eq:asymptotic_threshold_integral} yields
\[
F(x)
\geq
1+\frac{\delta(x)}{x+2}
=
1+\frac1{x^2}.
\]
Consequently,
\[
\lambda
=
F(\hat\gamma(\lambda))-1
\geq
\frac1{\hat\gamma(\lambda)^2},
\qquad
\hat\gamma(\lambda)\geq\frac1{\sqrt{\lambda}}.
\]
This completes the proof.
\end{proof}

\section{Proofs and Additional Details for Section~\ref{sec:ub}}

In this section, we present the proofs of the results in Section~\ref{sec:ub}.
\subsection{Proof of Theorem~\ref{thm:mTSS_conrob}}
\label{app:mTSS_conrob}

\begin{proof}
We first handle the case $\lambda=0$.
For $\lambda>0$, we establish consistency and then prove
robustness using Lemma~\ref{lem:detection_reduction}.

\paragraph{The endpoint $\lambda=0$.}
The calibration step gives $v=1$ and $r=0$, so Stage~1 has
zero duration.
When the predictions are accurate, Stage~2 processes the
jobs in nondecreasing order of their true processing times.
This order minimizes the total completion time, proving
that the algorithm is $1$-consistent.

For robustness, consider an arbitrary instance and let
$S=\sum_{i=1}^n x_i$.
Since the algorithm never idles while jobs remain unfinished,
every job completes by time $S$.
Moreover, $\opt(\mx)\geq S$, so
\[
\operatorname{ALG}(\mTSS \, ,\mx)
\leq n \cdot S
\leq n\cdot \opt(\mx)=\Psi_n(0)\cdot \opt(\mx).
\]
This proves the algorithm is $\Psi_n(0)$-robust. 

We henceforth assume $\lambda>0$ in the subsequent analysis..

\paragraph{Consistency.}
Assume that $\mathbf x=\mathbf y$, where the jobs are indexed so that
$0<y_1\le\cdots\le y_n$, and let
$\alg=\operatorname{ALG}(\mTSS,\mathbf y)$.
The offline optimum is
\[
\opt(\mathbf y)=\sum_{i=1}^n(n+1-i)y_i.
\]
We first record the cost incurred during Stage~1
(i.e., the $\rr$ stage; see Algorithm~\ref{alg:mTSS}),
which will also be useful in the robustness analysis.
For an index $v\in\{1,\ldots,n\}$ and a service level
$a\in[y_{v-1},y_v]$, with $y_0=0$, let $\mathcal R_v(a)$
be the cost accumulated by $\rr$ when every job $i$ has received
$\min\{y_i,a\}$ units of service.
During a service interval with $h$ unfinished jobs,
increasing their common service by $\delta$ takes
$h\delta$ time and contributes $h^2\delta$ to the holding cost.
Therefore,
\begin{align}
\mathcal R_v(a)
&\coloneq
\sum_{i=1}^{v-1}(n+1-i)^2(y_i-y_{i-1})
+(n+1-v)^2(a-y_{v-1})
\label{eq:RR-prefix-uniform}\\
&=
\sum_{i=1}^{v-1}(2n+1-2i)y_i+(n+1-v)^2a.
\label{eq:RR-prefix-telescoped}
\end{align}
Both formulas are valid for $v=1$: the sums are then empty
and $\mathcal R_1(a)=n^2a$.
Consequently, for $\ell\geq v$, if the algorithm next processes
jobs $v,\ldots,\ell$ in their predicted order, the total cost
accumulated up to the completion of job $\ell$ is
\begin{equation}
\mathcal R_v(a)+\sum_{i=v}^{\ell}(n+1-i)(y_i-a)
=\varGamma_\ell^{(v)}(\mathbf y,a).
\label{eq:Gamma-cost-identity}
\end{equation}

We then distinguish the following two cases.
If
$\varGamma_n^{(n)}(\mathbf y,y_n)<(1+\lambda)\opt(\mathbf y)$,
the algorithm runs $\rr$ to completion.
Taking $v=n$ and $a=y_n$ in
\eqref{eq:Gamma-cost-identity} gives
\[
\alg=\varGamma_n^{(n)}(\mathbf y,y_n)
<(1+\lambda)\opt(\mathbf y).
\]

Otherwise, $v$ is the first index for which
$\varGamma_n^{(v)}(\mathbf y,y_v)\geq(1+\lambda)\opt(\mathbf y)$,
and $r$ is the service level chosen by calibration.
Thus
\begin{equation}
\varGamma_n^{(i)}(\mathbf y,y_i)<(1+\lambda)\opt(\mathbf y)
\quad(i<v),
\qquad
\varGamma_n^{(v)}(\mathbf y,r)=(1+\lambda)\opt(\mathbf y).
\label{eq:calibration}
\end{equation}
Because
$\varGamma_n^{(n-1)}(\mathbf y,y_{n-1})
=\varGamma_n^{(n)}(\mathbf y,y_n)$, we have $v\leq n-1$.
Rearranging the definition of $\varGamma_n^{(v)}$ gives
\begin{equation}
\varGamma_n^{(v)}(\mathbf y,a)
=
\sum_{i=1}^{v-1}(2n+1-2i)y_i
+\sum_{i=v}^n(n+1-i)y_i
+\frac{(n+1-v)(n-v)}2a.
\label{eq:Gamma-affine}
\end{equation}
Since $v\leq n-1$, the coefficient of $a$ is positive.
Thus $\varGamma_n^{(v)}(\mathbf y,a)$ is strictly increasing in $a$.
The stopping rule gives $r\leq y_v$.
If $v=1$, then
$\varGamma_n^{(1)}(\mathbf y,0)=\opt(\mathbf y)
<(1+\lambda)\opt(\mathbf y)$, so $r>0=y_0$.
If $v\geq2$, the identity
\[
\varGamma_n^{(v-1)}(\mathbf y,y_{v-1})
=
\varGamma_n^{(v)}(\mathbf y,r)
-\frac{(n+1-v)(n-v)}2(r-y_{v-1})
\]
together with \eqref{eq:calibration} implies $r>y_{v-1}$.
Hence, uniformly for every $v\geq1$,
\[
y_{v-1}<r\leq y_v.
\]
Stage~1 completes jobs $1,\ldots,v-1$ and gives $r$ units
of service to each job indexed at least $v$; any such job
with $y_i=r$ also completes at the transition and contributes
a zero residual term.
By \eqref{eq:Gamma-cost-identity} and \eqref{eq:calibration},
\[
\alg
=\mathcal R_v(r)+\sum_{i=v}^n(n+1-i)(y_i-r)
=\varGamma_n^{(v)}(\mathbf y,r)
=(1+\lambda)\opt(\mathbf y).
\]
This proves $(1+\lambda)$-consistency.
\paragraph{Reduction of the robustness analysis.}
By Lemma~\ref{lem:detection_reduction}, it suffices to bound
$2n/(n+1)$ and the ratios
\[
\frac{\varGamma_s^{(v)}(\mathbf y,r)}
{\opt(\bar{\mathbf y}^{(s)})},
\qquad v\leq s\leq n,
\]
over all predictions and calibrated parameters $v,r$. Thus both types of prediction error reduce to the same
family of truncated vectors $\bar{\mathbf y}^{(s)}$.
Since $y_{v-1}\leq r\leq y_v$, their offline costs are
\begin{equation}
\label{eq:boundary-OPT}
\opt(\bar{\mathbf y}^{(s)})
=
\sum_{i=1}^{v-1}(n+1-i)y_i
+
\frac{(n-s)(n+s-2v+3)}2\,r
+
\sum_{i=v}^{s}(s+1-i)y_i.
\end{equation}
For fixed $v$ and $s$, both this expression and
$\varGamma_s^{(v)}(\mathbf y,r)$ are linear in
$(\mathbf y,r)$.
The remaining robustness analysis therefore reduces to
a linear-fractional optimization problem subject to the
ordering constraints and the calibration equality.

\paragraph{Reducing to two positive gaps.}
Fix $v$ and $s$, and write
\[
\mathbf w=(y_1,\ldots,y_{v-1},r,y_v,\ldots,y_n),
\qquad
u_i=w_i-w_{i-1}\quad(i\in[n+1]),
\qquad w_0=0.
\]
Since $w_i=\sum_{k=1}^i u_k$, both costs are linear in
$\mathbf u=(u_1,\ldots,u_{n+1})$.
Define
\[
N(\mathbf u)=\varGamma_s^{(v)}(\mathbf y,r),
\qquad
D(\mathbf u)=\opt(\bar{\mathbf y}^{(s)}),
\]
and
\[
H(\mathbf u)
=
\varGamma_n^{(v)}(\mathbf y,r)
-(1+\lambda)\opt(\mathbf y).
\]
Here $\mathbf y$ and $r$ are expressed in terms of
$\mathbf u$, and $H(\mathbf u)=0$ is precisely the
calibration equality.

To obtain an upper bound, we allow zero predictions
and replace the strict ordering inequalities by weak ones:
\[
0\leq y_1\leq\cdots\leq y_{v-1}
\leq r\leq y_v\leq\cdots\leq y_n,
\qquad H(\mathbf u)=0.
\]
When $v=1$, the chain starts with $0\leq r\leq y_1$.
Every nonzero feasible vector $\mathbf u$ has $r>0$.
Indeed, if $r=0$, then $y_i=0$ for $i<v$, and
\[
H(\mathbf u)=-\lambda\opt(\mathbf y)=0.
\]
Since $\lambda>0$, all predictions must vanish,
contradicting $\mathbf u\neq\mathbf0$.
Consequently, $D(\mathbf u)>0$ for every nonzero feasible
$\mathbf u$.

The ordering is equivalent to $u_i\ge0$.
Since $H$ is linear, calibration becomes a single equation
$h^\top\mathbf u=0$ for a fixed coefficient vector $h$.
We will show that it suffices to consider at most two
positive gaps.
The following lemma gives the required decomposition
and explains why it also preserves an upper bound on
the cost ratio.

\begin{lemma}[Restatement of Lemma~\ref{lem:gap-decomposition}]
\label{lem:gap-decomposition_restate}
Let $h\in\mathbb R^q$ and define the cone
$K=\{\mathbf{u}\in\mathbb R_{\ge0}^q:h^\top \mathbf{u}=0\}$.
For every $\mathbf{u}\in K\setminus\{\mathbf 0\}$, there exist an integer
$L\ge1$ and vectors
$\mathbf{u}^{(1)},\ldots,\mathbf{u}^{(L)}\in K\setminus\{\mathbf 0\}$ such that
\[
\mathbf{u}=\sum_{\ell=1}^L \mathbf{u}^{(\ell)},
\qquad
\bigl|\operatorname{supp}(\mathbf{u}^{(\ell)})\bigr|\le2
\quad\text{for every }\ell\in\{1,\ldots,L\},
\]
where $\operatorname{supp}(\mathbf{w})=\{i:w_i\ne0\}$.
Moreover, if $A,B:\mathbb R^q\to\mathbb R$ are linear and
$B(w)>0$ for every $w\in K\setminus\{\mathbf 0\}$, then
\[
\frac{A(\mathbf{u})}{B(\mathbf{u})}
\le
\max_{1\le\ell\le L}
\frac{A(\mathbf{u}^{(\ell)})}{B(\mathbf{u}^{(\ell)})}.
\]
\end{lemma}

\begin{proof}
Fix $\mathbf{u}\in K\setminus\{\mathbf 0\}$ and define
\[
\mathcal H_0=\{i:h_i=0\},\qquad
\mathcal H_+=\{i:h_i>0\},\qquad
\mathcal H_-=\{i:h_i<0\}.
\]
Since $h^\top \mathbf{u}=0$,
\[
T\coloneq\sum_{i\in\mathcal H_+}h_i u_i
=\sum_{j\in\mathcal H_-}(-h_j)u_j\ge0.
\]
Let $e_i$ denote the $i$th standard basis vector.
For each $i\in\mathcal H_0$, the vector $u_i e_i$ belongs to $K$
and has at most one nonzero coordinate.

If $T=0$, nonnegativity implies that $u_i=0$ for every
$i\in\mathcal H_+\cup\mathcal H_-$.
Hence
\[
\mathbf{u}=\sum_{i\in\mathcal H_0}u_i e_i,
\]
which gives the required decomposition after omitting zero vectors.

Suppose $T>0$. For each
$(i,j)\in\mathcal H_+\times\mathcal H_-$, define
\[
w^{ij}
\coloneq
\frac{(h_i u_i)(-h_j u_j)}{T}
\left(\frac{e_i}{h_i}+\frac{e_j}{-h_j}\right).
\]
Then we have $w^{ij}\ge\mathbf 0$ with $h^\top w^{ij}=0$ and $\operatorname{supp}(w^{ij})\subseteq\{i,j\}$.
To verify that these vectors recover all coordinates indexed by
$\mathcal H_+\cup\mathcal H_-$, expand and regroup the double sum:
\begin{align*}
\sum_{(i,j)\in\mathcal H_+ \times \mathcal H_-}w^{ij}
&=
\sum_{i\in\mathcal H_+}\sum_{j\in\mathcal H_-}
\left(
\frac{u_i(-h_j)u_j}{T}e_i
+
\frac{h_i u_i u_j}{T}e_j
\right)\\
&=
\sum_{i\in\mathcal H_+}
\frac{u_i}{T}
\left(\sum_{j\in\mathcal H_-}(-h_j)u_j\right)e_i
+
\sum_{j\in\mathcal H_-}
\frac{u_j}{T}
\left(\sum_{i\in\mathcal H_+}h_i u_i\right)e_j\\
&=
\sum_{i\in\mathcal H_+}u_i e_i
+
\sum_{j\in\mathcal H_-}u_j e_j.
\end{align*}
Since $\mathcal H_0,\mathcal H_+,\mathcal H_-$ partition
$\{1,\ldots,q\}$, it follows that
\[
\mathbf{u}=
\sum_{i\in\mathcal H_0}u_i e_i
+
\sum_{(i,j)\in\mathcal H_+ \times \mathcal H_-}w^{ij}.
\]
Omitting zero vectors and indexing the remaining vectors as
$\mathbf{u}^{(1)},\ldots,\mathbf{u}^{(L)}$ proves the first assertion.

Finally, by linearity,
\[
\frac{A(\mathbf{u})}{B(\mathbf{u})}
=
\frac{\sum_{\ell=1}^LA(\mathbf{u}^{(\ell)})}{B(\mathbf{u})}
=
\sum_{\ell=1}^L
\left(\frac{B(\mathbf{u}^{(\ell)})}{B(\mathbf{u})}\right)
\frac{A(\mathbf{u}^{(\ell)})}{B(\mathbf{u}^{(\ell)})}.
\]
The coefficients $B(\mathbf{u}^{(\ell)})/B(\mathbf{u})$ are positive and sum to one.
Therefore,
\[
\frac{A(\mathbf{u})}{B(\mathbf{u})}
\le
\max_{1\le\ell\le L}
\frac{A(\mathbf{u}^{(\ell)})}{B(\mathbf{u}^{(\ell)})},
\]
as claimed.
\end{proof}

Applying Lemma~\ref{lem:gap-decomposition} with $A=N$
and $B=D$, it suffices to consider at most two positive gaps.
We first compute $h_i=H(e_i)$.

For a unit gap at index $i\leq v$, we have $r=1$ and
exactly $c=n+1-i$ predictions equal to one.
For a unit gap at index $i>v$, we have $r=0$ and
exactly $z=n+2-i$ predictions equal to one.
Thus
\[
h_i=
\begin{cases}
\displaystyle
\frac c2\bigl(c-1-(c+1)\lambda\bigr),
& i\leq v,\quad c=n+1-i,\\[6pt]
\displaystyle
-\lambda\frac{z(z+1)}2,
& i>v,\quad z=n+2-i.
\end{cases}
\]

If all positive gaps have indices at most $v$, then
$y_v=\cdots=y_n=r$.
The truncated vector equals the prediction vector, and
\[
N(\mathbf u)
=
\varGamma_n^{(v)}(\mathbf y,r)
=
(1+\lambda)D(\mathbf u).
\]
These cases therefore have ratio $1+\lambda$.

Otherwise, feasibility requires two positive gaps of the form
\[
\mathbf u=a e_i+b e_j,
\qquad
i\leq v<j,
\qquad a,b>0,
\qquad H(\mathbf u)=0.
\]
Define $c=n+1-i$ and $z=n+2-j$.
The predictions then have the form
\[
\mathbf y=
\bigl(
\underbrace{0,\ldots,0}_{n-c\text{ copies}},
\underbrace{a,\ldots,a}_{c-z\text{ copies}},
\underbrace{a+b,\ldots,a+b}_{z\text{ copies}}
\bigr),
\qquad r=a,
\]
where $1\leq z\leq c\leq n$.
By linearity, calibration becomes
\begin{equation}
\label{eq:pair-calibration}
c\bigl(c-1-(c+1)\lambda\bigr)a
=
\lambda z(z+1)b.
\end{equation}
In particular, $c\geq2$ and
$\lambda<(c-1)/(c+1)$.

If $s\leq n-z$, the truncated vector has $c$ positive
processing times, all equal to $a$, and
\[
N(\mathbf u)=c^2a,
\qquad
D(\mathbf u)=\frac{c(c+1)}2a.
\]
Its ratio is $2c/(c+1)\leq2n/(n+1)$.

Otherwise, define $d=s-(n-z)\in[z]$.
Exactly $d$ processing times in the truncated vector
equal $a+b$, while the other positive processing times
equal $a$.
Substitution gives
\[
N(\mathbf u)
=
c^2a+\frac{d(2z+1-d)}2b,
\qquad
D(\mathbf u)
=
\frac{c(c+1)}2a+\frac{d(d+1)}2b.
\]

\paragraph{Reduction to a one-dimensional maximization.}
Define
\begin{equation}
\label{eq:ratio-parameters}
\rho_c=\frac{2c}{c+1},
\qquad
\tau_c=\frac{c-1}{\lambda(c+1)}-1,
\end{equation}
and, for $1\leq d\leq z$,
\begin{equation}
\label{eq:UV-coefficients}
U_{z,d}=\frac{d(2z+1-d)}{z(z+1)},
\qquad
V_{z,d}=\frac{d(d+1)}{z(z+1)}.
\end{equation}
Notice that $0<V_{z,d}\leq U_{z,d}\leq1$.
By \eqref{eq:pair-calibration},
\[
\frac ba=\frac{c(c+1)\tau_c}{z(z+1)},
\]
so the remaining ratios are
\begin{equation}
\label{eq:common-ratio}
R_{c,z,d}
\coloneq
\frac{N(\mathbf u)}{D(\mathbf u)}
=
\frac{\rho_c+\tau_c U_{z,d}}
{1+\tau_c V_{z,d}}.
\end{equation}
We bound these ratios through two comparisons.

\emph{First comparison: increasing $c$ to $n$.}
Fix $z,d$, and write $U=U_{z,d}$ and $V=V_{z,d}$.
The identity $\rho_c=1+\lambda+\lambda\tau_c$ gives
\begin{equation}
\label{eq:dimension-monotonicity}
R_{c,z,d}
=
1+\lambda+
\frac{\tau_c\bigl(\lambda(1-V)+U-V\bigr)}
{1+\tau_c V}.
\end{equation}
Since $0<V\leq U\leq1$, the coefficient
$\lambda(1-V)+U-V$ is nonnegative.
Moreover, $\tau_c$ increases with $c$, and
$t/(1+tV)$ increases for $t\geq0$.
Therefore,
\[
R_{c,z,d}\leq R_{n,z,d}.
\]

\emph{Second comparison: increasing $z$ to $n$.}
For $z<n$, define $\theta=d/z\in(0,1]$.
Direct substitution gives
\[
(U_{z,d},V_{z,d})
=
(1-\theta)(U_{z+1,d},V_{z+1,d})
+\theta(U_{z+1,d+1},V_{z+1,d+1}).
\]
Consequently, $R_{n,z,d}$ is a weighted average of
$R_{n,z+1,d}$ and $R_{n,z+1,d+1}$, with weights proportional
to their positive denominators.
Hence
\[
R_{n,z,d}
\leq
\max\{R_{n,z+1,d},R_{n,z+1,d+1}\}.
\]
Iterating increases $z$ to $n$ and yields
\[
R_{c,z,d}\leq\max_{s\in[n]}R_{n,n,s}.
\]

This maximum also dominates the remaining ratios:
$1+\lambda\leq\rho_n$, and $R_{n,n,1}$ is a weighted
average of $\rho_n$ and
$U_{n,1}/V_{n,1}=n\geq\rho_n$.
Thus $R_{n,n,1}\geq\rho_n=2n/(n+1)$.

Finally, substituting $c=z=n$ and $d=s$ into
\eqref{eq:common-ratio} gives
\begin{equation}
\label{eq:one-dimensional-bound}
\frac{\operatorname{ALG}(\mTSS\, ,\mx)}{\opt(\mx)}
\leq
\max_{s\in[n]}
\frac{
2\lambda n^2
+\bigl(\tfrac{n-1}{n+1}-\lambda\bigr)s(2n+1-s)
}{
\lambda n(n+1)
+\bigl(\tfrac{n-1}{n+1}-\lambda\bigr)s(s+1)
}
=
\Psi_n(\lambda).
\end{equation}
This completes the robustness proof.
\end{proof}

\subsection{Proof of Theorem~\ref{thm:structural-three-halves}}\label{appendix:approximation}
\begin{proof}
At $\lambda=0$ both bounds equal $n$, and at
$\lambda=(n-1)/(n+1)$ both equal $2n/(n+1)$.
It remains to consider $0<\lambda<(n-1)/(n+1)$.
We write $t=\frac{n+1}{n-1}\lambda\in(0,1)$.
By Theorem~\ref{thm:mTSS_conrob},
\[
\Psi_n(\lambda)=\max_{i\in[n]}U_i(t),
\qquad
U_i(t)=
\frac{
2tn^2+(1-t)i(2n+1-i)
}{
tn(n+1)+(1-t)i(i+1)
}.
\]
Let $s$ be the smallest index attaining this maximum.

Let $(\Phi^{(n)}(\lambda), p^{(1)},\ldots,p^{(\alpha)})$ be the solution to the corresponding Order-$\alpha$ system,
and extend it by setting $p^{(j)}=p^{(\alpha)}$
for $\alpha<j\le n-1$.
The sequence is nondecreasing, and the consistency budget equation (see \eqref{eq:order_alpha_2}) yields
\[
\sum_{j=1}^{n-1}(n-j)p^{(j)}=
\sum_{j=1}^{\alpha-1}(n-j)p^{(j)}
+\frac{(n+1-\alpha)(n-\alpha)}2p^{(\alpha)}
=
\frac{n(n+1)}2 \lambda
=
\frac{n(n-1)}2t.
\]
Thus $p^{(\alpha)}\ge t$.
Combining the $\alpha$-th attack constraint equation \eqref{eq:order_alpha_1} with the consistency budget equation \eqref{eq:order_alpha_2}
gives
\[
\begin{aligned}
\Phi^{(n)} (\lambda) \Bigl[
\alpha(\alpha+1)
+(n-\alpha)(n+\alpha+1)p^{(\alpha)}
\Bigr]
={}&
\alpha(2n+1-\alpha)+tn(n-1)
+(n-\alpha)(n-\alpha+1)p^{(\alpha)}.
\end{aligned}
\]
Since $\Phi^{(n)} (\lambda)>1$ and $p^{(\alpha)}\ge t$, it follows that
\[
\Phi^{(n)} (\lambda)
\le
\frac{
\alpha(2n+1-\alpha)
+t\bigl[n(n-1)+(n-\alpha)(n-\alpha+1)\bigr]
}{
\alpha(\alpha+1)+t(n-\alpha)(n+\alpha+1)
}
=
U_\alpha(t).
\]
Therefore,
\[
\Phi^{(n)} (\lambda)  \leq U_\alpha (t) \leq \max_{i \in [n]} U_i (t) = \Psi_n (\lambda).
\]
This proves the first inequality.

For the second inequality,
we first show that $s\leq \alpha$. The selection rule of $\alpha$ (see \eqref{choose_alpha}) gives 
\[
\Phi^{(n)} (\lambda) \ge\frac{n-\alpha}{\alpha+1}.
\]

The numerator and denominator increments from $U_i$ to
$U_{i+1}$ have ratio $(n-i)/(i+1)$,
which strictly decreases with $i$.
Consequently, $U_{i+1}$ is a weighted average of $U_i$ and
$(n-i)/(i+1)$.
Once $U_i\ge(n-i)/(i+1)$, the sequence is nonincreasing
thereafter.
Since
\[
U_\alpha\ge\Phi^{(n)} (\lambda)\ge\frac{n-\alpha}{\alpha+1},
\]
we conclude that $s\le\alpha$.

The $s$-th attack constraint equation \eqref{eq:order_alpha_1} is therefore available, which yields
\begin{equation}
\label{Phi_n>=}
\Phi^{(n)} (\lambda)
\ge
\frac{
s(2n+1-s)+2(n-s)(n-s+1)p^{(s)}
}{
s(s+1)+(n-s)(n+s+1)p^{(s)}
}.
\end{equation}
Moreover, monotonicity of the extended sequence implies
\[
t\frac{n(n-1)}2
\ge
\sum_{j=s}^{n-1}(n-j)p^{(j)}
\ge
\frac{(n-s)(n-s+1)}2\,p^{(s)}.
\]
Equivalently, $p^{(s)} \leq \frac{n(n-1)}{(n-s)(n-s+1)}t$.
The derivative of the right-hand side of \eqref{Phi_n>=}
with respect to $p^{(s)}$ has the same sign as
\[
-s(n-s)\bigl(2n^2-ns+n+s^2-1\bigr)<0.
\]
Thus the right-hand side of \eqref{Phi_n>=} is decreasing with respect to $p^{(s)}$. It follows that
\[
\Phi^{(n)}(\lambda)\ge L_s(t)
\coloneq
\frac{
s(2n+1-s)+2tn(n-1)
}{
s(s+1)+
\frac{tn(n-1)(n+s+1)}{n-s+1}
}.
\]

Suppose first that $s>(n+3)/4$.
Then $s\ge2$, and maximality gives $U_s\ge U_{s-1}$.
The weighted-average observation above implies
\[
\Psi_n(\lambda) =
U_s\le\frac{n+1-s}{s}.
\]
Combining this with $\Phi^{(n)}(\lambda)\ge 2n/(n+1)$ gives
\[
\frac{\Psi_n(\lambda)}{\Phi^{(n)}(\lambda)}
\le
\frac{(n+1)(n+1-s)}{2ns}
<
\frac{(n+1)(3n+1)}{2n(n+3)}
\le\frac32.
\]

Now suppose that $s\le(n+3)/4$.
If $s=1$, then $U_s=L_s$, so the claim follows.
For $s\ge2$, define
\[
a\coloneq\frac{n(n-1)}{(n-s)(n-s+1)}-1,
\quad
q\coloneq\frac{n(n-1)(s+1)}
{(2n+1-s)(n-s)(n+s+1)},
\quad
w\coloneq\frac{(n-s)(n+s+1)}{s(s+1)}t.
\]
Define $r\coloneq(s-1)/(n-1)\in(0,1/4]$.
Then
\[
1+a
=
\frac{n}
{(1-r)((1-r)(n-1)+1)}
\le\frac1{(1-r)^2},
\]
so
\[
a\le\frac{r(2-r)}{(1-r)^2}\le\frac79.
\]
Also,
\[
q=
\frac{n(r(n-1)+2)}
{(1-r)((2-r)(n-1)+2)((1+r)(n-1)+3)}
\ge
\frac{r}{(2-r)(1-r^2)}.
\]
Indeed, after cross-multiplication, the difference is
\[
(4-4r-r^3)(n-1)+4-4r-2r^2\ge0.
\]
Since
\[
(2-r)^2(1+r)=4-3r^2+r^3\le4,
\]
we obtain
\[
q
\ge
\frac{1-r}{(2-r)^2(1+r)}
\frac{r(2-r)}{(1-r)^2}
\ge
\frac3{16}\frac{r(2-r)}{(1-r)^2}
\ge\frac a6.
\]
Therefore, we have
\[
0<a\le\frac79,
\qquad
q\ge\frac a6.
\]

Direct simplification now gives
\[
\frac{U_s(t)}{L_s(t)}
=
1+
\frac{aw\bigl(1-q/(1+a)+qw\bigr)}
{(1+2qw)(1+w)}.
\]
Thus $U_s/L_s\le3/2$ is equivalent to
\[
1+
\left[
1-2a+2q\left(1+\frac a{1+a}\right)
\right]w
+
2q(1-a)w^2
\ge0.
\]
If $a\le1/2$, all coefficients are nonnegative.
Otherwise, $1/2\le a\le7/9$, so $q\ge1/12$ and
$a/(1+a)\ge1/3$. Therefore,
\[
1-2a+2q\left(1+\frac a{1+a}\right)\ge-\frac13,
\qquad
2q(1-a)\ge\frac1{27}.
\]
The left-hand side is consequently at least
\[
1-\frac w3+\frac{w^2}{27}
=
\frac{(w-9/2)^2}{27}+\frac14
>0.
\]
Combining this with $\Phi^{(n)}(\lambda)\ge L_s(t)$ yields
\[
\frac{\Psi_n(\lambda)}{\Phi^{(n)}(\lambda)}
=
\frac{U_s(t)}{\Phi^{(n)}(\lambda)}
\le
\frac{U_s(t)}{L_s(t)}
\le\frac32,
\]
completing the proof.
\end{proof}

\subsection{Proof of Theorem~\ref{thm:analytic_upper}}
\label{app:analytic_upper}
\begin{proof}
Fix $0<\lambda\leq(n-1)/(n+1)$ and define
\[
R\coloneq{\left(\tfrac12+\sqrt{{\tfrac54}+1/\lambda}\right)}.
\]
By Theorem~\ref{thm:mTSS_conrob}, it suffices to prove
$\Psi_n(\lambda)\leq R$.
Note that $R>2$ and
\[
\lambda(R-2)=\frac{1-\lambda}{R+1}.
\]
For every $s\in[n]$, 
\[
\begin{aligned}
s(2n+1-s)-Rs(s+1)
&=2ns-(R+1)s^2-(R-1)s\leq 2ns-(R+1)s^2\\
&=\frac{n^2}{R+1}
-(R+1)\left(s-\frac{n}{R+1}\right)^2\leq\frac{n^2}{R+1}.
\end{aligned}
\]
Since $0\leq\tfrac{n-1}{n+1}-\lambda\leq1-\lambda$,
we obtain
\[
\begin{aligned}
&\left(\frac{n-1}{n+1}-\lambda\right)
\bigl[s(2n+1-s)-Rs(s+1)\bigr]\\
&\qquad\leq
\frac{(1-\lambda)n^2}{R+1}
=\lambda n^2(R-2)
\leq R\lambda n(n+1)-2\lambda n^2.
\end{aligned}
\]
Rearranging and dividing by the positive denominator shows
that every ratio in the maximization defining $\Psi_n(\lambda)$
is at most $R$.
Thus $\Psi_n(\lambda)\leq R$, proving the robustness guarantee.
Consistency follows from Theorem~\ref{thm:mTSS_conrob}.
\end{proof}

\subsection{Proof of Theorem~\ref{thm:tolerance_simulation}} \label{app:tolerance_simulation_proof}

\begin{proof}
Define the virtual processing times
\[
\widehat x_i\coloneq
\begin{cases}
x_i,
& x_i<y_i/(1+\tau),\\[2pt]
y_i,
& y_i/(1+\tau)\leq x_i\leq(1+\tau)y_i,\\[2pt]
x_i-\tau y_i,
& x_i>(1+\tau)y_i.
\end{cases}
\]
By construction, a job below the tolerance interval is
reported as completed at $s_i=x_i$, whereas a job within
the interval is reported as completed at $s_i=y_i$.
For a job above the interval, the additional $\tau y_i$
units of processing are not counted in $s_i$, so its
completion is reported at $s_i=x_i-\tau y_i$.
Thus, in every case, job $i$ is reported as completed
at simulated service $\widehat x_i$.
Consequently, $\mA$ receives exactly the same completion
feedback as in its execution on
$(\widehat{\mathbf x},\mathbf y)$.

Let $\widehat C_i$ denote the completion time of job $i$
under $\mA$ on $(\widehat{\mathbf x},\mathbf y)$, and let
$C_i^{\mathrm{tol}}$ denote its completion time under
$\mA^\tau_{\mathrm{tol}}$ on $I=(\mathbf x,\mathbf y)$.
We first compare these completion times.

Fix a job $i$, and consider $\mA$'s execution on
$(\widehat{\mathbf x},\mathbf y)$ up to its completion
time $\widehat C_i$.
Let $\mathcal T_i$ denote the set of jobs $j$ to which
$\mA$ has allocated at least $y_j$ units of processing
by time $\widehat C_i$ in its execution on
$(\widehat{\mathbf x},\mathbf y)$.
Since $\mA$ can perform at most $\widehat C_i$ units
of processing by time $\widehat C_i$, we have $\sum_{j\in\mathcal T_i}y_j\leq\widehat C_i$.
To follow this part of $\mA$'s execution,
$\mA^\tau_{\mathrm{tol}}$ takes $\widehat C_i$ time,
plus the additional processing performed while $\mA$ is paused.
Only jobs in $\mathcal T_i$ can require such additional
processing, and each receives at most $\tau y_j$.
After following this part of $\mA$'s execution and performing
the associated additional processing,
$\mA^\tau_{\mathrm{tol}}$ has completed job $i$.
Therefore,
\[
C_i^{\mathrm{tol}}
\leq
\widehat C_i+\tau\sum_{j\in\mathcal T_i}y_j
\leq
(1+\tau)\widehat C_i.
\]
Summing over all jobs and using the robustness of $\mA$,
we obtain
\[
\begin{aligned}
\operatorname{ALG}(\mA^\tau_{\mathrm{tol}},I)
&\leq(1+\tau)
\operatorname{ALG}(\mA,(\widehat{\mathbf x},\mathbf y))\leq(1+\tau)\gamma
\operatorname{OPT}(\widehat{\mathbf x})\leq(1+\tau)^2\gamma
\operatorname{OPT}(\mathbf x),
\end{aligned}
\]
where the last inequality follows from
$\widehat x_i\leq(1+\tau)x_i$ for every $i$.

Now suppose $\varepsilon\leq\tau$.
Then $\widehat{\mathbf x}=\mathbf y$, so the execution
observed by $\mA$ is its execution under accurate predictions.
Moreover, each additional-processing phase takes at most
$\varepsilon y_i$ time.
The preceding time comparison therefore holds with
$1+\varepsilon$ in place of $1+\tau$.
Using $\beta$-consistency of $\mA$ and
$y_i\leq(1+\varepsilon)x_i$, we obtain
\[
\begin{aligned}
\operatorname{ALG}(\mA^\tau_{\mathrm{tol}},I)
&\leq(1+\varepsilon)
\operatorname{ALG}(\mA,(\mathbf y,\mathbf y))\leq\beta(1+\varepsilon)
\operatorname{OPT}(\mathbf y)\leq\beta(1+\varepsilon)^2
\operatorname{OPT}(\mathbf x).
\end{aligned}
\]
Taking $\varepsilon=0$ proves $\beta$-consistency.
\end{proof}


\end{document}